\documentclass[letter]{statsoc}

\usepackage{amsmath}
\usepackage{amssymb}
\usepackage[dvips]{epsfig}
\usepackage[dvips]{graphics}
\usepackage{graphicx}
\usepackage{dsfont}
\usepackage{natbib}
\usepackage{graphicx}
\usepackage{xcolor}
\usepackage{tabularx}

\newcommand{\balpha}{\mbox{\boldmath $\alpha$}}
\newcommand{\bcdot}{\mbox{\boldmath $\cdot$}}

\newcommand{\bd}{\mbox{\boldmath $d$}}

\title[Adversarial and Amiable Inference in Medical Diagnosis, Reliability, and Survival Analysis]{Adversarial and Amiable Inference in Medical Diagnosis, \\Reliability, and Survival Analysis}
\author{Nozer D. Singpurwalla
\footnote{{\it {Address for correspondence}}: Nozer D. Singpurwalla,
Department of Systems Engineering and Engineering Management, The City University of Hong Kong, Tat Chee Avenue, Kowloon, Hong Kong.  
E-mail: {\tt {nsingpur@cityu.edu.hk}}}}
\address{City University of Hong Kong, Hong Kong.}
\author{Barry C. Arnold}
\address{University of California, Riverside, California, USA.}
\author{Joseph L. Gastwirth}
\address{The George Washington University, Washington D.C., USA.}
\author{Anna S. Gordon}
\address{Mindjet Corporation, San Francisco, California, USA.}
\author[Singuprwalla {\it et al.}]{Hon Keung Tony Ng}
\address{Southern Methodist University, Dallas, Texas, USA.}

\begin{document}
\maketitle
\begin{abstract}
In this paper, we develop a family of bivariate beta distributions that encapsulate both positive and negative correlations, and which can be of general interest for Bayesian inference. We then invoke a use of these bivariate distributions in two contexts. The first is diagnostic testing in medicine, threat detection, and signal processing. The second is system survivability assessment, relevant to engineering reliability, and to survival analysis in biomedicine. 
In diagnostic testing one encounters two parameters that characterize the efficacy of the testing mechanism, {\it test sensitivity}, and {\it test specificity}. These tend to be adversarial when their values are interpreted as utilities. In system survivability, the parameters of interest are the component reliabilities, whose values when interpreted as utilities tend to exhibit co-operative (amiable) behavior. 
Besides probability modeling and Bayesian inference, this paper has a foundational import. Specifically, it advocates a conceptual change in how one may think about reliability and survival analysis. The philosophical writings of de Finetti, Kolmogorov, Popper, and Savage, when brought to bear on these topics constitute the essence of this change. Its consequence is that we have at hand a defensible framework for invoking Bayesian inferential methods in diagnostics, reliability, and survival analysis. Another consequence is a deeper appreciation of the judgment of independent lifetimes. Specifically, we make the important point that independent lifetimes entail at a minimum, a two-stage hierarchical construction. 
\end{abstract}

\setcounter{section}{-1}
\section{Introduction: Motivation and Overview}

The work described here is motivated by two scenarios, one from the medical sciences, and the other from systems science. A resolution of the problems posed by these scenarios entails probability modeling and statistical inference, the former spawned by the needs of the latter. This paper demonstrates an interplay between these two specialties as they come together to constitute the spirit of {\it statistical science}. 


In medical diagnosis, and also matters of national security -- like threat detection and signal processing -- testing for abnormalities using diagnostic instruments plays a key role. Diagnostic tests are characterized by two parameters, {\it test sensitivity}, $\eta$, and {\it test specificity}, $\theta$. There are different ways to interpret these parameters, and interpretation dictates methodology. We interpret $\eta$ and $\theta$ as the degree of {\it propensity} [or {\it chance}] in the sense of \cite{popp59} [\cite{defi37}], and not as probabilities, as is currently done. 

Specifically, $\eta$ is to be seen as the ``tendency" of a diagnostic instrument declaring a diseased individual as diseased, and $\theta$ as the ``tendency" of the instrument declaring a healthy person as healthy. 
Looking at $\eta$ and $\theta$ as propensities provides a proper foundation for making Bayesian inferences about these parameters via probability statements, where probability to us is a two-sided bet in the sense of \cite{defi37}. Were $\eta$ and $\theta$ be interpreted as probabilities, and probability interpreted as a two-sided bet, then endowing a prior probability to $\eta$ and $\theta$ for the purpose of Bayesian inference is tantamount to assigning personal probabilities to a personal probability, and this could be a matter of debate, if not flawed. 

Irrespective of interpretation, one would want diagnostic instruments with $\eta$ and $\theta$ equal to one. However, such instruments are expensive to build, and thus one settles for $\eta$ and $\theta$ being close to one. But a caveat here is that $\eta$ and $\theta$ are {\it adversarial} in the sense that an increase in $\eta$ leads to a decrease in $\theta$, and mutatis-mutandis. Indeed, were one to equate the value of a propensity with utility, then the behavior of the two parameters can be likened to a two-person zero-sum game; thus our use of the term {\it adversarial inference}.  

Noting the fact that $(1 - \theta)$ and $(1 - \eta)$ are akin to the Type-I and Type-II errors of a Neyman-Pearson test of a hypothesis, one may still be inclined to interpret $\eta$ and $\theta$ as probabilities, and declare propensity as another label for probability. However, there is an operational difference. In the Neyman-Pearson scheme, one designs tests with specified errors, whereas a diagnostic test's specificity and sensitivity are the consequences of an instrument's engineered design, which a user likes to estimate. In a sense sensitivity and specificity are more akin to the $p$-values of Fisher's significance test. The difference is that whereas $p$-values help validate a hypothesis, inferences about $\eta$ and $\theta$ help assess the efficacy of an instrument. The underlying message here is that in the context of diagnostics, interpreting $\eta$ and $\theta$ as propensities and not as probabilities, makes sense because propensities encapsulate the physical characteristics of an instrument, and making inferences about them is a meaningful endeavor. However, in the context of testing hypotheses, such parameters are rightly interpreted as probabilities, because such probabilities encapsulate the consequences of ones inductive behavior in the sense of \cite{neym57}.  

Moving to systems science, assessing the {\it survivability} of multi-component systems -- like networks or a biological organism, is of importance. This is because survivability is a key ingredient of a system's performance, where survivability is ones personal probability of an item's propensity to survive \citep[c.f.][]{lisi02}. It is often the case that the components of complex systems experience commonalities with respect to attributes like materials, manufacture, the operating environment, etc. A consequence is that propensity to survive of each of the system's components experience a positive dependence. That is, the propensities are {\it amiable} in the sense that were the values of such propensities viewed as utilities  \citep{sing09}, then the propensities behave as if they are engaging in a two-person co-operative game. Assessing a system's survivability under the premise of such co-operative behavior, motivates us to label inferences associated with such scenarios as {\it amiable inference}. Even though our motivation for amiable inference is systems science, we foresee other scenarios wherein positively dependent (unknown) parameters on the unit square may arise. 

For statistical inference under the premise of adversarial or amiable parameters, a Bayesian approach seems natural. But to invoke this approach, suitable probability models that encapsulate the game theoretic character of the parameters are needed. There appears to be a dearth of  such models, and a purpose of this paper is to fill this void. 
One such family of models is proposed here in Section 2. But before discussing these models, a perspective on the notion of propensity and its relevance to reliability and survival analysis is appropriate, and this is done in Section 1 below. 

The rest of the paper is devoted to invoking the mechanics of the ideas presented in Sections 1 and 2 to the specific problems mentioned above. 

\section{Foundational Issues in Reliability and Risk Analysis} 

The relevance of the notion of propensity in the context of diagnostics has been discussed in the previous section of this paper. Here, we discuss its appropriateness in the contexts of reliability theory and survival analysis. There is much precedence to what we say below dating back to the writings of de Finetti, Jeffreys, Kolmogorov, Popper, and Savage.   Relating the essence of these writings to the above disciplines is a contribution of this paper to the foundational aspects of reliability, risk, and survival analysis. 

The conventional approach for assessing the performance of an item is based on the metric of {\it reliability}, where reliability is defined as a probability. This is also true of survival analysis. But there are many ways to interpret probability and this is where the foundational issues matter. Irrespective of interpretation, probability models entail unknown parameters that are estimated using frequentist or Bayesian methods. In the paragraph which follows, we make a case for the latter based on relevance to a user. With Bayesian methods, the unknown parameters are endowed with a prior distribution, and the prior to posterior conversion made via Bayes' Law. A justification for introducing priors and invoking Bayes' Law is de Finetti's celebrated theorem on exchangeable sequences, and without the judgment of exchangeability, inductive inference is not possible. Inherent to de Finetti's theorem is the notion of chance (or propensity), and the essence of the theorem is the existence of a personal probability on this chance. Thus appears the role of propensity in reliability and survival analysis in particular, and more generally, in {\it applied probability} as well. 

To make a case for using Bayesian methods in the contexts of reliability and survival analysis, we start with the claim that it is an item's {\it survivability} - not its reliability - that should be the metric of performance; see \citet[p.289]{sing06}. This is because survivability can be operationalized via a two-sided bet, whereas reliability, which entails unknown parameters, cannot be operationalized. Reliability is better interpreted as a {\it propensity} (or {\it chance}). Conversationally, by propensity we mean a tendency, or a dispositional property of the physical world which manifests itself as a relative frequency: see \cite{popp59} or Pierce [in \cite{mill75}]; also \citet[p.230]{kolm69}. Despite Popper's endless efforts to equate propensity with probability, it has been argued that propensity is not a probability \cite[c.f.][]{hump85}. Probability, to \citet[][p.239]{kolm69} is an undefined primitive, to \citet[p.51]{jeff61} is a logical argument, to \cite{defi37} is a disposition towards a two-sided bet, and to \cite{sava54} is a strength of belief. Because notions like disposition and tendency are metaphysical, the term propensity, like {\it experience}, may best be seen as an undefined primitive. Support for this point of view is borne out in actuality, because layperson and even engineers are able to conceptualize and talk about reliability on intuitive grounds without an appreciation of its mathematical definition \cite[c.f.][]{sing01}. The distinction between probability and propensity is best exposited in de Finetti's theorem on exchangeability mentioned before, wherein one's uncertainty about propensity is encapsulated via a personal probability (which can be operationalized). 

The above lines of thinking motivate us to suggest a paradigm change in reliability and survival analysis, by proposing that an item's reliability (as an undefined primitive) be interpreted as its tendency to survive (under specified conditions for a specified period of time), and that its survivability as ones personal probability about this propensity. Survivability being devoid of conditioning parameters is more appealing to a user because its operationalization via a two-sided bet induces transparency. Because survivability is obtained by averaging out reliability over its underlying parameters, reliability now becomes a stepping stone to survivability, and it is survivability that conveys an actionable import. The proposed shift in focus from reliability to survivability constitutes a perspective change in the assurance sciences. 

But interpreting reliability (with its unknown parameters) as a propensity does something more fundamental, at least from a conceptual point of view. It paves the path for invoking de Finetti's theorem, and in so doing provides a proper foundation for using Bayesian inferential and decision theoretic methods in reliability and survival analysis. Since the initial writings of  \cite{corn69}, \cite{arja93}, \cite{bape93}, and \cite{spiz93}, such methods have continued to gain much traction. Also noteworthy are several contributions in the monograph edited by \cite{bacl93}.

\section{A Family of Bivariate Distributions on the Unit Square} 

The diagnostic testing scenario of the introductory section spawns the need for a family of bivariate distributions on the unit square with negative dependence. By contrast, the system survivability scenario calls for multivariate distributions on the unit hypercube with positive dependence. Such multivariate distributions are in general difficult to construct, especially if the number of parameters is to be kept down to a minimum. Consequently, we restrict attention to bivariate distributions on the unit square with either positive and/or negative dependence. There is a price to be paid for doing so because it is unclear to us as to how a pairwise modularization of a multi-component system can be constituted to make inference about the entire system. All the same, a consideration of the bivariate case is instructional, and could serve as a first step towards addressing the multi-component case. 




An archetypal strategy for constructing bounded bivariate distributions is the one adopted by \cite{olli03} -- henceforth $(OL)^{+}$, and also by \cite{arng11} -- henceforth $AN(n)$. Here, one starts with $n$ independent gamma distributed random variables $U_{i}$, with a scale (shape) parameter 1 ($\alpha_{i}$), $i = 1, 2, \ldots, n$. The $(OL)^{+}$ construction has $n = 3$; it starts by considering the bivariate random variable $(X, Y)$, where  
\begin{eqnarray*}
X \stackrel{def.}{=} \frac{U_{1}}{U_{1} + U_{3}}~~{\mbox {and }} Y \stackrel{def.}{=} \frac{U_{2}}{U_{2} + U_{3}}.
\end{eqnarray*}
The distribution of $X [Y]$ is a beta (distribution of the first kind) on $[0, 1]$ with parameters $(\alpha_{1}, \alpha_{3})$ [$(\alpha_{2}, \alpha_{3})$]; it is denoted ${\cal B}(\alpha_{1}, \alpha_{3})$ [${\cal B}(\alpha_{2}, \alpha_{3})$]. The joint distribution of $(X, Y)$ is a bivariate beta on the unit square with a positive correlation, covering the entire range of 0 and 1. \cite{jone01} has also obtained this distribution but via a different line of construction. Other examples of bivariate distributions with beta marginals appear in \cite{guwo85}, \cite{nako05}, \cite{nada09}, \linebreak \cite{bala09} and \cite{guor11}. Like the bivariate beta distribution of Jones-Olkin-Liu, these distributions also exhibit positive dependence. Whereas such distributions are suitable for what we have labeled amiable inference, adversarial inference requires bivariate distributions with negative dependence. The well-known bivariate Dirichlet distribution has a negative correlation and beta marginals, but the range of possible values that the variables take is restricted, because of the requirement that they sum to one. Its suitability as a vehicle for inference in diagnostic testing is therefore limited. Fortunately, by a simple {\it complementation} of one of the variables in the $(OL)^{+}$ construction, we can obtain a bivariate beta distribution on the unit square with a negative correlation. We denote this distribution  $(OL)^{-}$; the superscripts $+ (-)$ associated with the  $(OL)$ style constructions denote positive (negative) correlation. To summarize, the bivariate random variable $(X, 1-Y)$ [or equivalently $(1-X, Y)$] has the  $(OL)^{-}$ distribution, and this distribution is suitable candidate for the adversarial inference needs of diagnostic testing. 

Since $(1 - Y) = U_{3}/(U_{2} + U_{3})$, we are motivated to consider expanded versions of the  $(OL)$ strategy to construct bivariate beta distributions on the unit square with both positive and negative correlations. This is indeed the motivation behind the \cite{arng11}, $AN(5)$, construction for which $n = 5$. Here, 
\begin{eqnarray*}
X_{1} \stackrel{def.}{=} \frac{U_{1} + U_{3}}{U_{1} + U_{3} + U_{4} + U_{5}}~~{\mbox {and }} Y_{1} \stackrel{def.}{=}  \frac{U_{2} + U_{4}}{U_{2} + U_{3} + U_{4} + U_{5}},
\end{eqnarray*}
from which it follows that $X_{1} \sim {\cal B}(\alpha_{1} + \alpha_{3}, \alpha_{4} + \alpha_{5})$ and $Y_{1} \sim {\cal B}(\alpha_{2} + \alpha_{4}, \alpha_{3} + \alpha_{5})$. Furthermore, by a judicious choice of $\alpha_{i}$, $i = 1, \ldots, 5$, both positive and negative correlations can be induced. 

The $AN(5)$ distribution introduced above has some attractive properties. For example, with $\alpha_{1} = \alpha_{2} = 0$, it yields the bivariate Dirichlet distribution of \linebreak \cite{bala09}, and for $\alpha_{3} = \alpha_{4} = 0$, it collapses to the $(OL)^{+}$ distribution. A drawback of the $AN(5)$ construction is that it does not encompass the $(OL)^{-}$ distribution, nor will it encompass, for $n = 3$, constructions based on complementation of the type $(1- X, Y)$ and $(1-X, 1-Y)$. Such lack of closure motivates us to seek a further expansion of the $AN(5)$ construction in the hope of generating a more encompassing family of bivariate beta distributions. 
This topic is delegated to the Appendix, with the comment that closure properties are relevant to Bayesian inference because they provide flexibility regarding the choice of prior distributions. 

In the next two sections, we show how the material of this section can be used to discuss inference in diagnostics and system survivability. Whereas out approach to inference is Bayesian, it is not our intent to convey the impression that Bayesian approaches are the only viable ones for addressing the matter of adversarial and amiable inference. Our choice for a Bayesian approach is dictated by the considerations of Section 1.

\section{Adversarial Inference in Diagnostic Testing}



Diagnostic tests which play a key role in matters ranging from medicine to mathematical finance, are imperfect. For example, a test to detect antibodies of AIDS will occasionally diagnose a disease-free individual as being diseased, or will fail to diagnose a diseased individual  \citep[c.f.][]{gast87}. Diagnostic tests that perform perfectly are sometimes available, such tests are called {\it confirmatory tests}. The less reliable, but cost-effective tests, are called {\it screening tests}. Our focus here is on screening tests whose efficacy we endeavor to address. As stated, the efficacy of screening tests is characterized by the two adversarial parameters, sensitivity $\eta$, and specificity $\theta$. 
Notwithstanding the several foundational issues on diagnostics raised by \cite{davi76} in his insightful paper, our goal is to develop an inferential mechanism for these parameters accounting for their adversarial nature. 

\subsection{{Notation and Terminology}}

Let $D$ denote the event that an individual is actually diseased, and ${\bar D}$ the complement of $D$. The truth or falsity of $D$ is affirmed by a confirmatory test. Let $S$ (${\bar S}$) denote the event that a screening test declares an individual diseased (not diseased). Let $\pi = \Pr(D)$ be the probability (propensity) that a randomly chosen individual has the disease in question; $\pi$ is called the {\it disease prevalence}. Were $\eta$ and $\theta$ be interpreted as probabilities, then $\Pr(S|D)$ and $\Pr({\bar S}| {\bar D})$ are indeed $\eta$ and $\theta$, the sensitivity and specificity, respectively, of the screening test. Whereas $\eta$ and $\theta$ encapsulate the quality of the screening test, $\Pr(D|S) = \Lambda$ and $\Pr({\bar D}| {\bar S}) = \Psi$, encapsulate the predictive ability of the screening test. The parameters $\Lambda$ ($\Psi$) are known as the {\it predictive value} of {\it a positive (negative) screening test}. 

Interpreting $\eta$ and $\theta$ as probabilities, a Bayesian approach for inference about $\pi$, $\eta$ and $\theta$, was proposed by \cite{gajo91} -- henceforth GJR (1991) -- based on independent beta prior distributions on $\eta$ and $\theta$. A similar approach was taken by \cite{pepe90}. Besides a philosophical objection to endowing probabilities to probabilities, such priors do not account for the adversarial character of $\eta$ and $\theta$. The posterior distributions of $\pi$, $\eta$ and $\theta$ are then used by GJR (1991) approach to induce the posterior distribution of $\Lambda$ and $\Psi$. The GJR (1991) analysis based on screening data from AIDS patients, concluded that unless $\eta$ and $\theta$ are precisely known, inferences about $\pi$, $\Lambda$ and $\Psi$ are not credible. Thus, credible inferences about $\eta$ and $\theta$ are important, and our aim here is to work towards this goal. Our point of departure from the GJR (1991) is the use of joint priors on $\eta$ and $\theta$ that incorporate negative dependence, and the use of confirmatory (instead of screening) test data to develop our inferential mechanism. With $\eta$ and $\theta$ interpreted as propensities, inferences about $\Lambda$ and $\Psi$ entail a strategy that is different from that of GJR (1991), and also that of \cite{pepe90}. This is articulated later in Section 3.5. 

The prior distributions chosen here are the $(OL)^{-}$ distribution and the $AN(5)$ distribution with parameters so chosen that $\eta$ and $\theta$ are negatively correlated. Recall that the $AN(5)$ family of distributions does not encompass the $(OL)^{-}$ distribution. We have chosen the $AN(5)$ distribution instead of the more generalized bivariate beta $(GBB)$ distribution -- namely the $AN(8)$ family of the Appendix -- because the more parsimonious nature of the $AN(5)$ construction eases the computational burden. For convenience, we shall denote by $h(\eta, \theta; \bcdot)$ the joint probability density function of the $AN(5)$ bivariate beta distribution, and by $h_{1}(\eta; \bcdot)$ and 
$h_{2}(\theta; \bcdot)$ its marginal density functions. This is because a closed form expression for the joint density function is not available. The marginal densities $h_{1}(\eta; \bcdot)$ and $h_{2}(\theta; \bcdot)$ are of course beta. The joint probability density function of the $(OL)^{-}$ bivariate beta distribution is available in closed form; it will be given in Section 3.3. 

\subsection{Test Design and Likelihood}

Suppose that a sample of size $n$ is chosen at random from a population of interest, and a confirmatory test given to each subject. This is in contrast to the sampling plan considered by GJR (1991) who give a confirmatory test only to individuals who screen positive. 

Let $N_{1}[N_{2}]$ be the number of individuals declared $D[{\bar D}]$ by the confirmatory test. Note that $N_{1} + N_{2} = n$, and that 
\begin{eqnarray*} 
\Pr(N_{1} = n_{1} | \pi; n) = {n \choose n_{1}} \pi^{n_{1}} (1 - \pi)^{n - n_{1}},~~0 < \pi < 1, n_{1} = 0, 1, \ldots, n, 
\end{eqnarray*} 
or equivalently 
\begin{eqnarray*} 
\Pr(N_{2} = n - n_{1} | \pi; n) = {n \choose n - n_{1}} (1 - \pi)^{n - n_{1}} \pi^{n_{1}},~~0 < \pi < 1, n_{1} = 0, 1, \ldots, n.  
\end{eqnarray*} 


The $N_{1}$ individuals who test positive are then given the screening test whose efficacy we wish to assess. Let $K_{1} [N_{1} - K_{1}]$ be the number of diseased individuals who are declared diseased [not diseased] by the screening test. Clearly, 
\begin{eqnarray*} 
\Pr(K_{1} = k_{1} | N_{1} = n_{1}; \eta) =  {n_{1} \choose k_{1}} \eta^{k_{1}} (1 - \eta)^{n_{1} - k_{1}},~~0 < \eta < 1, k_{1} = 0, 1, \ldots, n_{1}.  
\end{eqnarray*} 
Similarly, let $K_{2} [n-N_{1}-K_{2}]$ be the number of non-diseased individuals who are declared non-diseased [diseased] by screening test. Then, 
\begin{eqnarray*} 
\Pr(K_{2} = k_{2} | N_{1} = n_{1}; \theta) =  {n - n_{1} \choose k_{2}} \theta^{k_{2}} (1 - \theta)^{n - n_{1} - k_{2}},~~0 < \theta < 1, k_{2} = 0, 1, \ldots, n-n_{1}.  
\end{eqnarray*} 
Thus, the results of the confirmatory and the screening tests given to $n$ individuals, yield as data the quantities $n_{1}$, $k_{1}$ and $k_{2}$. Both $K_{1}$ and $K_{2}$ depend on $N_{1}$, where $N_{1}$ is random. Thus, $K_{1}$ and $K_{2}$ are {\it conditionally} (given $N_{1}$) independent, but unconditionally, they are dependent, this being the case irrespective of whether we condition or not on $\eta$ and $\theta$. In what follows, we assume the independence of $K_{1}$ and $K_{2}$ conditional on $N_{1}$, $\eta$, and $\theta$; this assumption is meaningful. Under the above assumption of conditional independence, a likelihood for $\pi$, $\eta$, and $\theta$, with $n_{1}$, $k_{1}$ and $k_{2}$ as observed data, can be inducted via the following probability model (under the philosophical {\it principle of conditionalization}): 
\begin{eqnarray*} 
& & \Pr(K_{2} =  k_{2}, K_{1} = k_{1}, N_{1} = n_{1}| \pi, \eta, \theta) \\
& = & \Pr(K_{2} = k_{2} | N_{1} = n_{1}, \theta) \bcdot \Pr(K_{1} = k_{1} | N_{1} = n_{1}, \eta) \bcdot \Pr(N_{1} = n_{1} | \pi)\\
& = & {n \choose n_{1}} {n_{1} \choose k_{1}} {n - n_{1} \choose k_{2}} \theta^{k_{2}} (1 - \theta)^{n - n_{1} - k_{2}} \bcdot \eta^{k_{1}} (1 - \eta)^{n_{1} - k_{1}} \bcdot \pi^{n_{1}} (1 - \pi)^{n - n_{1}}.
\end{eqnarray*} 

\subsection{Posterior Analysis}

Since $\pi$, the disease prevalence parameter, is in no way influenced by the quality of the screening instrument, the prior on $\pi$ is assumed to be independent of the joint prior on $\eta$ and $\theta$. It is taken to be a beta distribution with parameters $\alpha$ and $\beta$. As mentioned before, the joint priors on $\eta$ and $\theta$ are 
\begin{itemize} 
  \item[(i)] the $(OL)^{-}$ distribution whose probability density function is:  
\begin{eqnarray*}
g(\eta, \theta; \alpha_{1}, \alpha_{2}, \alpha_{3}) \propto \frac{\eta^{\alpha_{1} - 1} (1 - \eta)^{\alpha_{2} + \alpha_{3} - 1} \theta^{\alpha_{1} + \alpha_{3} - 1} (1 - \theta)^{\alpha_{2} - 1}}{[1 - \eta(1 - \theta)]^{\alpha_{1} + \alpha_{2} + \alpha_{3}} },
\end{eqnarray*}
for $0 < \eta, \theta < 1$, with marginals $g_{1}(\eta; \alpha_{1}, \alpha_{3})$ and $g_{2}(\theta; \alpha_{2}, \alpha_{3})$, which are beta distributions, and  
  \item[(ii)]  the $AN(5)$ distribution with joint and marginal probability density functions $h(\eta, \theta; \bcdot)$, $h_{1}(\eta; \bcdot)$ and $h_{2}(\theta; \bcdot)$, respectively. 
\end{itemize} 

The above priors and the likelihood, yield the joint posterior distribution of $\pi$, $\eta$, and $\theta$ as: 
\begin{eqnarray*}
\Pr(\pi, \eta, \theta; \bd) & \propto & {\cal L}(\pi, \eta, \theta; \bd) g(\eta, \theta; \bcdot) {\cal B}(\pi; \alpha, \beta), {\mbox { or as}}   \\ 
& \propto & {\cal L}(\pi, \eta, \theta; \bd) h(\eta, \theta; \bcdot) {\cal B}(\pi; \alpha, \beta).  
\end{eqnarray*}
depending on whether the $AN(5)$ or the $(OL)^{-}$ distribution is used as prior; here, the data is $\bd = (n, n_{1}, k_{1}, k_{2})$, and ${\cal B}(\pi; \alpha, \beta)$ is the beta prior on $\pi$. 

Since $\pi$ factors out in the likelihood, and since its prior is assumed to be independent of joint prior of $\eta$ and $\theta$, the posterior distribution of $\pi$ falls out as ${\cal B}(\pi; \alpha + n_{1}, \beta + n - n_{1})$. Consequently, the joint posterior distribution of $\eta$ and $\theta$ is: 
\begin{eqnarray*}
\Pr(\eta, \theta; \bd) & \propto & \theta^{k_{2}} (1 - \theta)^{n - k_{1} - k_{2}} \eta^{k_{1}} (1 - \eta)^{n_1 - k_{1}} \bcdot g(\eta, \theta; \bcdot), {\mbox { or}} \\ 
& \propto & \theta^{k_{2}} (1 - \theta)^{n - k_{1} - k_{2}} \eta^{k_{1}} (1 - \eta)^{n_1 - k_{1}} \bcdot h(\eta, \theta; \bcdot). 
\end{eqnarray*}
This joint posterior distribution is not available in closed form, irrespective of the choice of the prior. Thus, $\Pr(\eta, \theta; \bd)$ is to be assessed numerically. A direct approach is to construct an $m \times m$ sized grid over the unit square, and denote the $(i, j)$-th member of the grid by $(\eta_{i}, \theta_{j})$, $i, j = 1, \ldots, m$. The joint posterior distribution of $\eta$ and $\theta$, at the point $(\eta_{i}, \theta_{j})$ is then approximated at all $i, j = 1, \ldots, m$, as: 
\begin{eqnarray*}
\frac{\Pr(\eta_{i}, \theta_{j}; \bd)}{\sum\limits_{i^{*}=1}^{m} \sum\limits_{j^{*} = 1}^{m} \Pr(\eta_{i^{*}}, \theta_{j^{*}}; \bd)}.
\end{eqnarray*}
The larger the $m$, the better is the approximation. Since the marginal posterior distributions of $\eta$ and $\theta$ are of practical interest, these will be approximated, for $\eta$ at the point $\eta_{i}$, $i = 1, \ldots, m$, as:
\begin{eqnarray*}
\frac{\sum\limits_{j^{*}=1}^{m} \Pr(\eta_{i}, \theta_{j^{*}}; \bd)}{\sum\limits_{i^{*}=1}^{m} \sum\limits_{j^{*} = 1}^{m} \Pr(\eta_{i^{*}}, \theta_{j^{*}}; \bd)} = \Pr(\eta_{i}; \bd), 
\end{eqnarray*}
and similarly, for $\theta = \theta_{j}$, $j = 1, \ldots, m$, we obtain $\Pr(\theta_{i}; \bd)$. 
This completes our discussion on posterior analysis. 

\subsection{Proof of Principle: Validation by Synthetic Data}

It is difficult, if not impossible, to assess the quality of a diagnostic test via assessed values of its adversarial parameters $\eta$ and $\theta$, unless these parameters are known with certainty. This means that it does not make sense to use actual (field) data to establish proof of principle of the  proposed approach. Consequently, we need to generate synthetic data with known properties, and how such data are generated is described below. The priors for $\eta$ and $\theta$ under scrutiny here are the $(OL)^{-}$ and the $AN(5)$ with negative correlation, and also the independent beta priors of GJR (1991). A use of this latter family of priors will enable us to assess the operational merit of incorporating a priori knowledge in the assessment exercise. 

\begin{figure}
\begin{center} 
  \includegraphics[scale = 0.35]{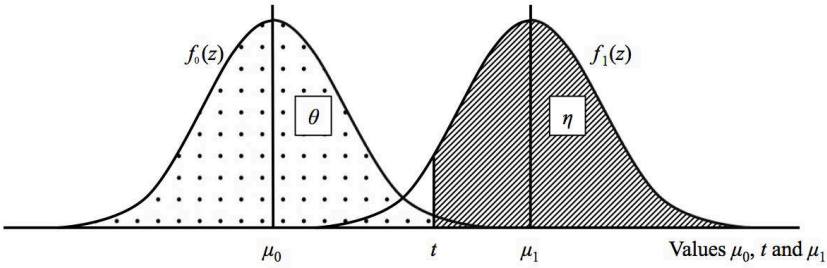}\\
\caption{Distributions of $Z_{0}$ and $Z_{1}$.}
\label{fig1}
\end{center}
\end{figure}

To generate $n_{1}$, $k_{1}$ and $k_{2}$ under known values of $\pi$, $\eta$, and $\theta$, we proceed as follows: Let $Z_{0}$ denote the values taken by an observable diagnostic variable for members belonging to disease-free class ${\bar D}$, and $Z_{1}$ denote the values taken by the variable for members belonging to the diseased class, $D$. Assume that $Z_{i}$ has a normal probability density function $f_{i}(z)$ with mean $\mu_{i}$ and variance 1, for $i = 0, 1$. Let $t$ be a ``threshold", where $t$ is between $\mu_{0}$ and $\mu_{1}$. Any individual whose diagnostic measurement value exceeds $t$ is classified as diseased; otherwise the individual is classified as non-diseased. See Figure \ref{fig1} which shows the dispositions of $\mu_{0}$, $\mu_{1}$ and $t$. Thus, given the distributions of $Z_{0}$ and $Z_{1}$, and the threshold $t$, $\theta$ and $\eta$ can be precisely determined. 

For any pre-specified sample size $n$, and a pre-specified disease prevalence probability $\pi$, we obtain $n_{1}$, the number of individuals in the sample that should belong to the class $D$, as $n_{1} = \lfloor \pi n \rfloor$, where $\lfloor a \rfloor$ denotes the integer part of $a$. 
We now generate $n_{1}$ observations from $f_{1}(z)$ and denote these as $D_{1}, D_{2}, \ldots, D_{n_{1}}$; similarly, we generate $(n - n_{1})$ observations from $f_{0}(z)$ and denote these as ${\bar D}_{1}, {\bar D}_{2}, \ldots, {\bar D}_{(n-n_{1})}$. Then, $k_{1}$ is the total number of $D_{i}$'s out of $n_{1}$ that are greater than $t$, and $k_{2}$ is the number of ${\bar D}_{i}$'s out of $(n - n_{1})$ that are less than or equal to $t$. A naive estimate of
$\eta$ is $k_{1}/n_{1}$, and that of $\theta$ is $k_{2}/(n - n_{1})$. Furthermore, if $k$ is the total number of $D_{i}$'s and ${\bar D}_{i}$'s that are greater than $t$, then a naive estimate of $\pi$ is $k/n$. Thus, for any specified
$n$ and $\pi$, we have generated $\bd = (n, n_{1}, k_{1}, k_{2})$ as our data with the true values of $\eta$ and $\theta$ determined via $Z$, the standard normal variate, as $\eta = \Pr(Z > t - \mu_{1})$ and $\theta = \Pr(Z \leq t - \mu_{0})$.

In what follows we set $\pi = 0.35$, and choose a range of values for $n$, namely $n = 15$, 30, 50 and 100. The means $\mu_{0}$ and $\mu_{1}$ are taken to be 3 and 4, respectively, and $t$ is set at 3.25. The above choices yield the true values for $\eta$ and $\theta$ as $\eta = 0.773$ and $\theta = 0.599$.

For the two families of joint prior distributions of $\eta$ and $\theta$, we choose ($\alpha_{1} = 10$, $\alpha_{2} = 2.5$, and $\alpha_{3} = 5$), and ($\alpha_{1} = \alpha_{2} = \alpha_{3} = \alpha_{4} = 5$, and $\alpha_{5} = 0.0001$), respectively. The former set yields a beta distribution with parameters $(10, 5)$ as the marginal prior for $\eta$, and a beta distribution with parameters $(5, 2.5)$ as the marginal prior for $\theta$. The latter set ensures that the marginals for $\eta$ and $\theta$ under the $AN(5)$ model are comparable to the beta distributions given above vis-{\`a}-vis their means, which happen to be about 0.67 for both $\eta$ and $\theta$. For the parameter choices given above, the theoretical correlation between  $\eta$ and $\theta$ under the $(OL)^{-}$ distribution is $-0.45$, and that under the $AN(5)$ distribution is $-0.65$. 

{\noindent {\underline {Comparative Analysis: Results of Simulation}}} 

Using the schemata for generating the data $\bd$ described above, along with the choice of prior parameters given above, we obtain the joint posterior distributions for $\eta$ and $\theta$, for $n = 15, 30, 50$, and 100. For purposes of calibration against the independence case, a joint prior that is the product of independent beta distributions, $\eta \sim {\cal B}(10, 5)$ and $\theta \sim {\cal B}(5, 2.5)$, denoted {\it Independent} was also considered. The contours of the joint priors and their corresponding posterior distributions for the four choices of $n$ are shown in the second, third, and fourth columns of Table 1. The solid lines in each plot show the location of the true
values of $\eta$ and $\theta$, namely, 0.773 and 0.599. These facilitate a comparison of the true against the centroids of the posterior contours, which are shown by dashed lines.

Tables 2 and 3 are replicas of Table 1 save for the fact that they pertain to the marginal priors and posterior distributions of $\eta$ and $\theta$ respectively for the distributional choices mentioned above. As with Table 1, the solid lines on these plots display the true values of $\eta$ and $\theta$; this facilitates a comparison with the modes of the prior and posterior marginal distributions, shown by the dashed lines. 


\begin{table}
\caption{Comparison of the Contours of the Joint Prior and Posterior Distributions of $\eta$ and $\theta$.}
\centering
\begin{tabular}{ | m{.6in} | m{1.6in} | m{1.6in} | m{1.6in}|}\hline
Values of $n$ & \textcolor{red}{Independent} \newline $\eta\sim {\cal B}(10,5)$ \newline $\theta\sim {\cal B}(5,2.5)$ \newline $\rho=0$ & \textcolor{blue}{$(OL)^-$}\newline $\alpha_1=10$, $\alpha_2=2.5$, \newline$\alpha_3=5$  \newline $\rho=-.45$ & \textcolor{violet}{$AN(5)$} \newline $\alpha_i=5,\ \ i=1,\dots,4$, $\alpha_5=0.0001$  \newline $\rho=-.65$\\ \hline
$n=0$ \newline (Prior)
&
\includegraphics[scale = .3]{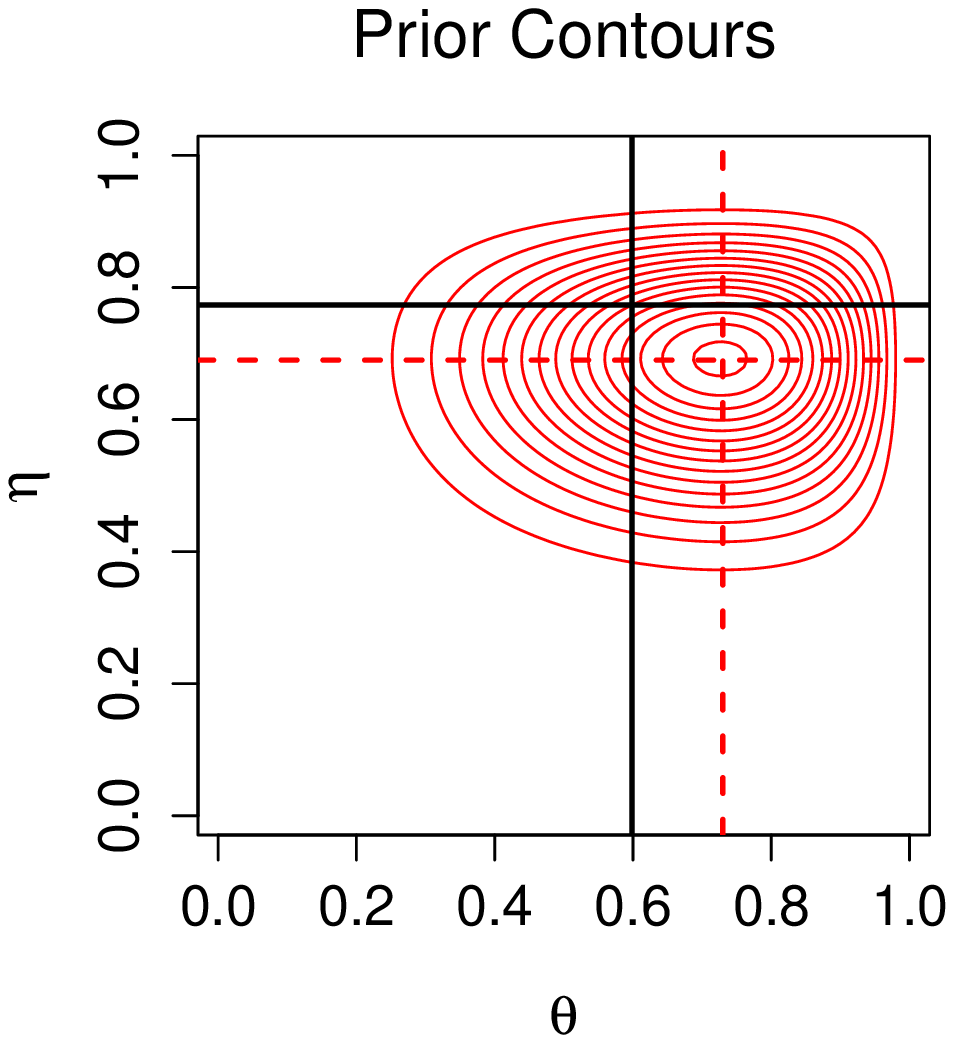}
& 
\includegraphics[scale = .3]{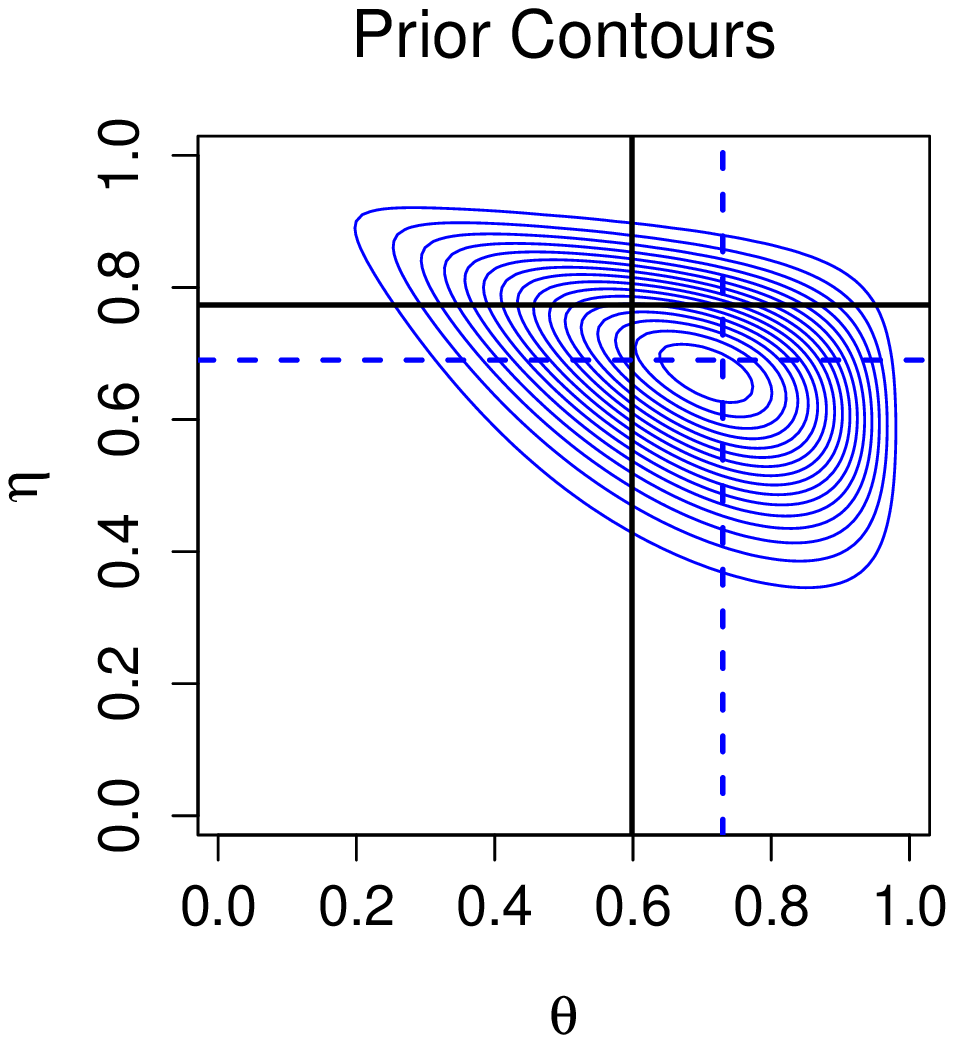} 
&
\includegraphics[scale = .3]{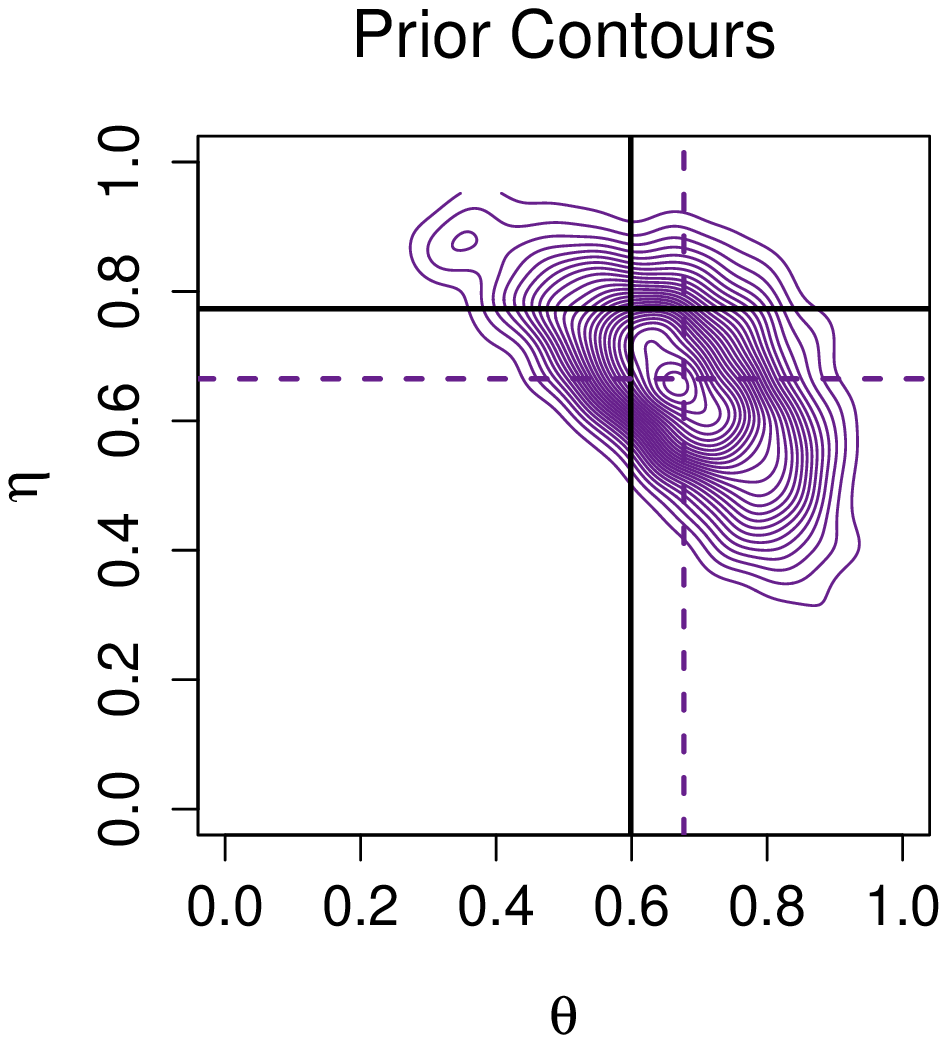} \\ \hline
$n=15$
&
\includegraphics[scale = .3]{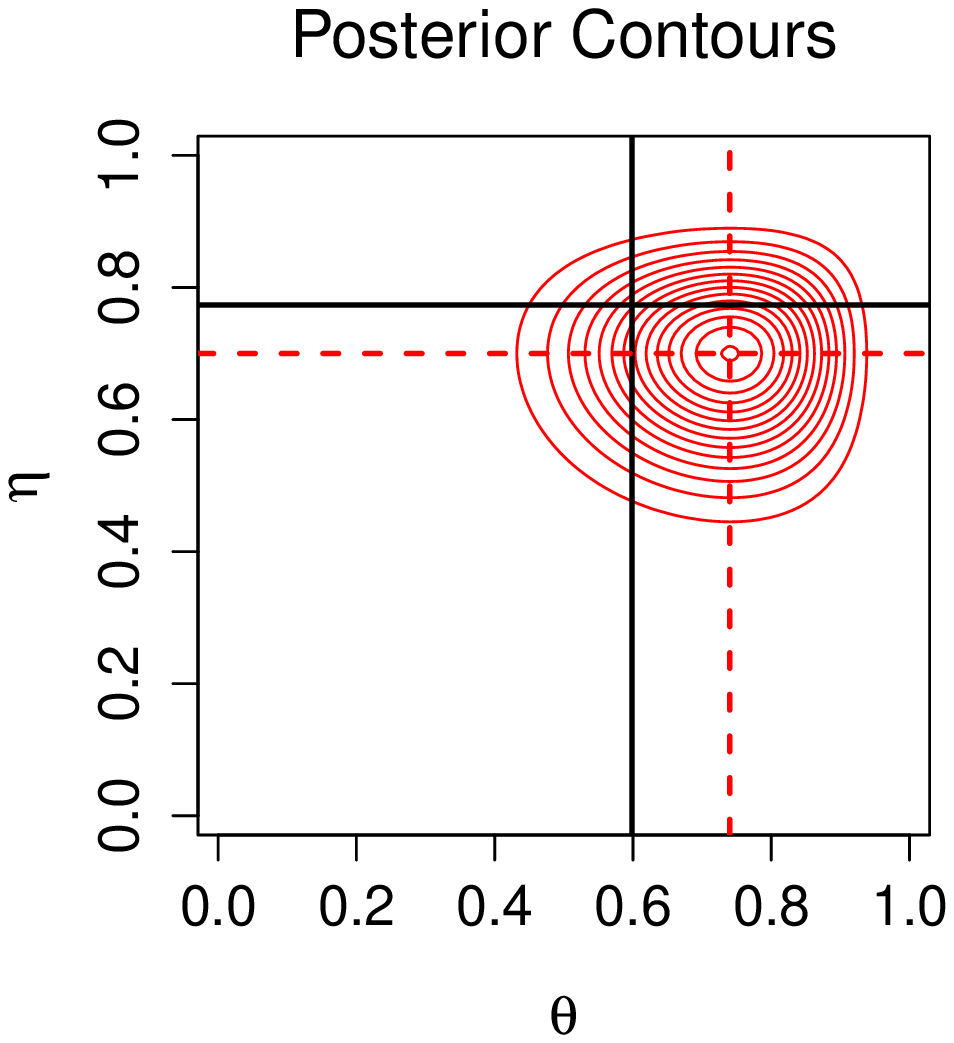}
& 
\includegraphics[scale = .3]{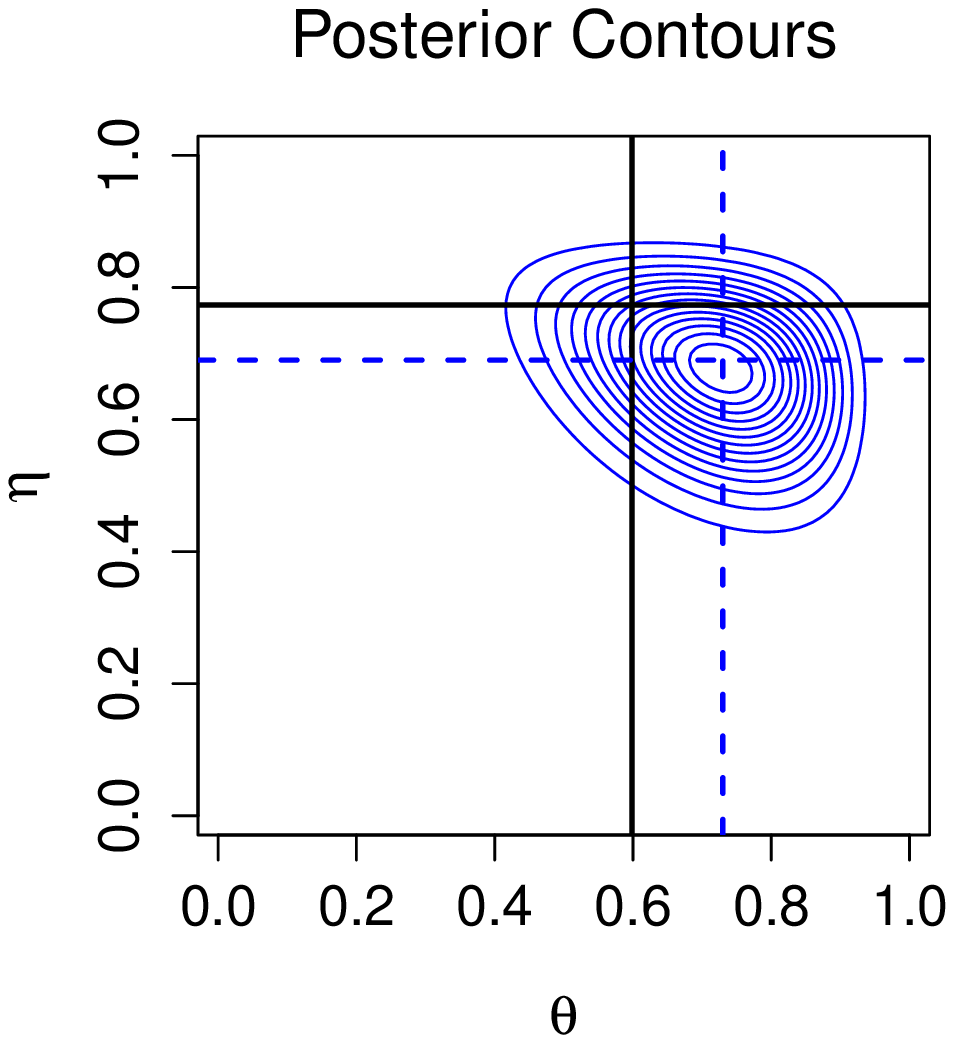} 
&
\includegraphics[scale = .3]{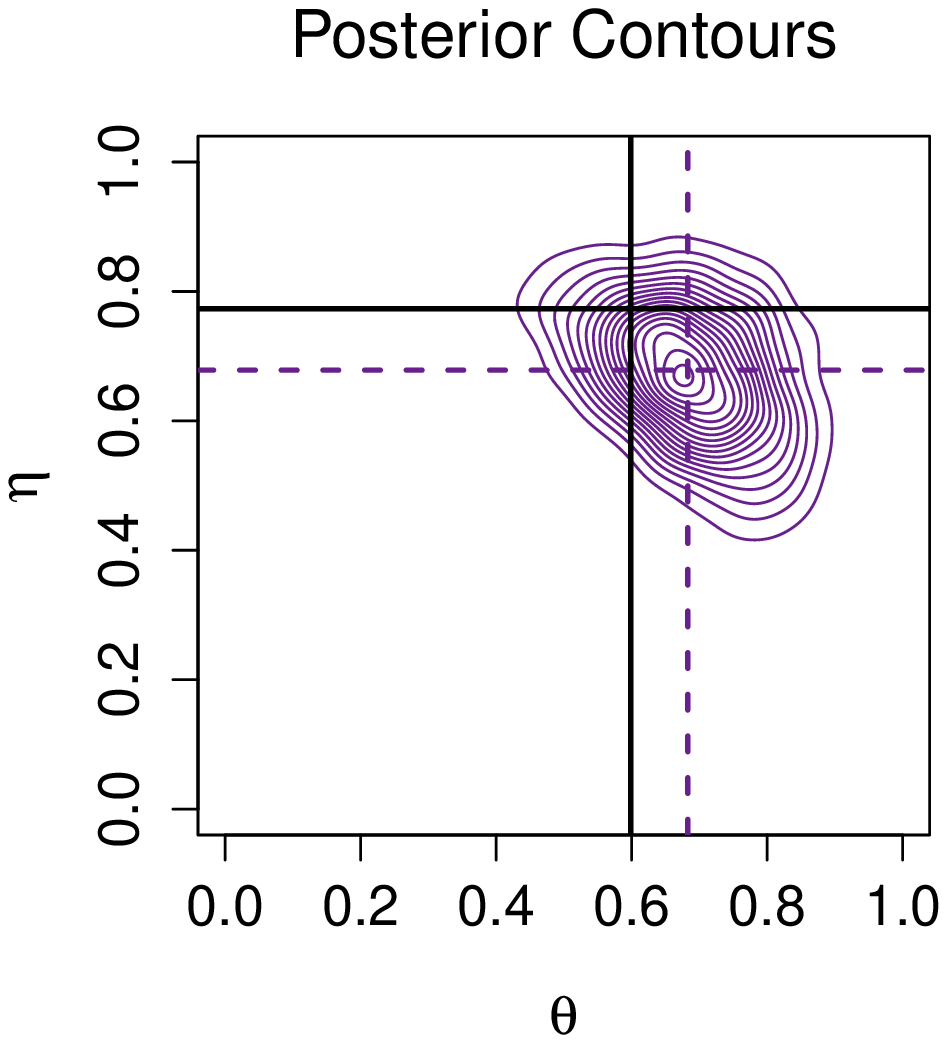}\\ \hline
$n=30$
&
\includegraphics[scale = .3]{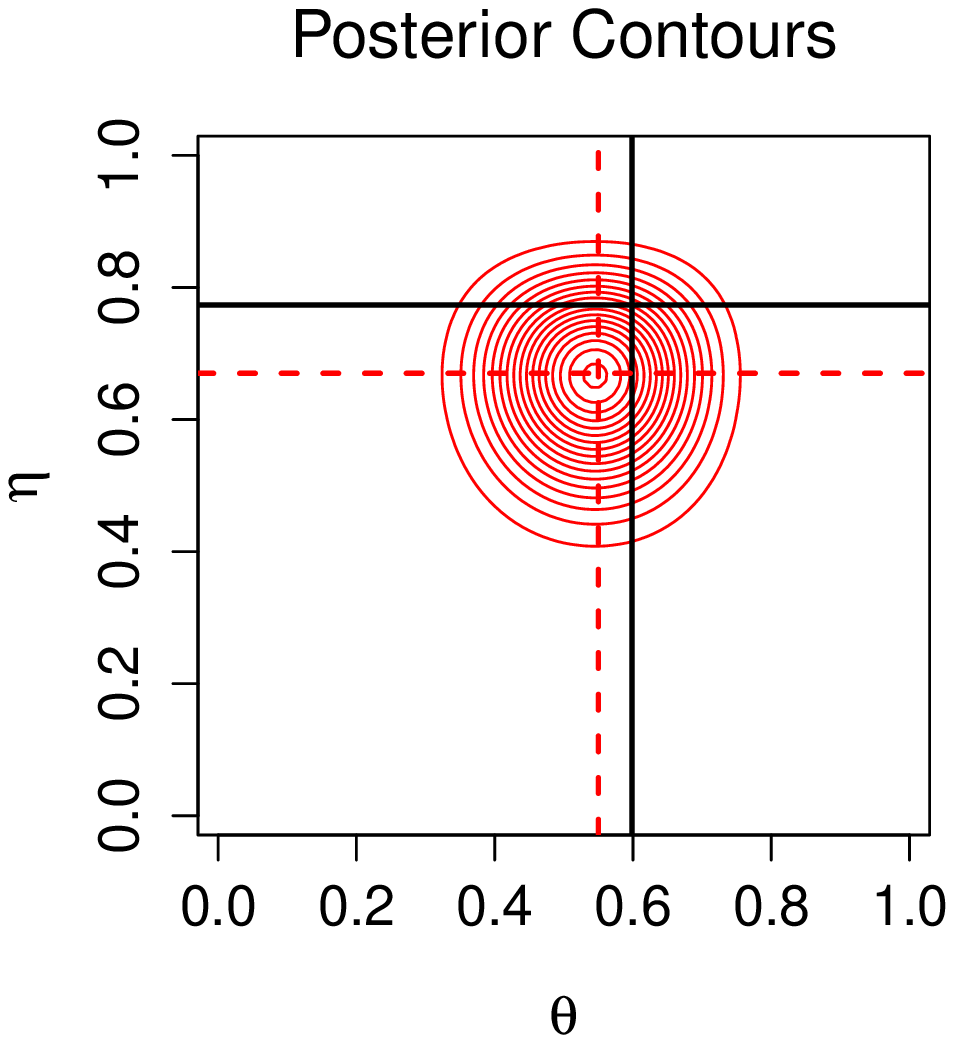}
& 
\includegraphics[scale = .3]{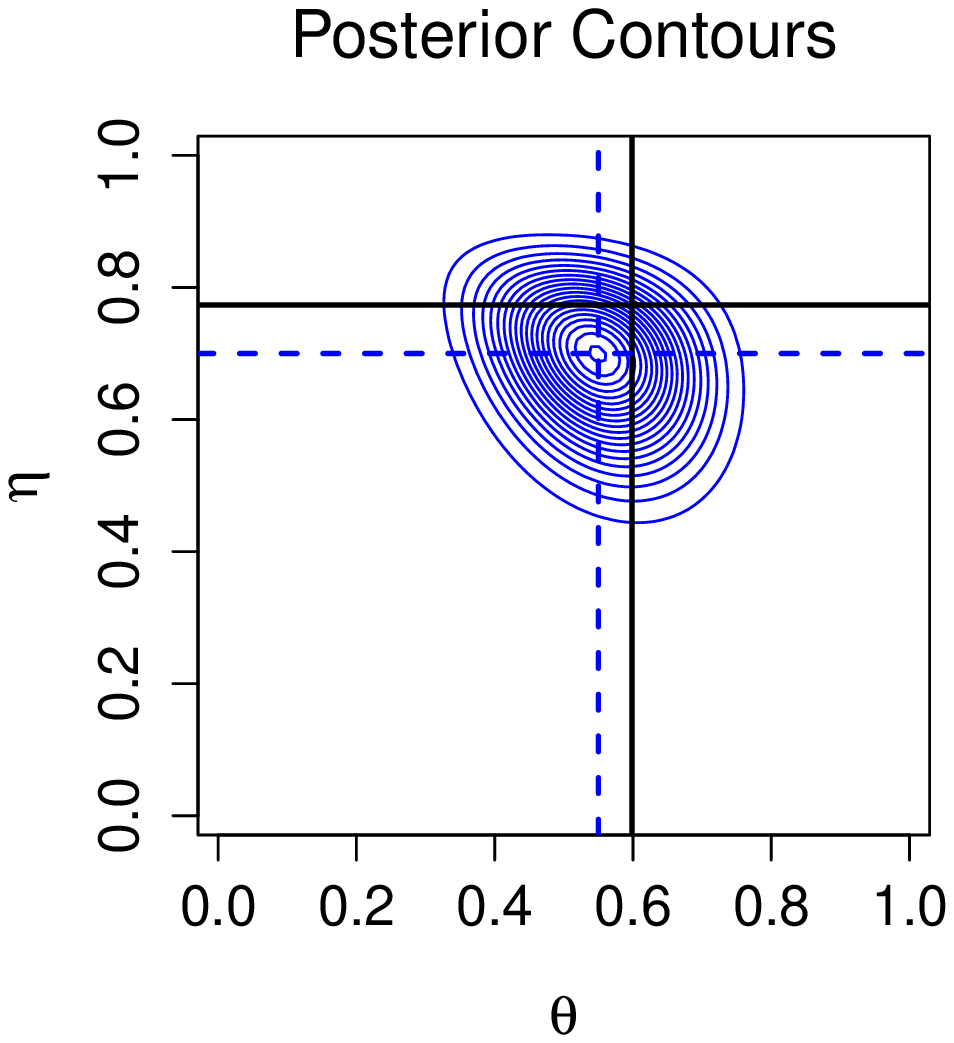} 
&
\includegraphics[scale = .3]{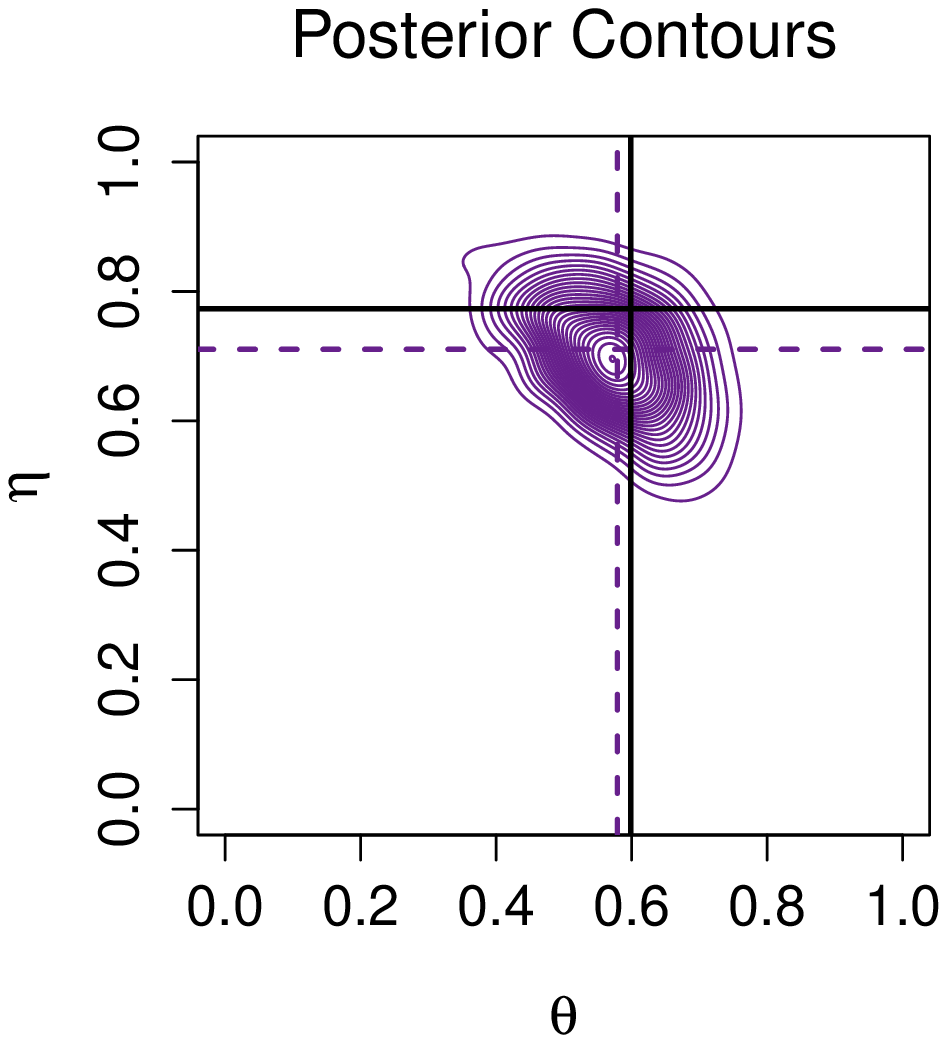}\\ \hline
$n=50$
&
\includegraphics[scale = .3]{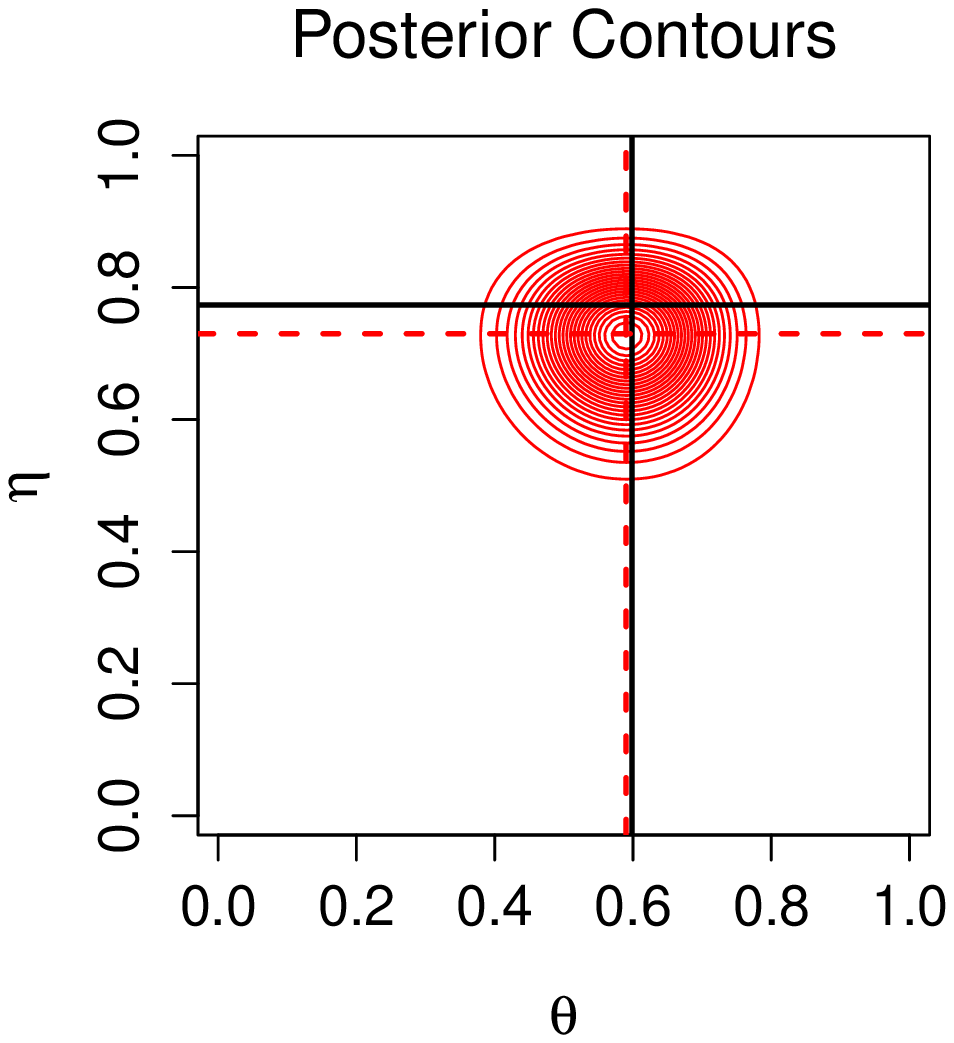}
& 
\includegraphics[scale = .3]{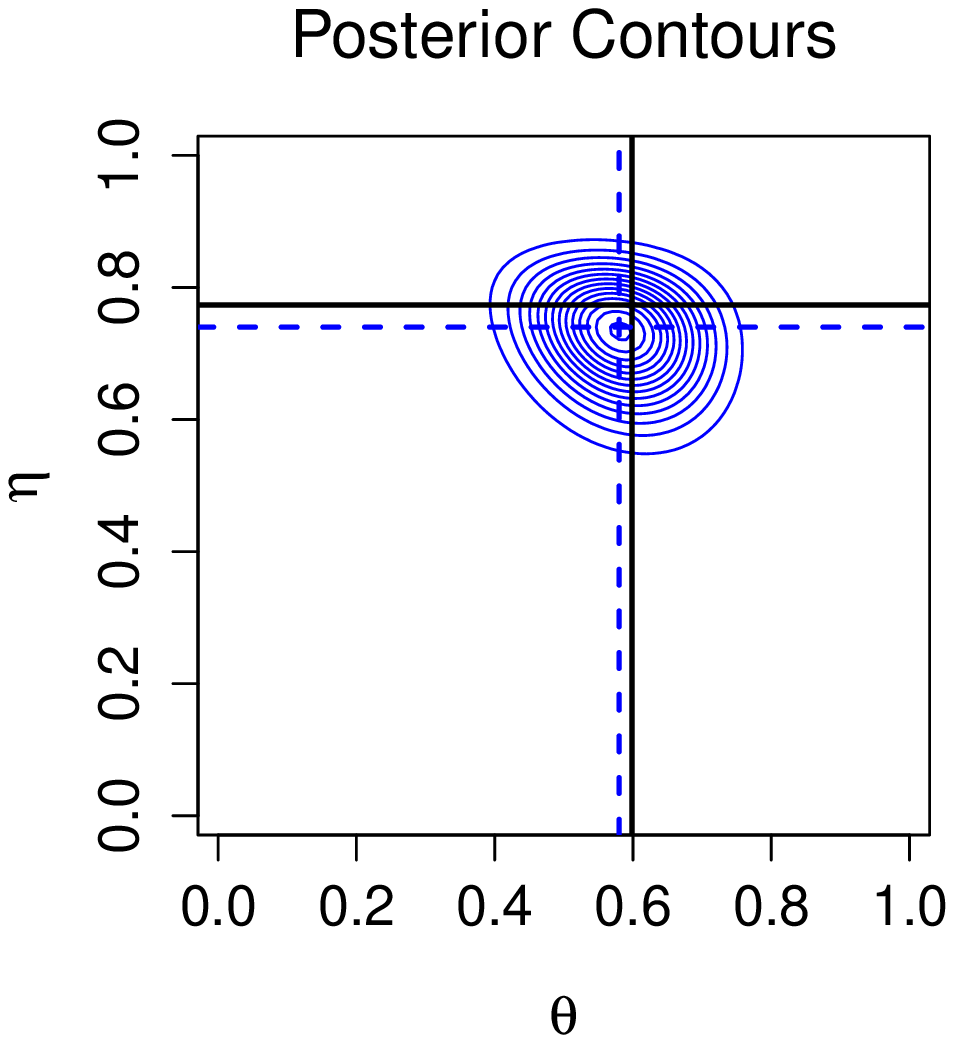} 
&
\includegraphics[scale = .3]{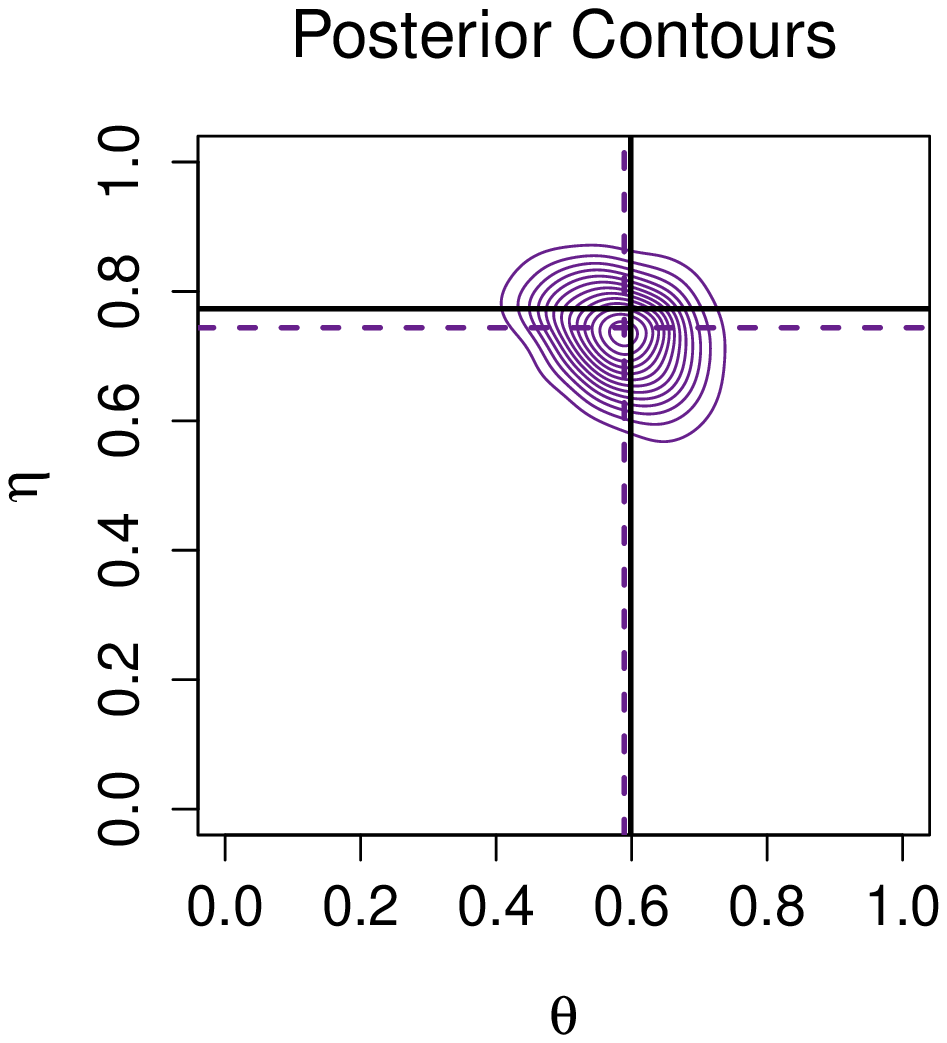}\\ \hline
$n=100$
&
\includegraphics[scale = .3]{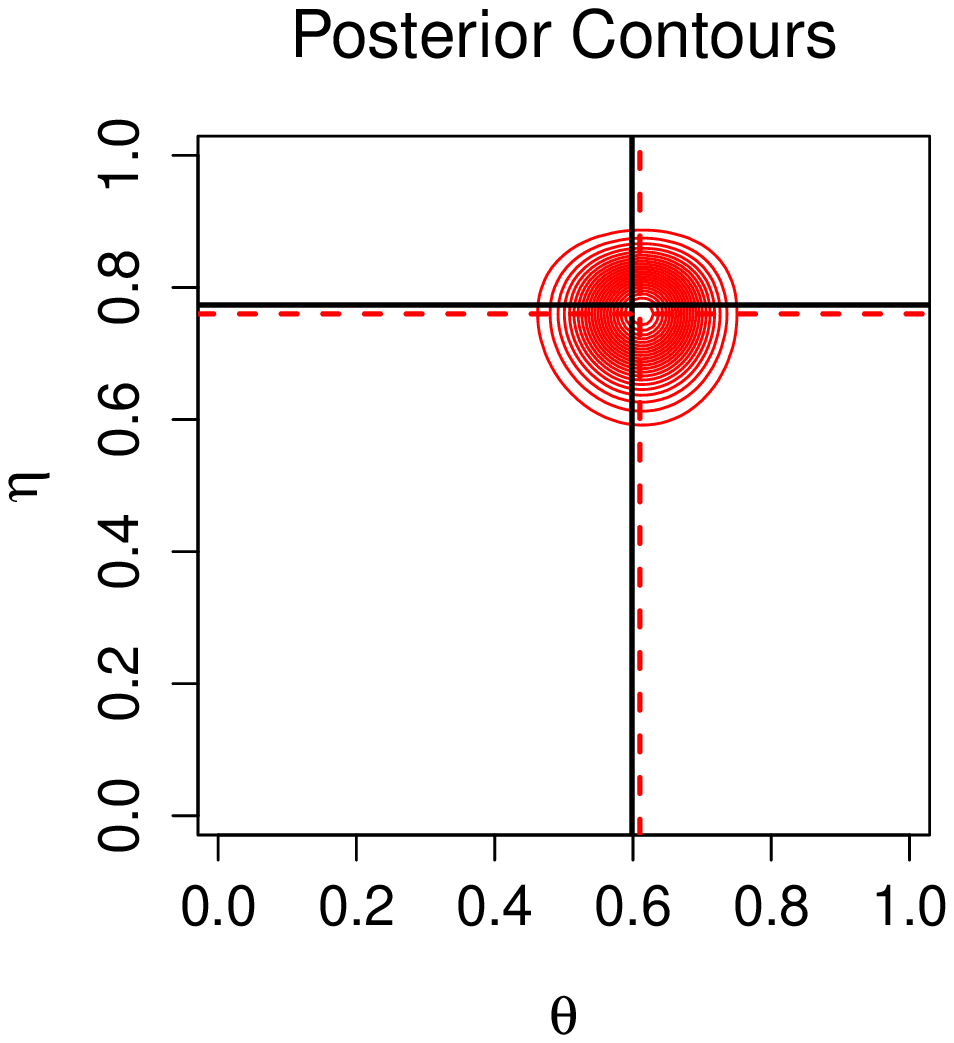}
& 
\includegraphics[scale = .3]{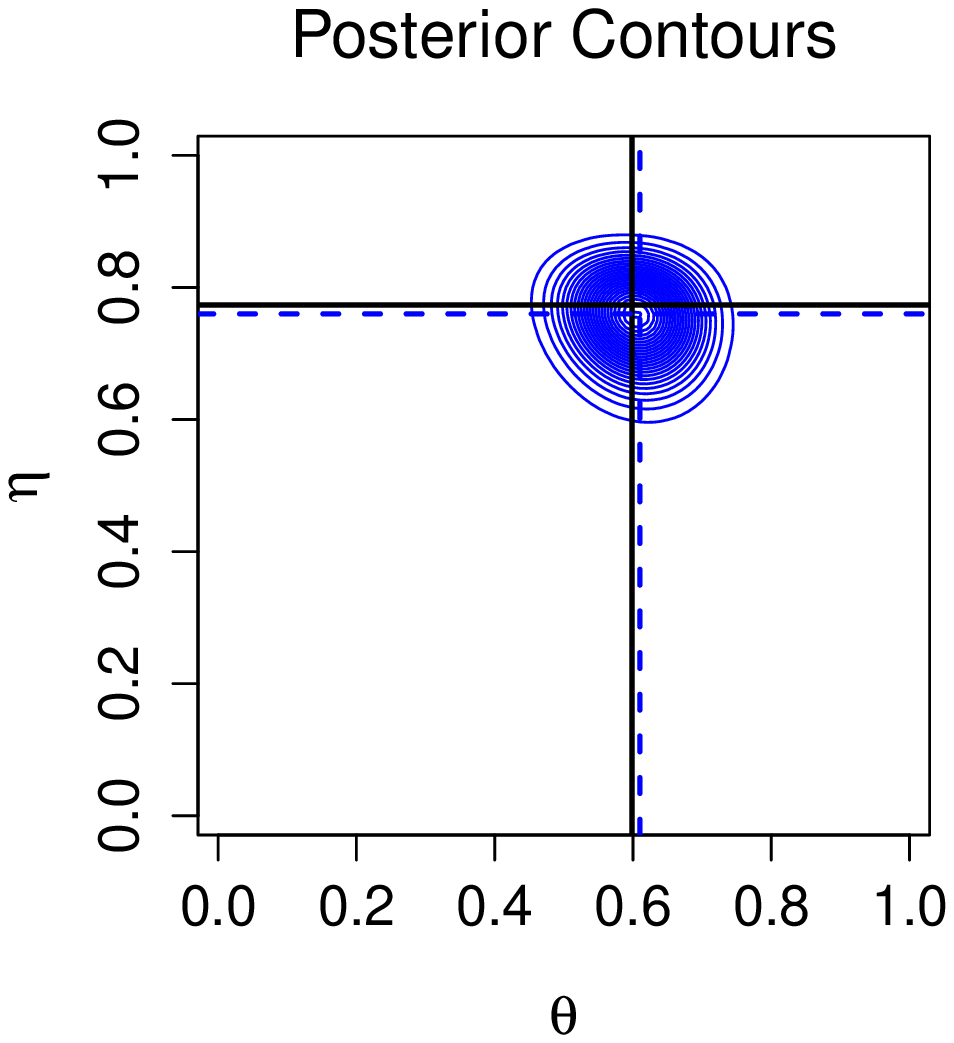} 
&
\includegraphics[scale = .3]{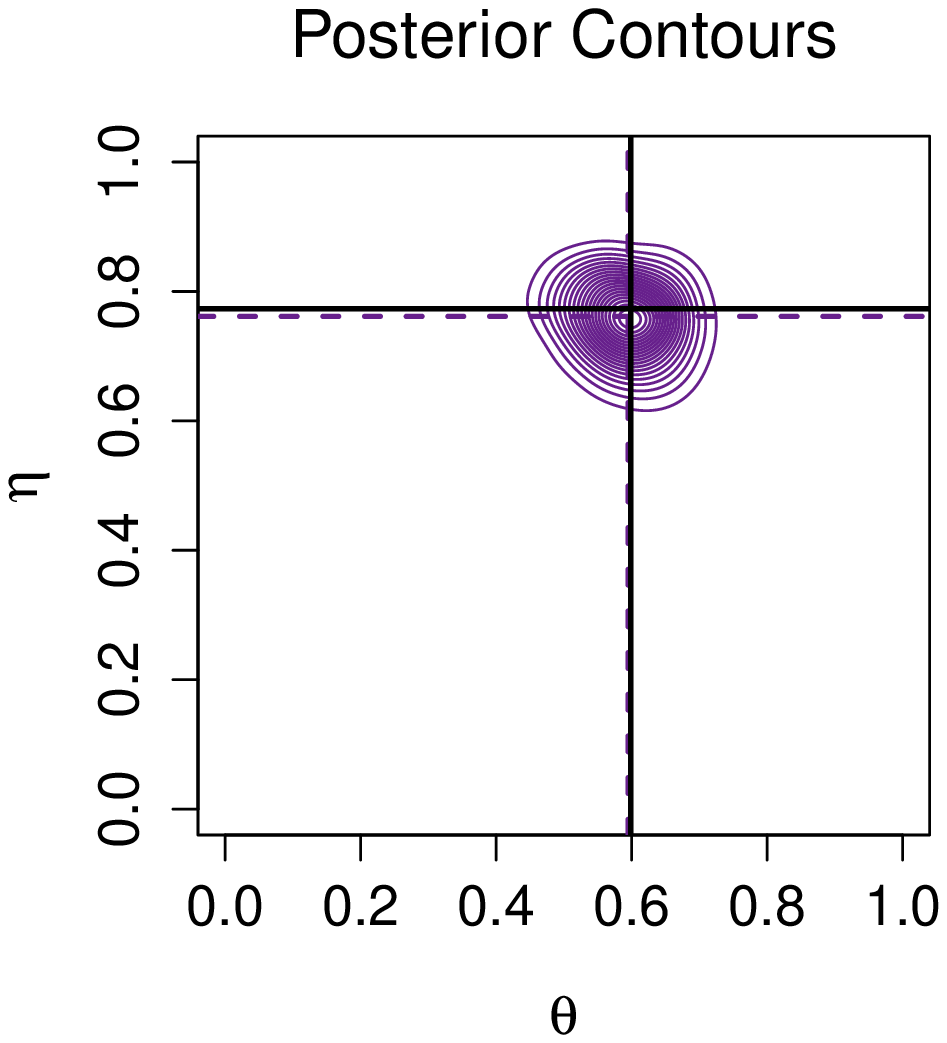}\\ \hline
\end{tabular}

\label{tbl:contourplots}
\end{table}

\begin{table}
\caption{Comparison of the Prior and Posterior Marginal Distributions of $\eta$.}
\centering
\begin{tabular}{ | m{.6in} | m{1.6in} | m{1.6in} | m{1.6in}|}
\hline
Values of $n$ & \textcolor{red}{Independent} \newline $\eta\sim {\cal B}(10,5)$ \newline $\theta\sim {\cal B}(5,2.5)$ \newline $\rho=0$ & \textcolor{blue}{$(OL)^-$}\newline $\alpha_1=10$, $\alpha_2=2.5$, \newline$\alpha_3=5$  \newline $\rho=-0.45$ & \textcolor{violet}{$AN(5)$} \newline $\alpha_i=5,\ \ i=1,\dots,4$, $\alpha_5=0.0001$  \newline $\rho=-0.65$\\ \hline
$n=0$ \newline (Prior)
&
\includegraphics[scale = .3]{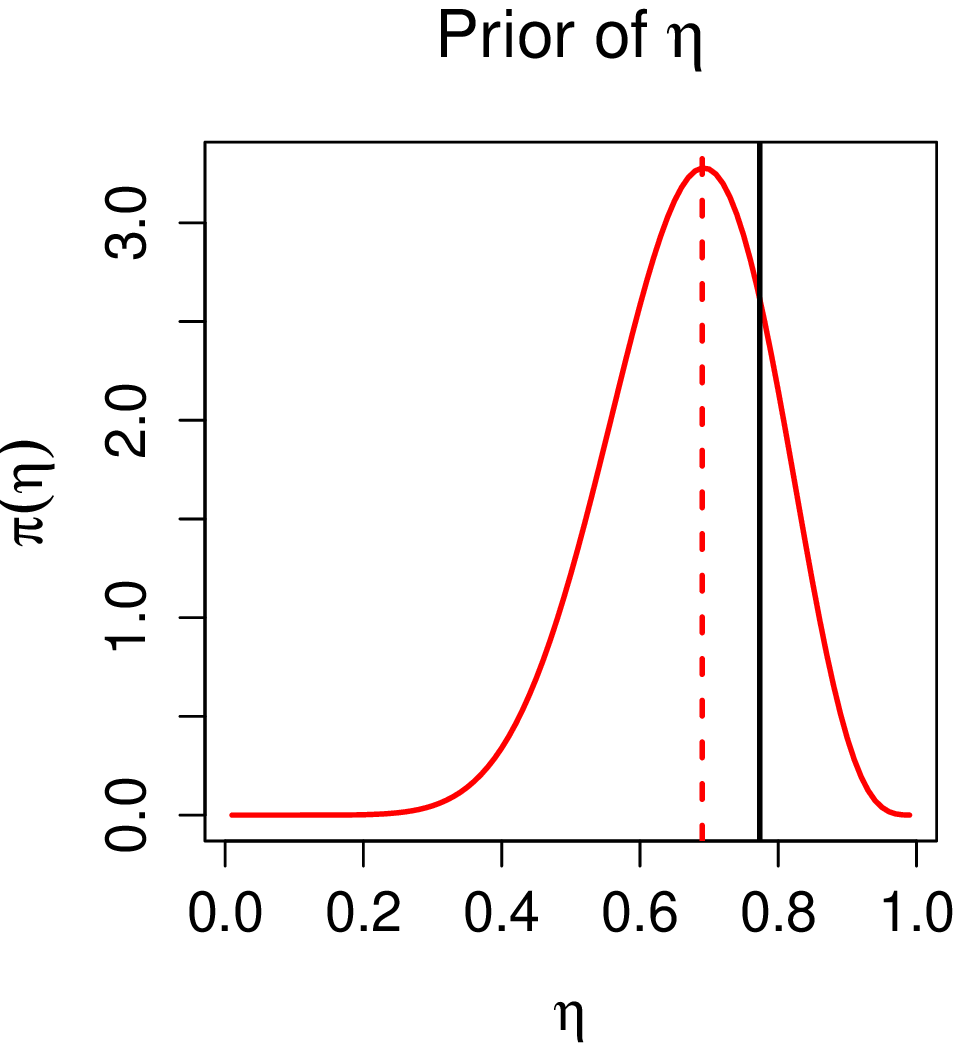}
& 
\includegraphics[scale = .3]{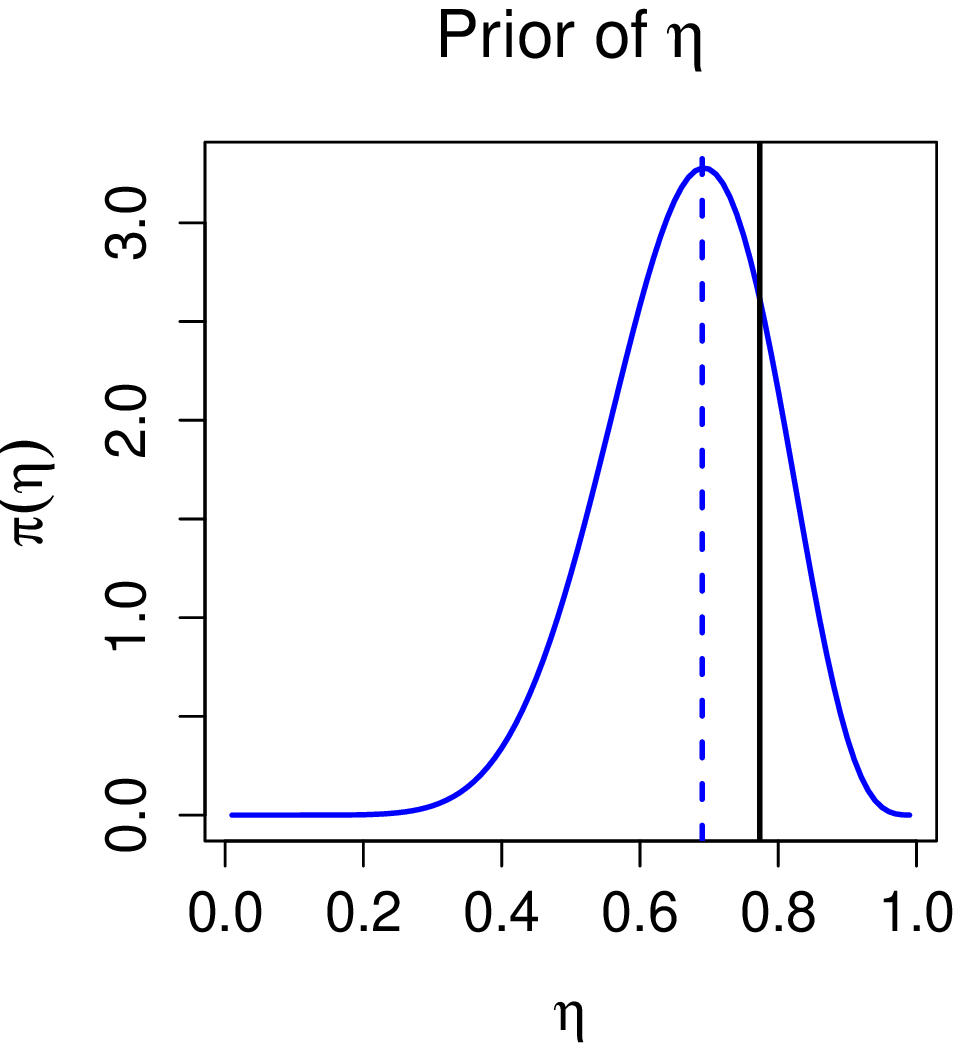} 
&
\includegraphics[scale = .3]{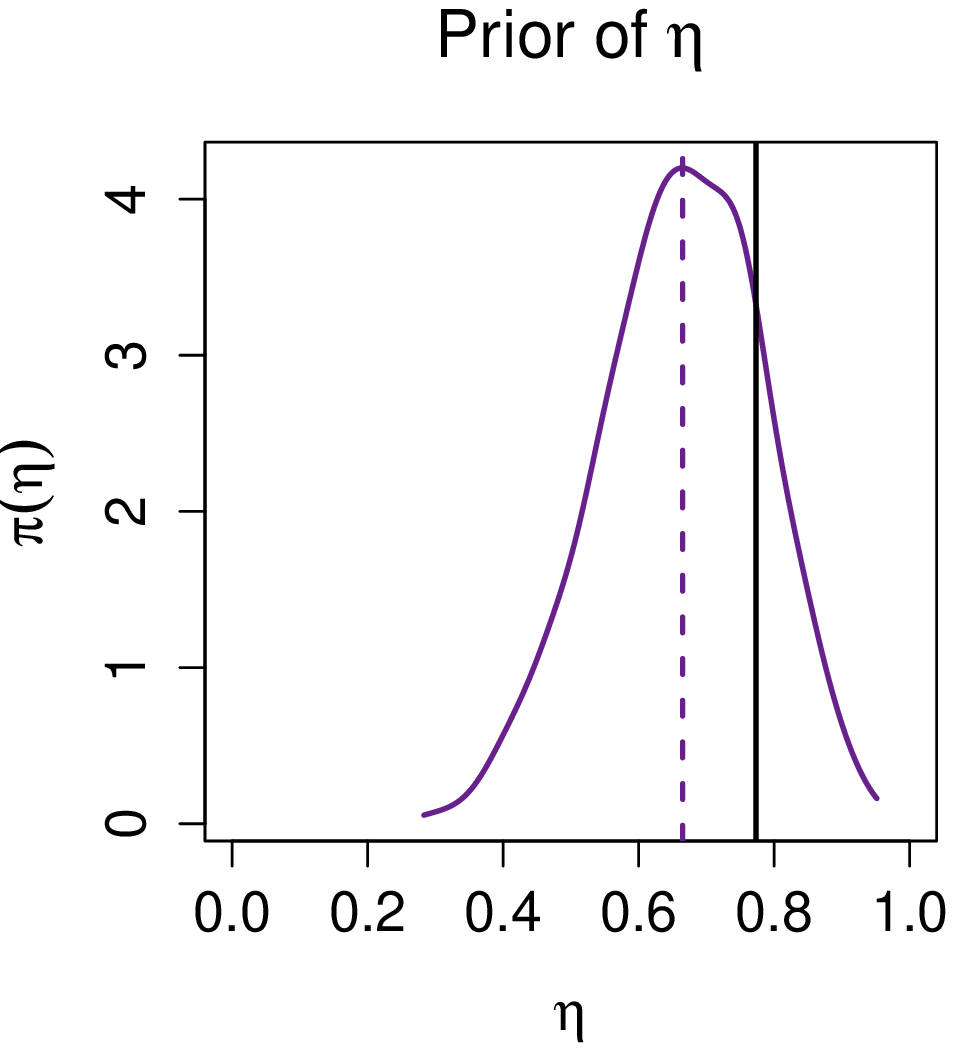} \\ \hline
$n=15$
&
\includegraphics[scale = .3]{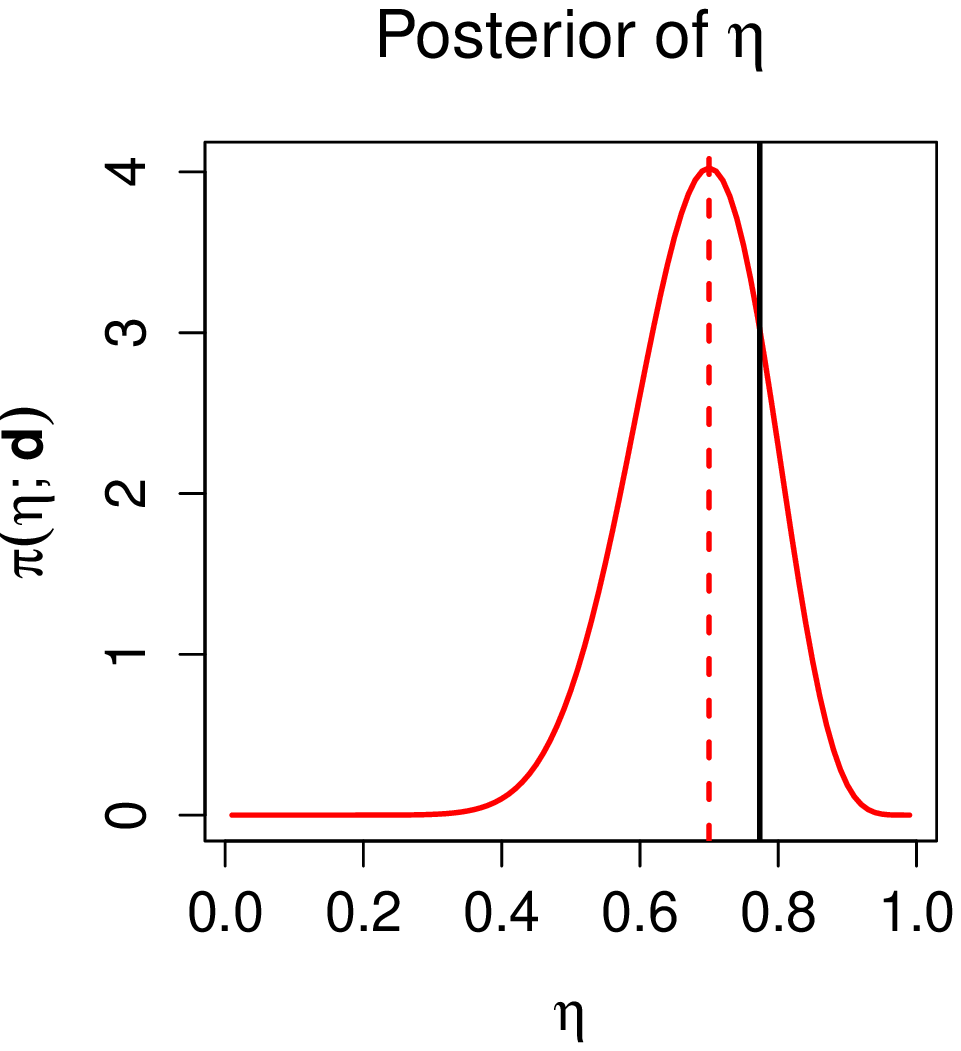}
& 
\includegraphics[scale = .3]{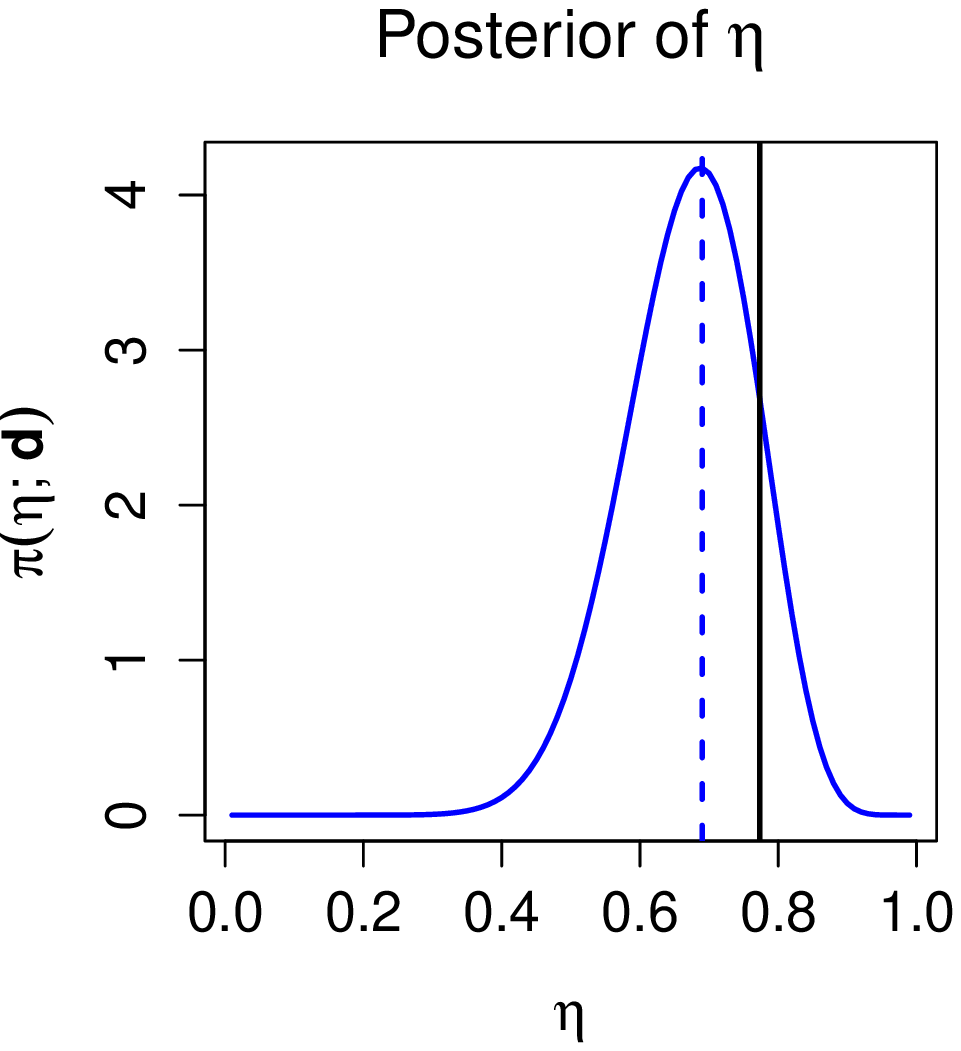} 
&
\includegraphics[scale = .3]{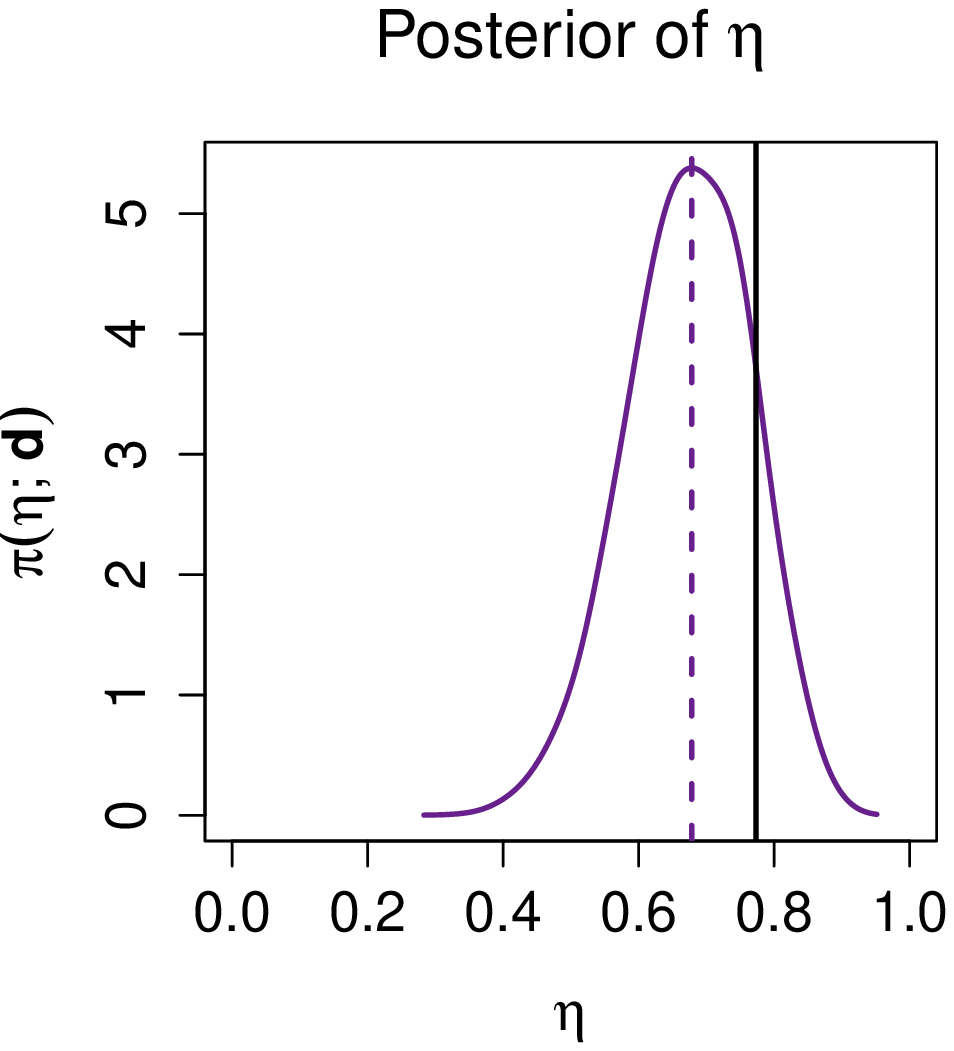}\\ \hline
$n=30$
&
\includegraphics[scale = .3]{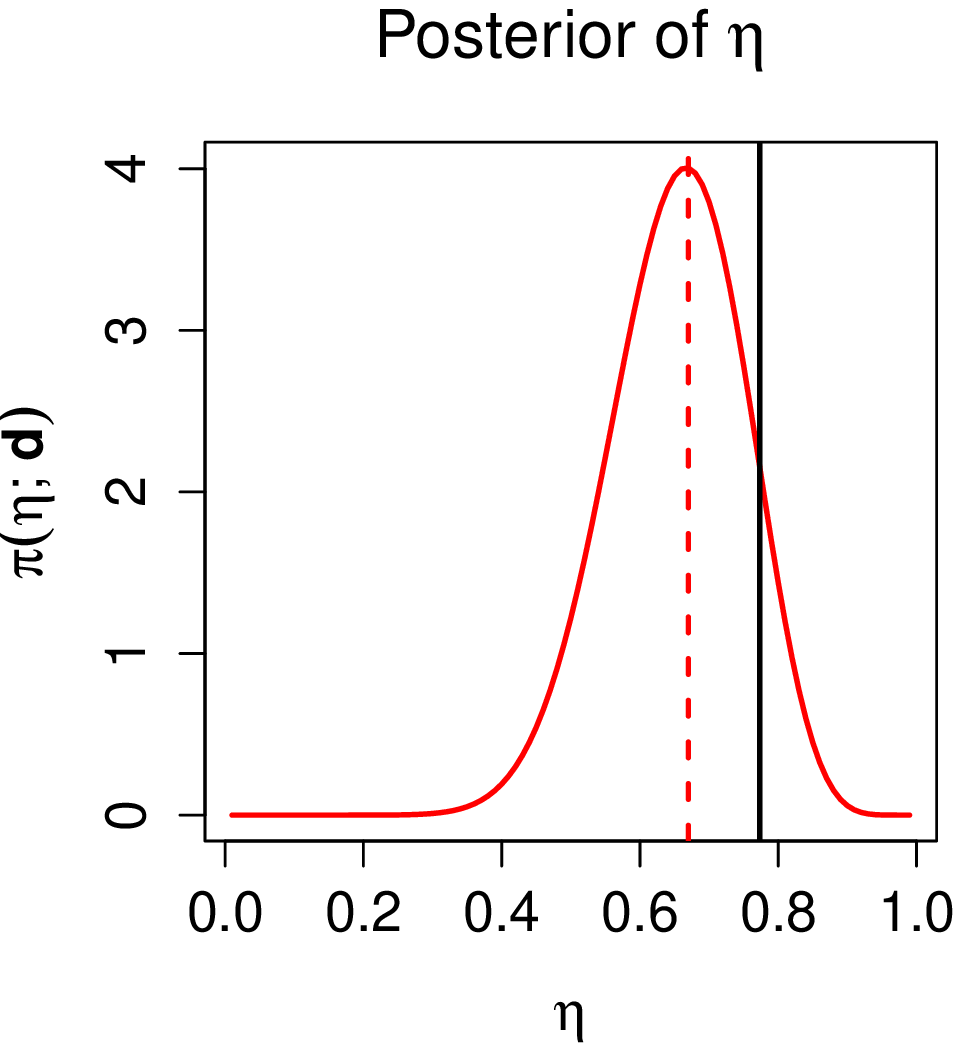}
& 
\includegraphics[scale = .3]{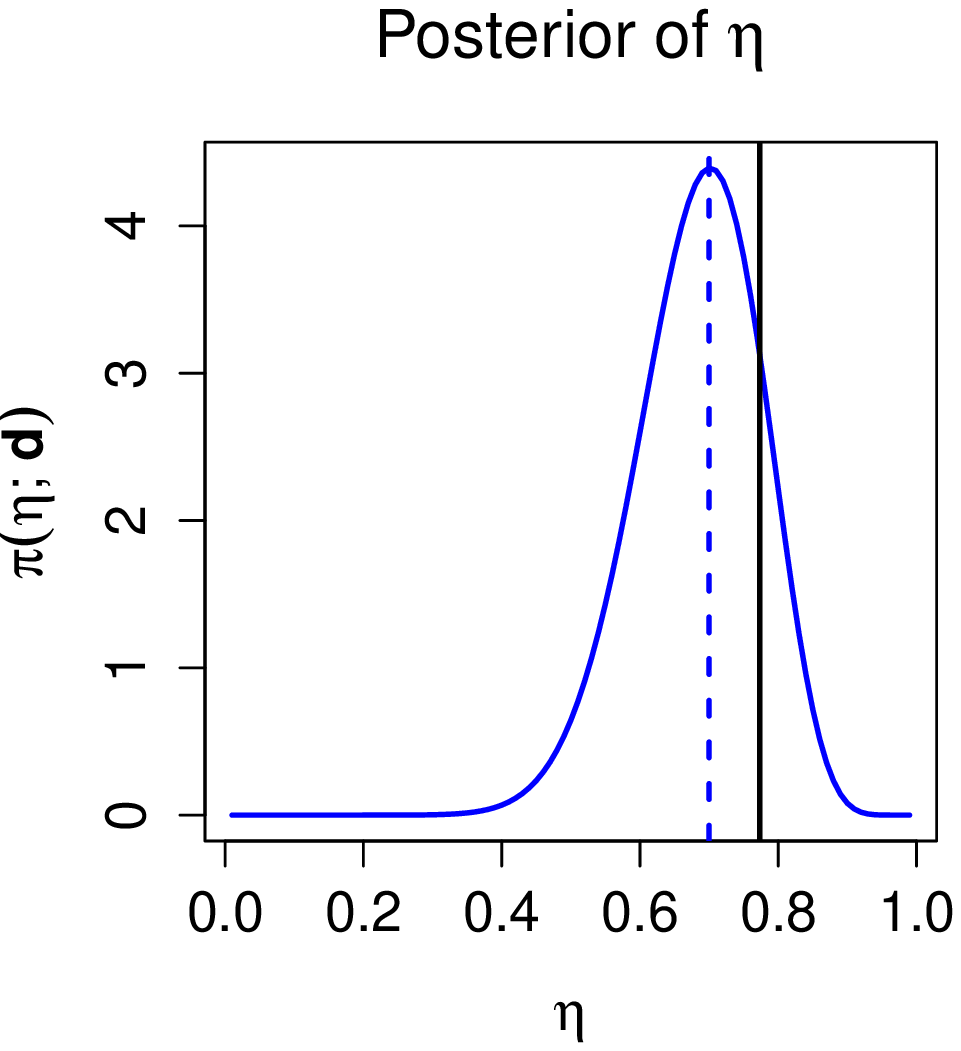} 
&
\includegraphics[scale = .3]{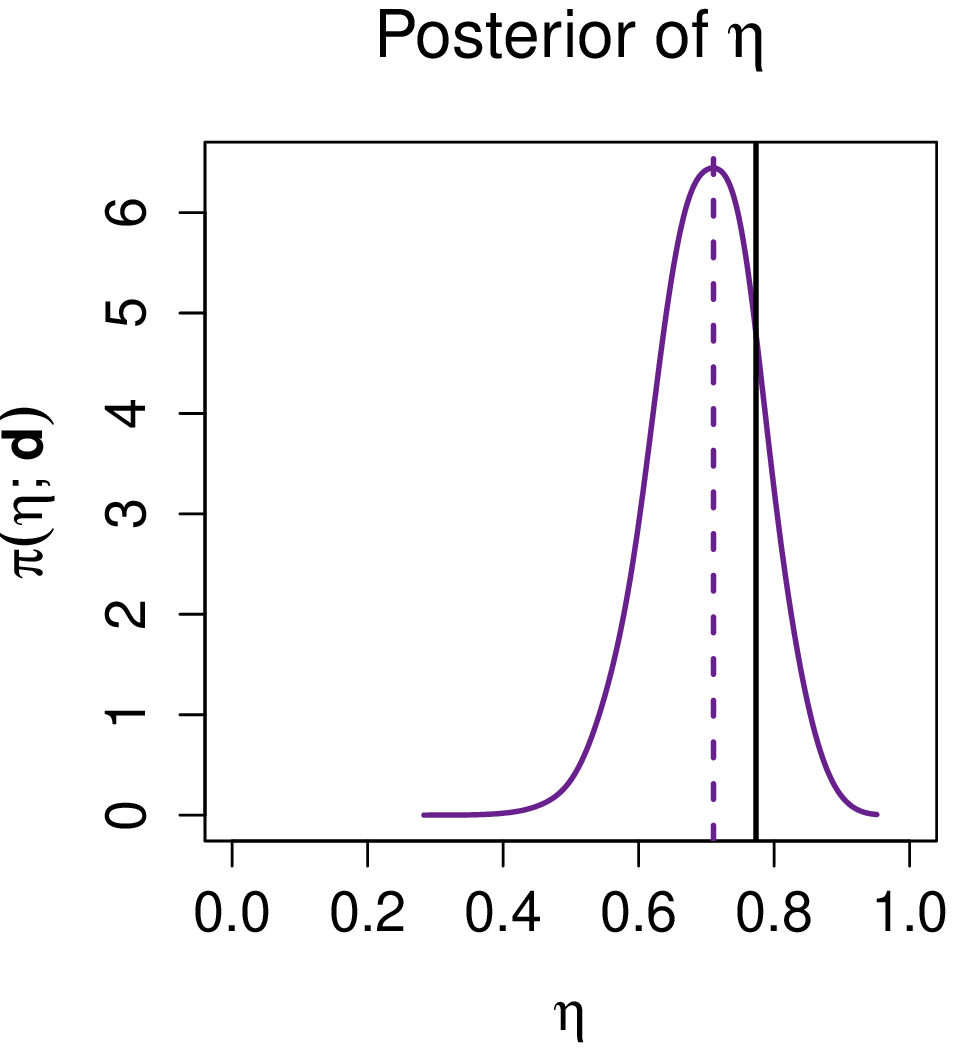}\\ \hline
$n=50$
&
\includegraphics[scale = .3]{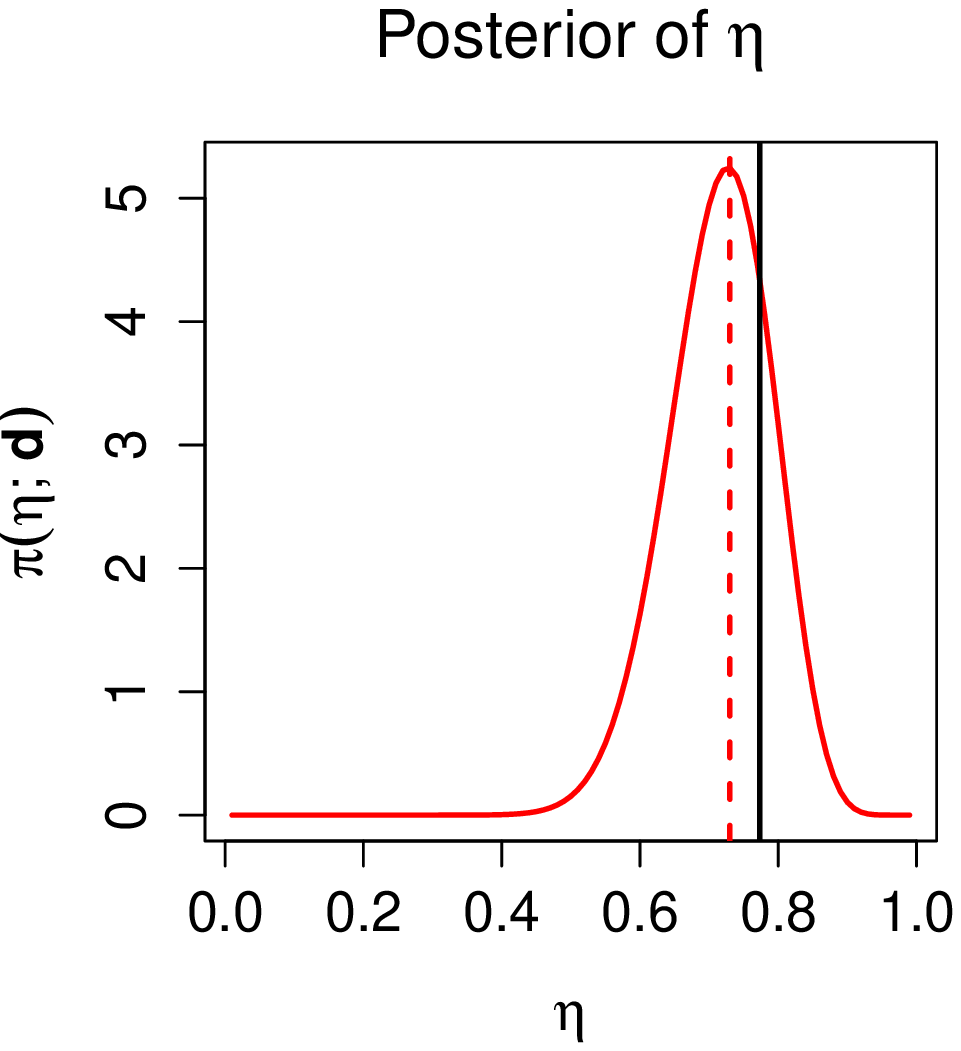}
& 
\includegraphics[scale = .3]{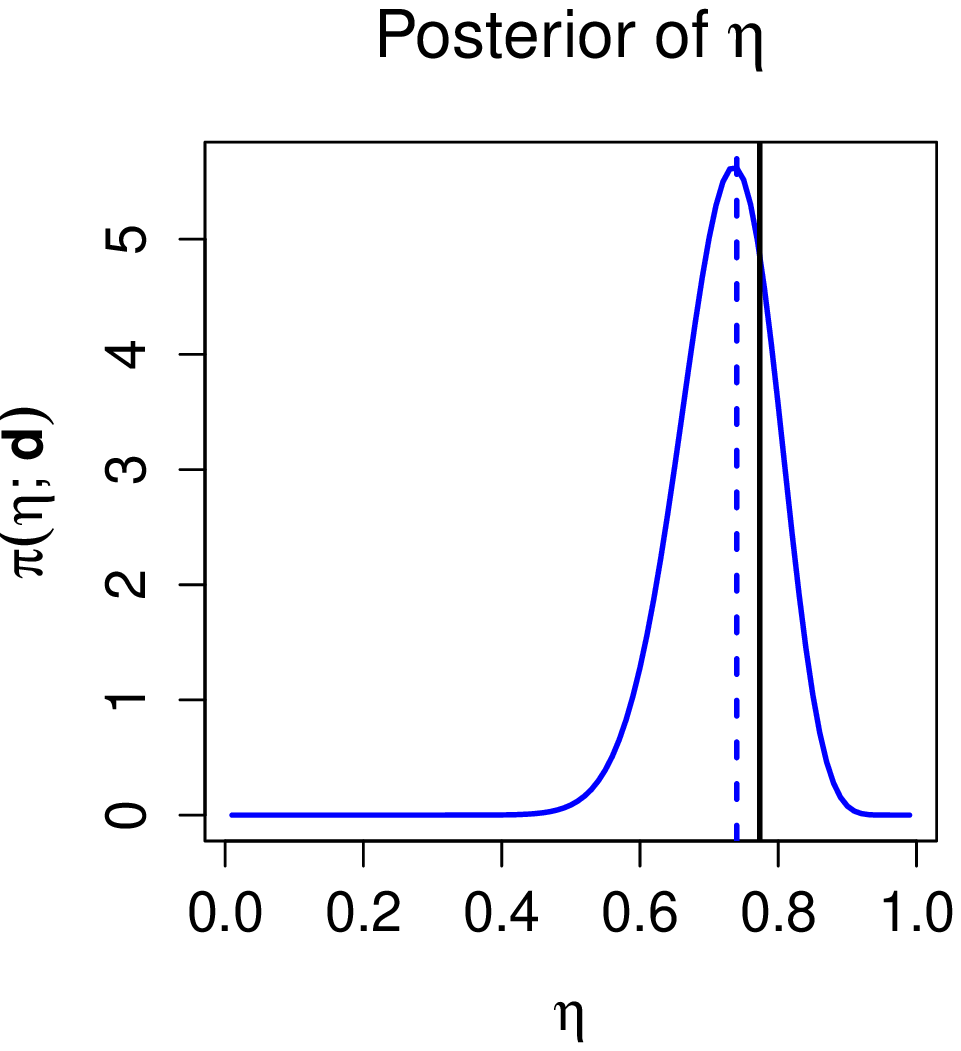} 
&
\includegraphics[scale = .3]{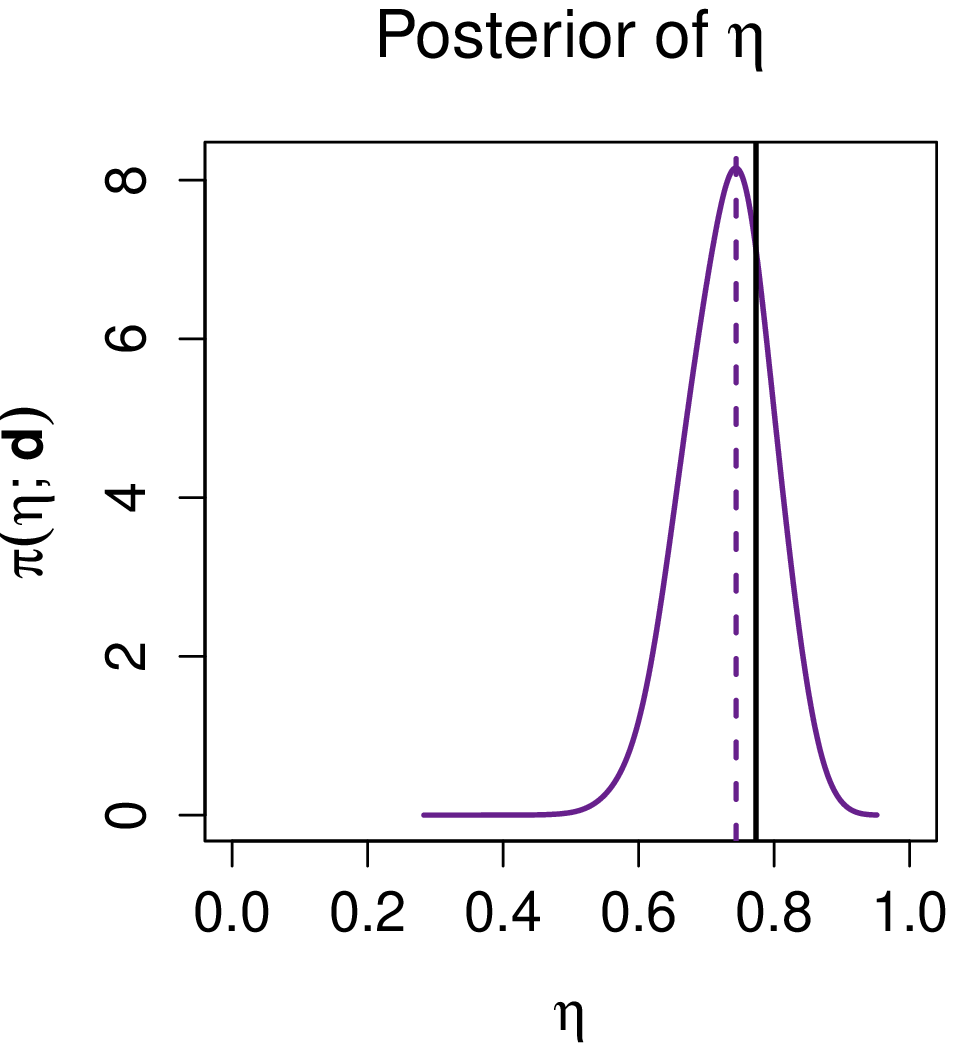}\\ \hline
$n=100$
&
\includegraphics[scale = .3]{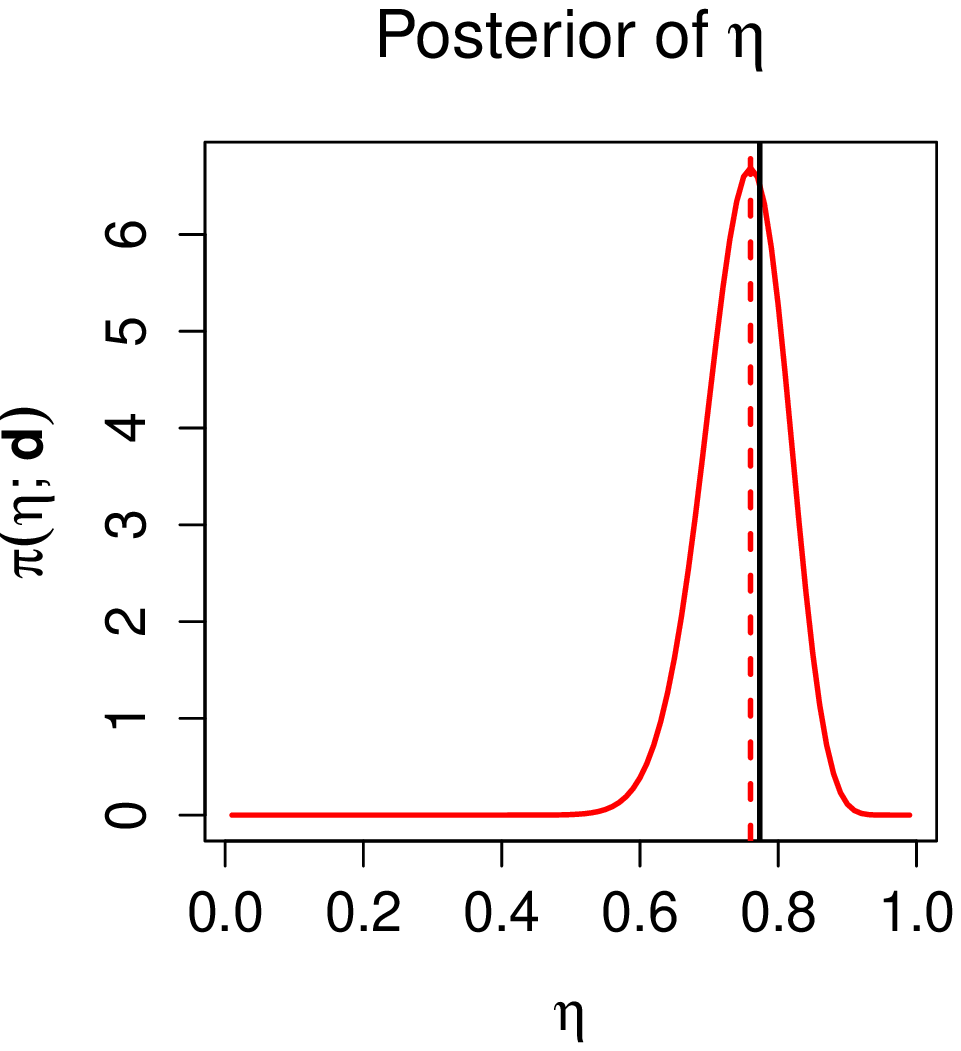}
& 
\includegraphics[scale = .3]{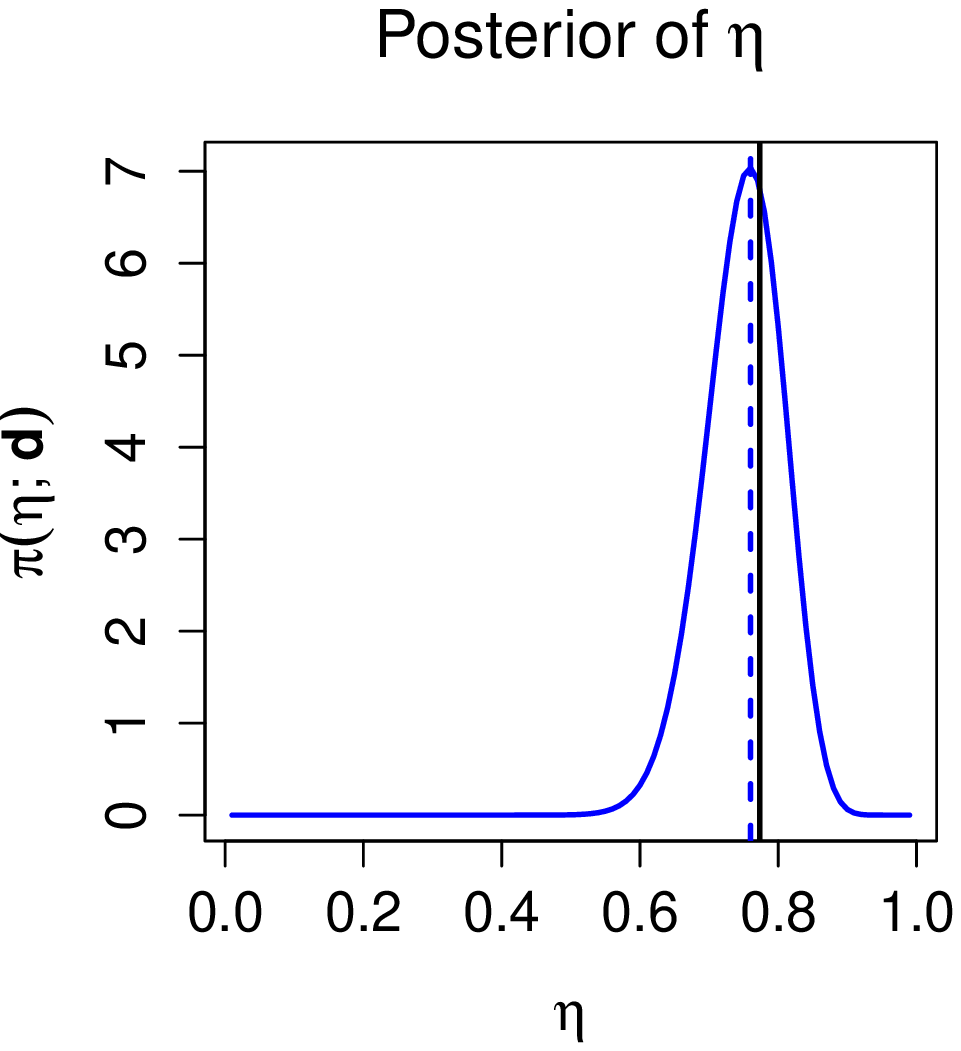} 
&
\includegraphics[scale = .3]{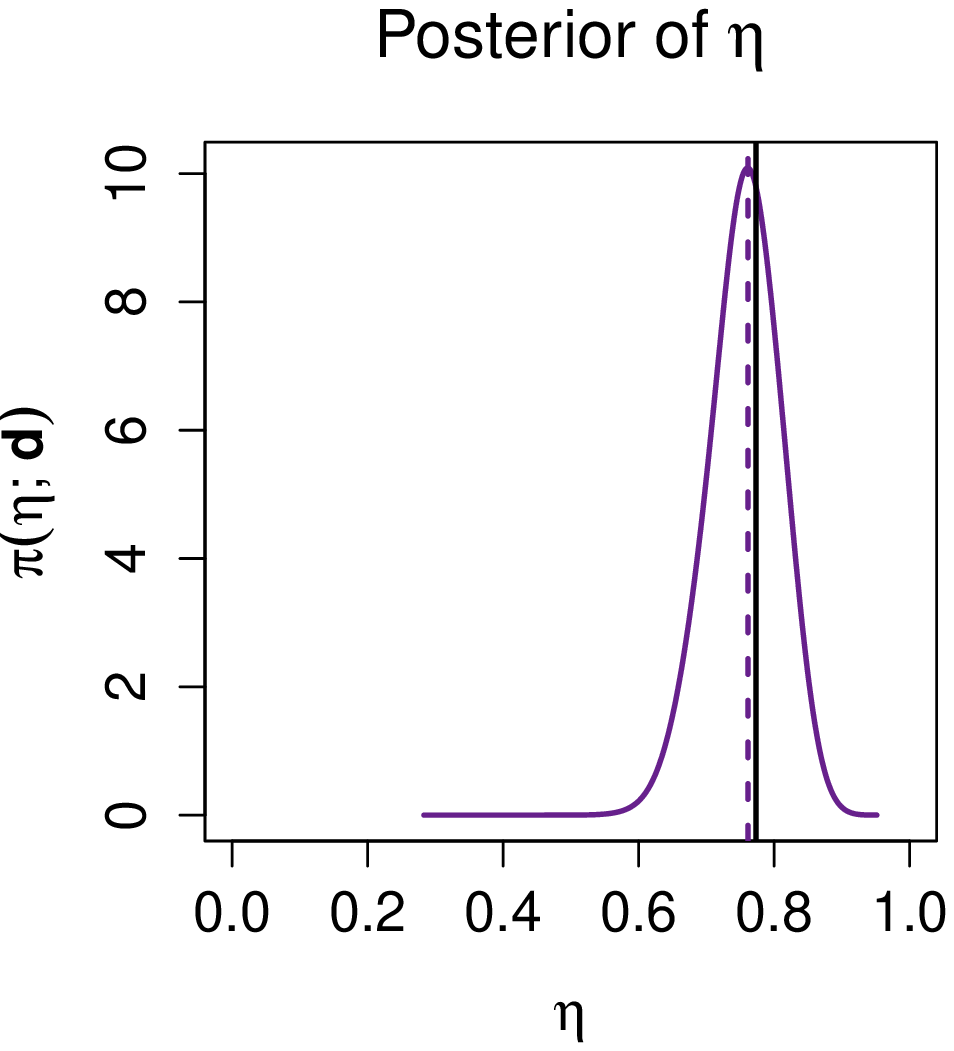}\\ \hline
\end{tabular}
\label{tbl:etaplots}
\end{table}

\begin{table}
\caption{Comparison of the Prior and Posterior Marginal Distributions of $\theta$.} 
\centering
\begin{tabular}{ | m{.6in} | m{1.6in} | m{1.6in} | m{1.6in}|}
\hline
Values of $n$ & \textcolor{red}{Independent} \newline $\eta\sim {\cal B}(10,5)$ \newline $\theta\sim {\cal B}(5,2.5)$ \newline $\rho=0$ & \textcolor{blue}{$(OL)^-$}\newline $\alpha_1=10$, $\alpha_2=2.5$, \newline$\alpha_3=5$  \newline $\rho=-0.45$ & \textcolor{violet}{$AN(5)$} \newline $\alpha_i=5,\ \ i=1,\dots,4$, $\alpha_5=0.0001$  \newline $\rho=-0.65$\\ \hline
$n=0$ \newline (Prior)
&
\includegraphics[scale = .3]{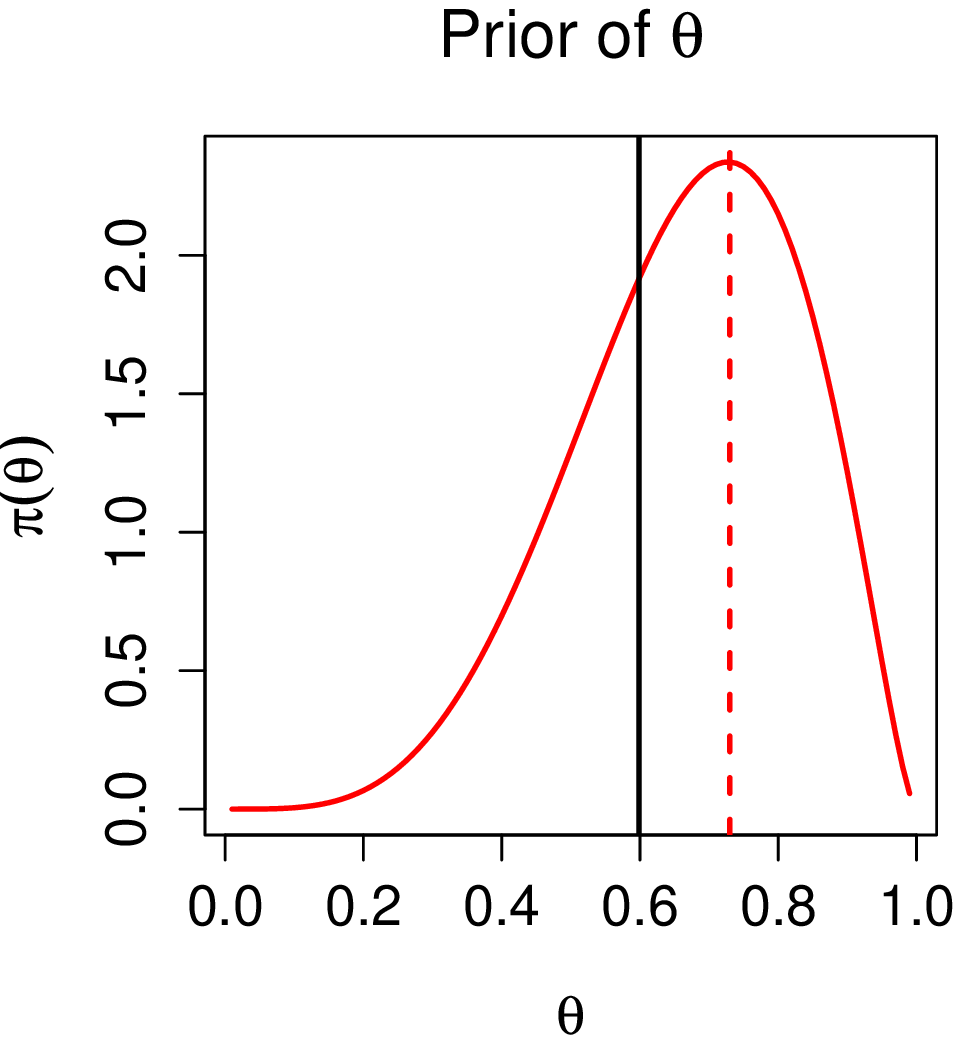}
& 
\includegraphics[scale = .3]{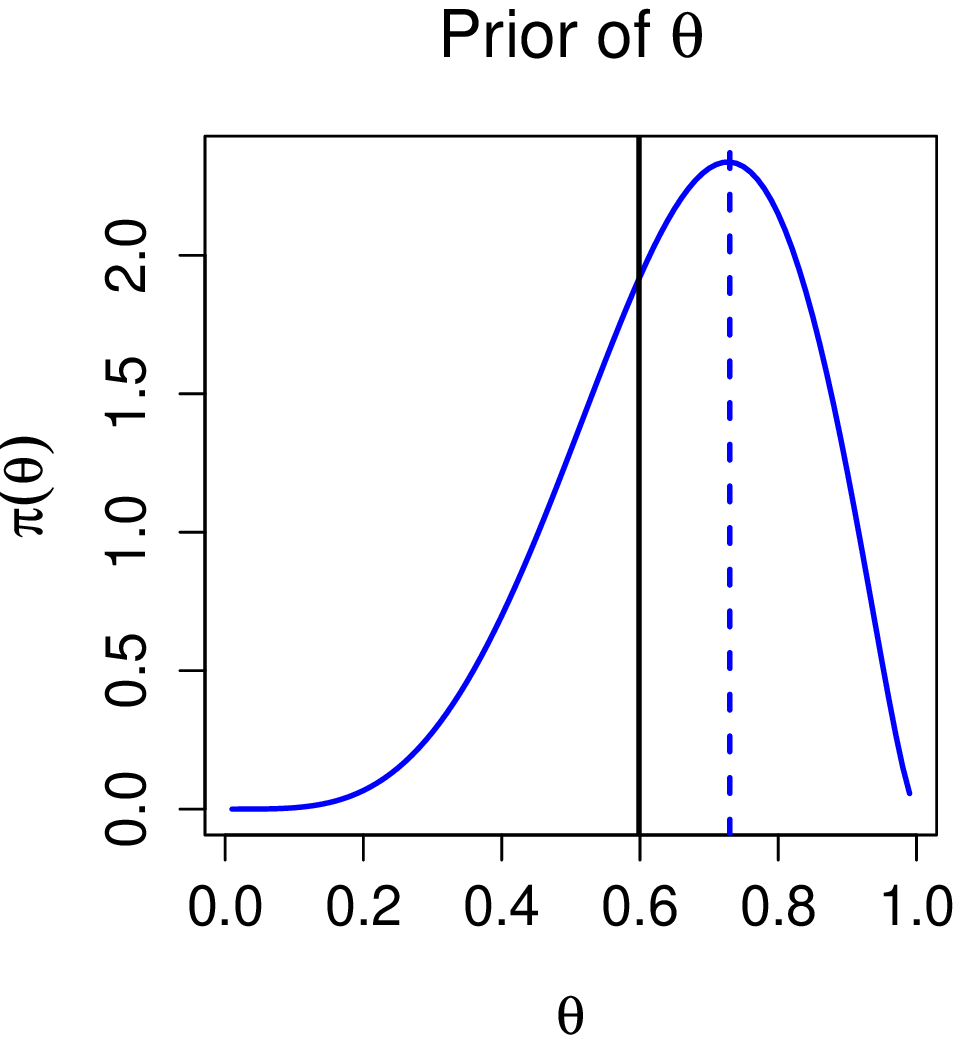} 
&
\includegraphics[scale = .3]{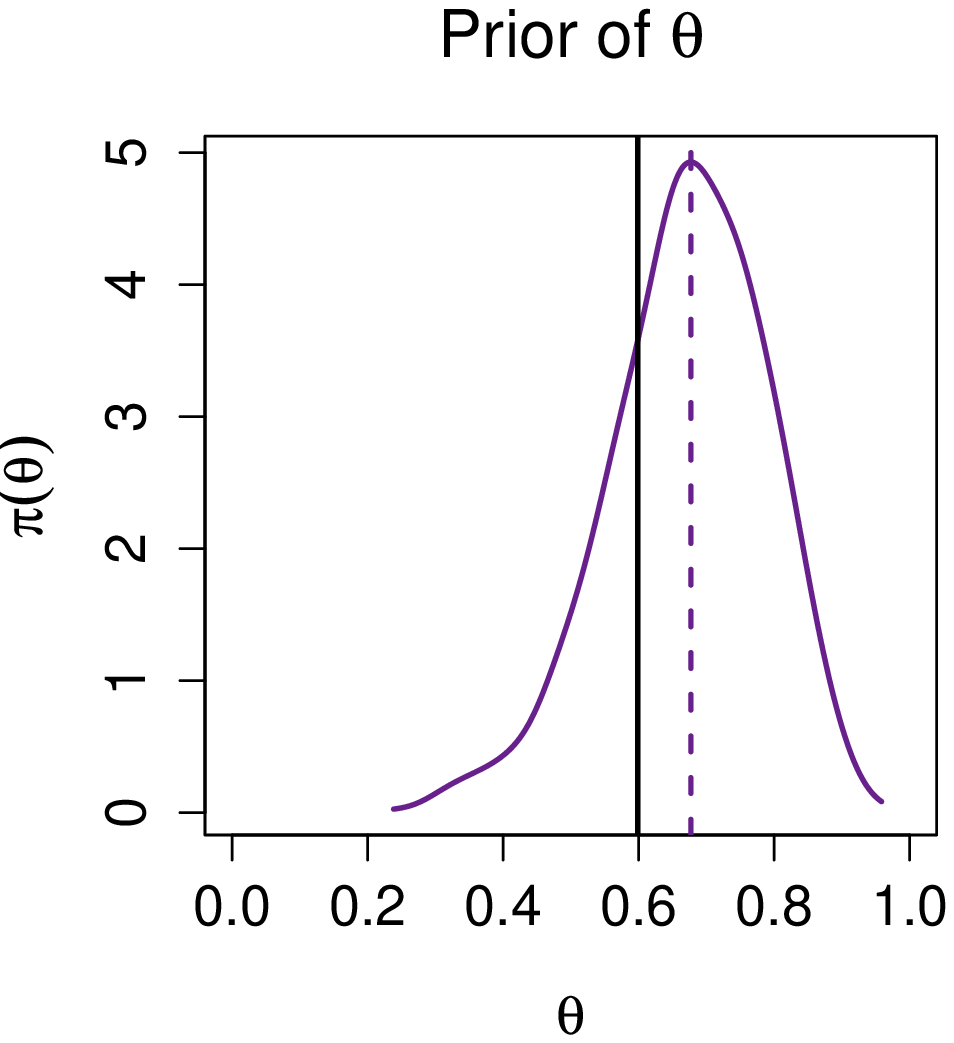} \\ \hline
$n=15$
&
\includegraphics[scale = .3]{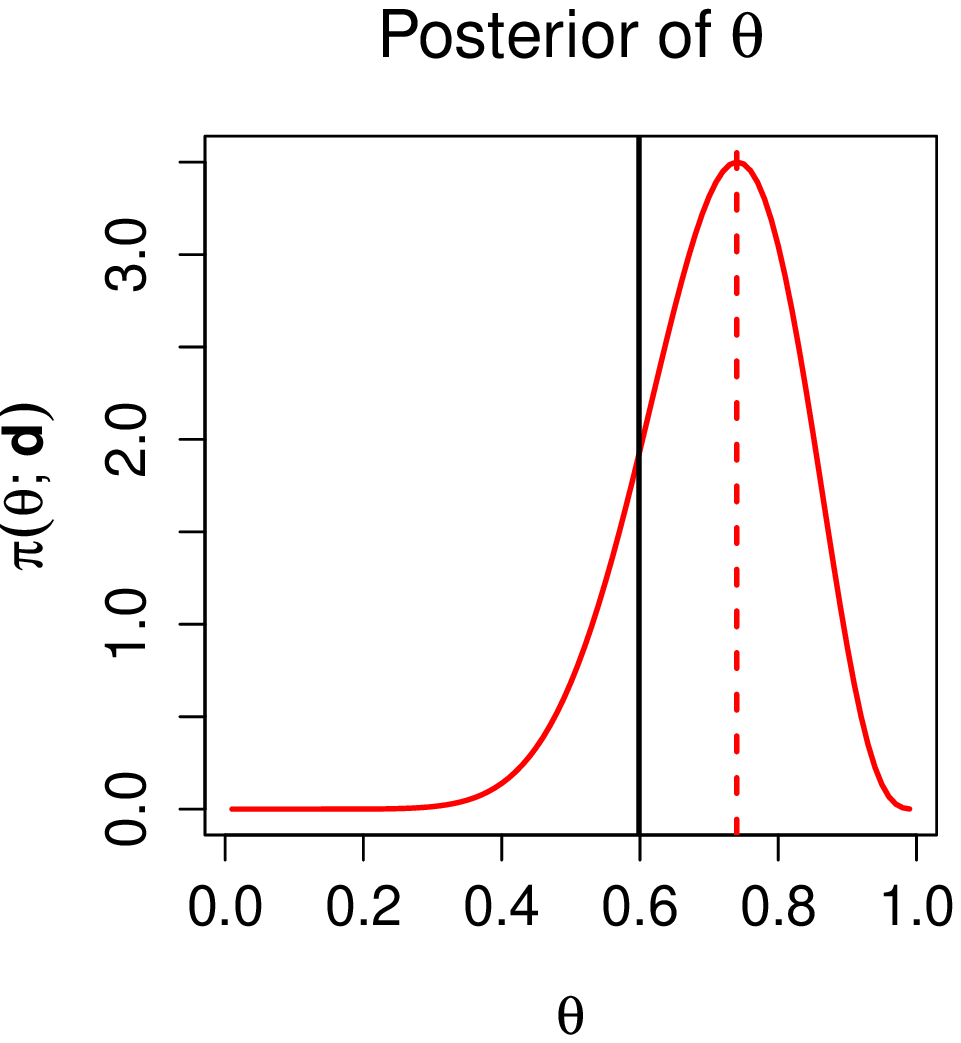}
& 
\includegraphics[scale = .3]{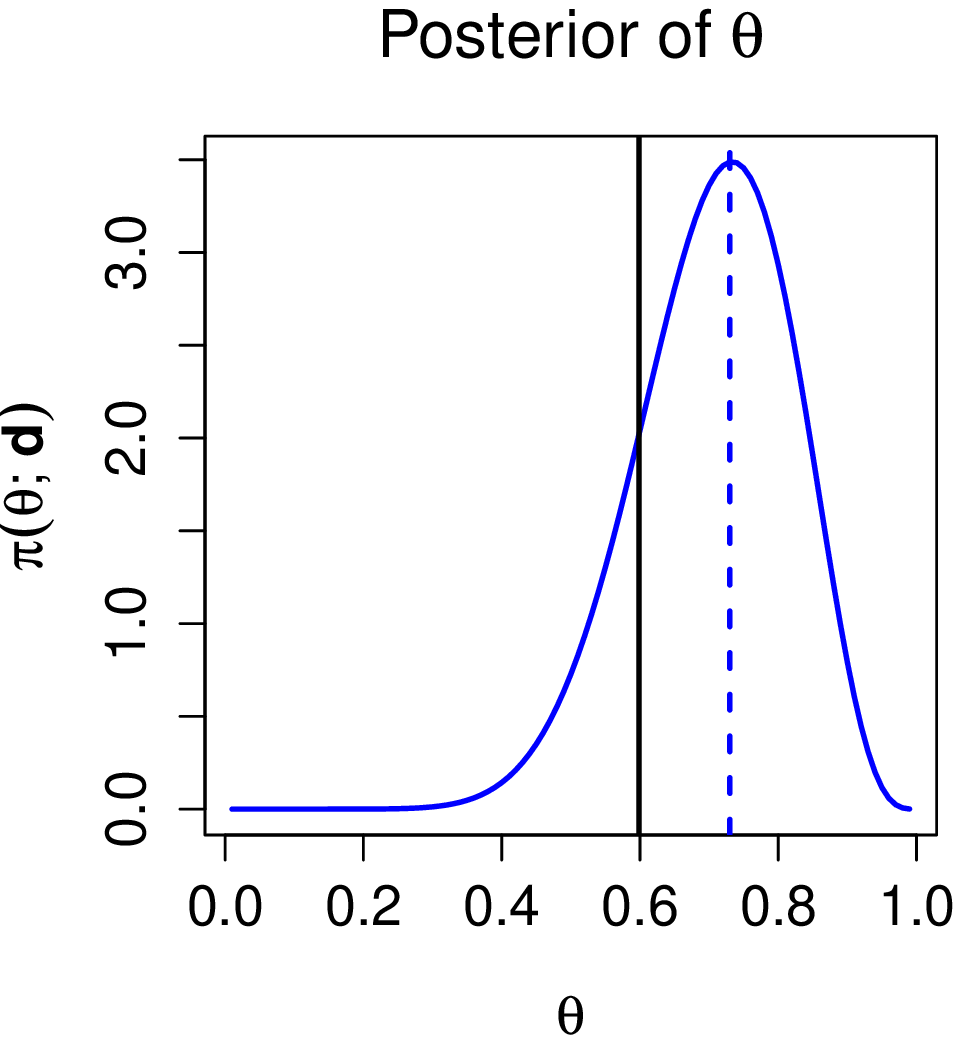} 
&
\includegraphics[scale = .3]{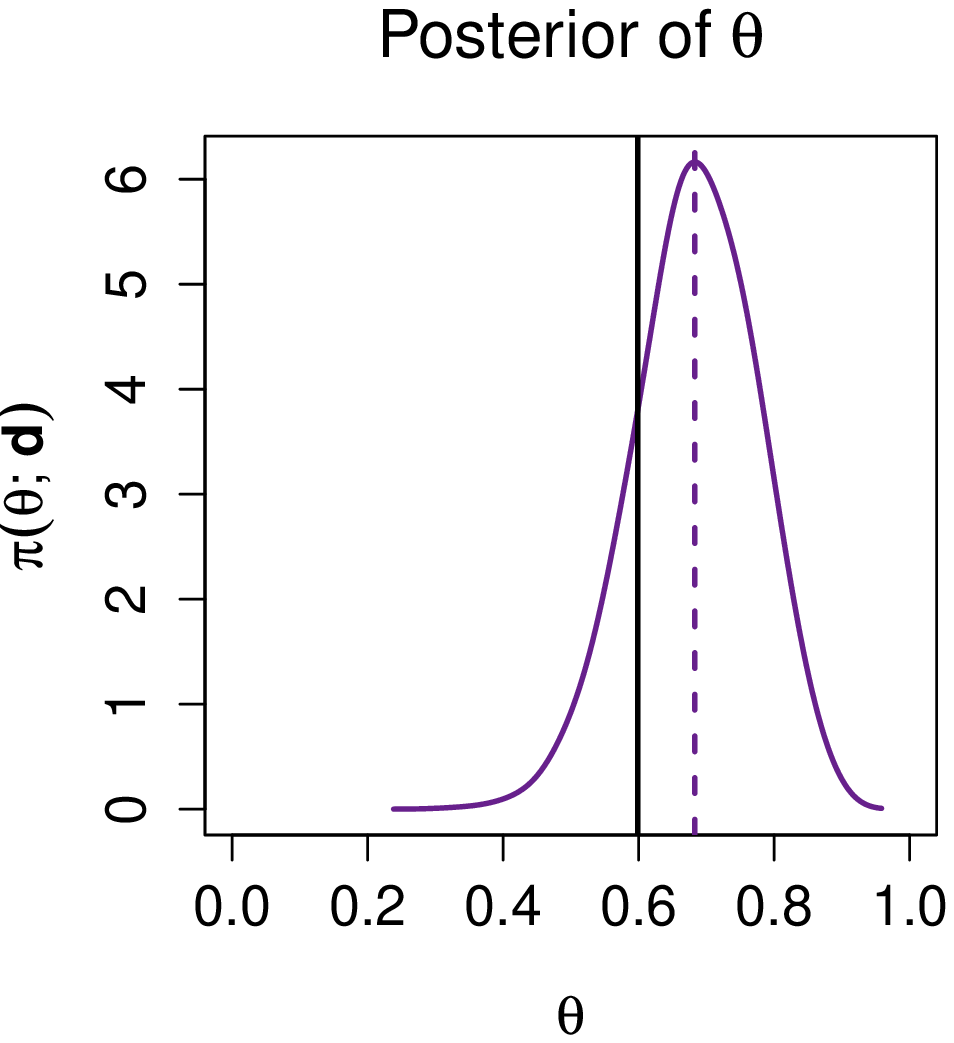}\\ \hline
$n=30$
&
\includegraphics[scale = .3]{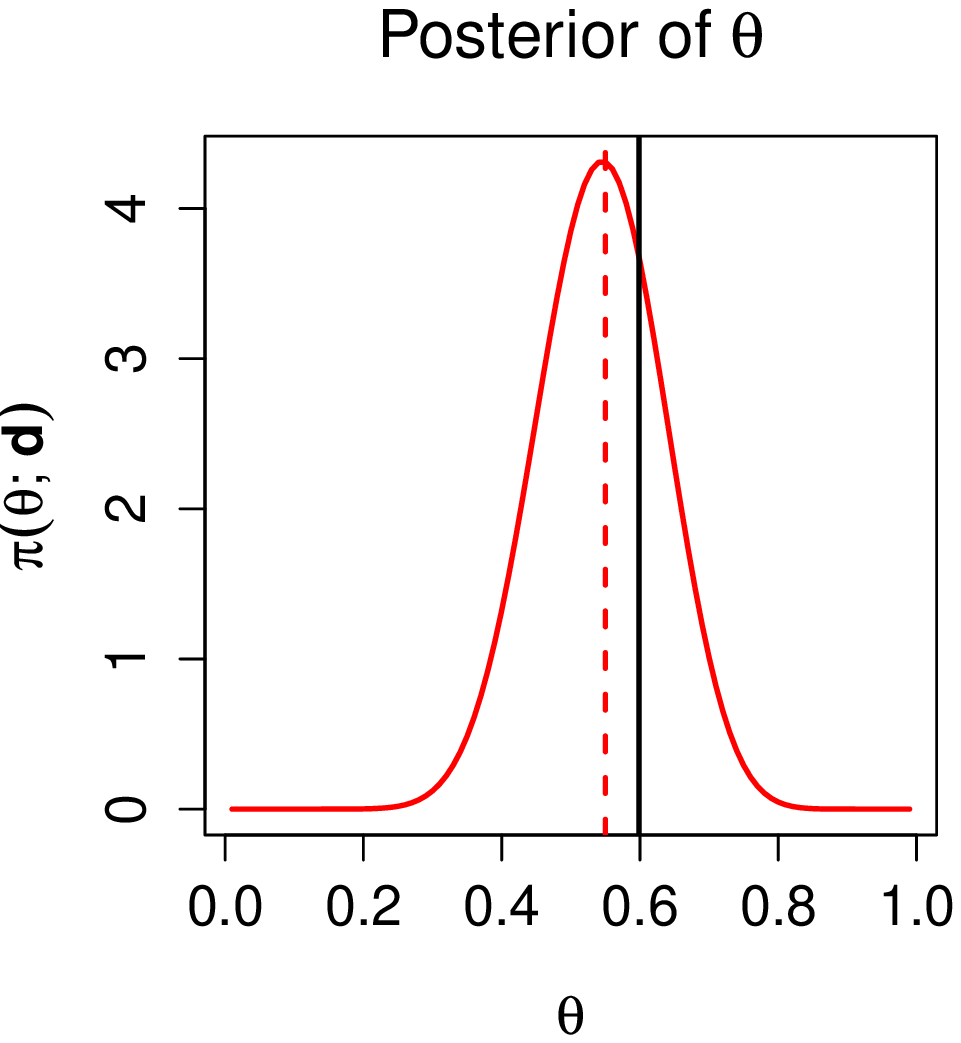}
& 
\includegraphics[scale = .3]{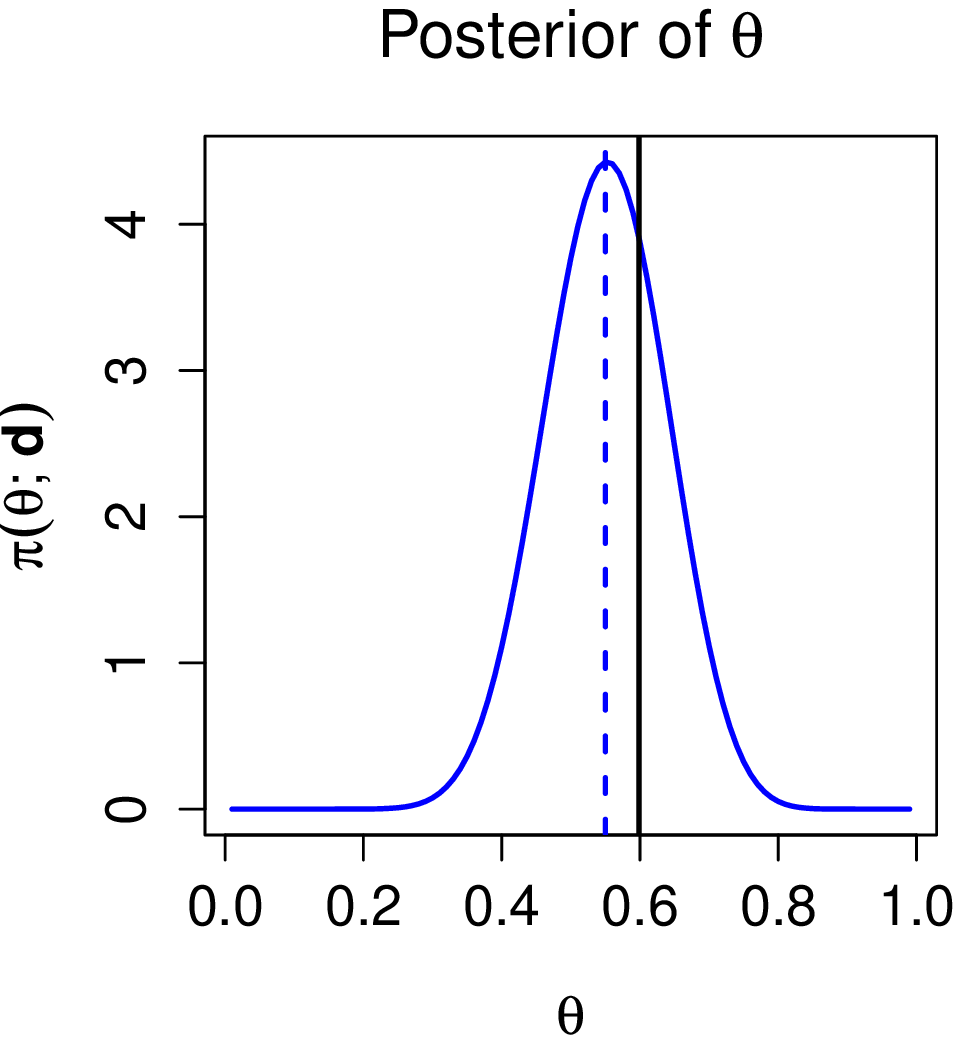} 
&
\includegraphics[scale = .3]{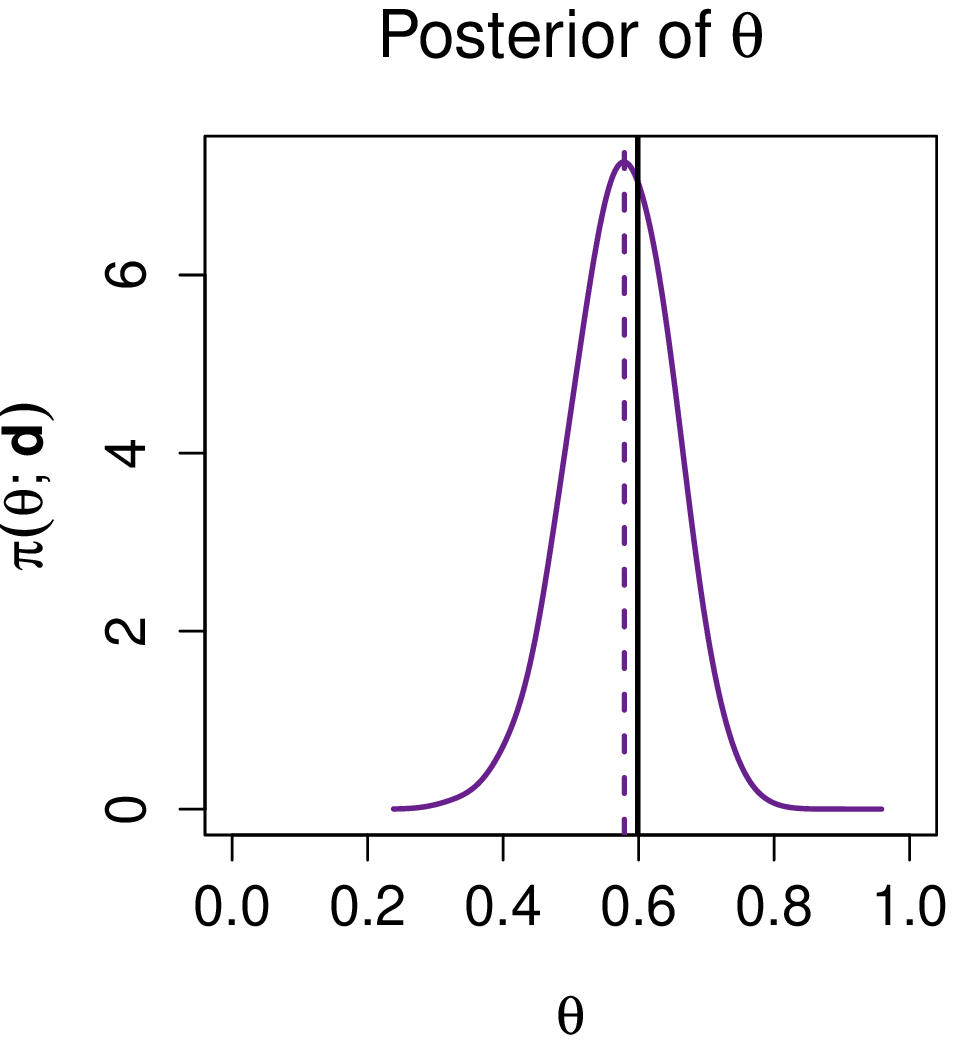}\\ \hline
$n=50$
&
\includegraphics[scale = .3]{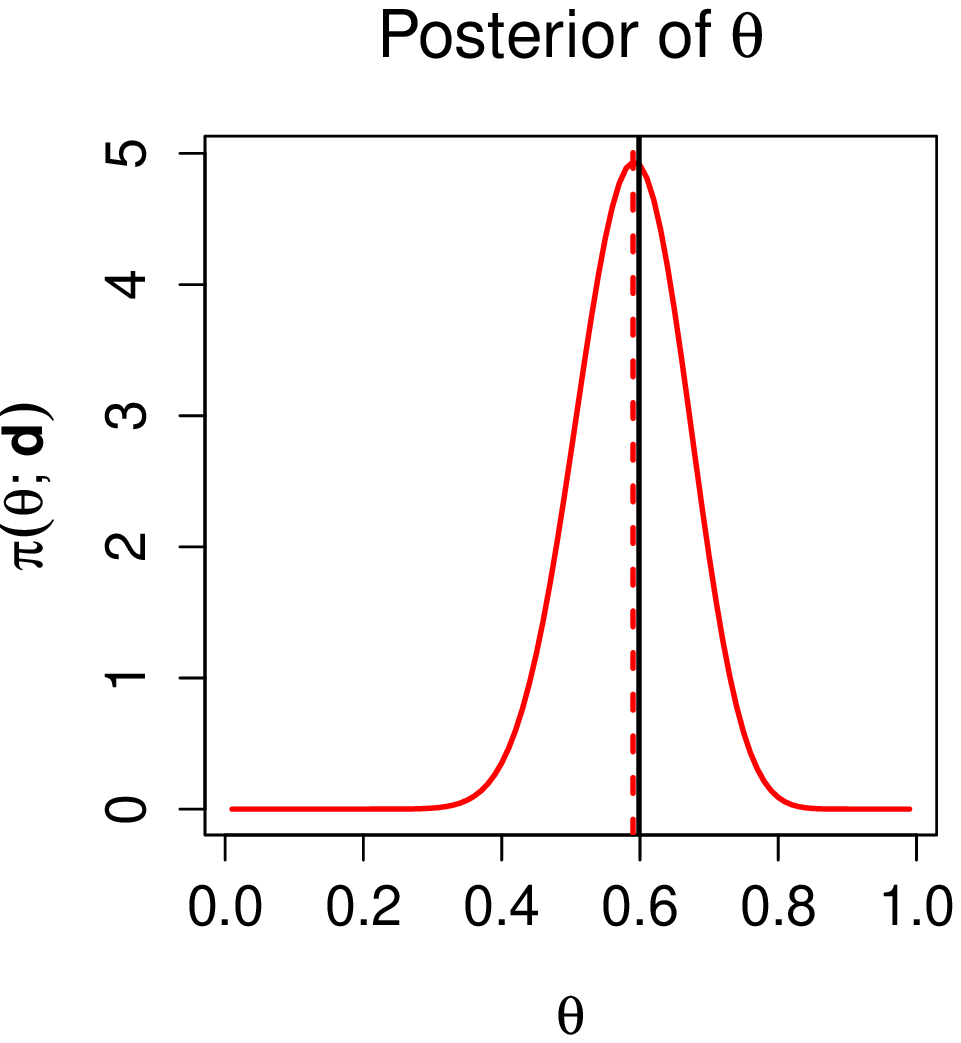}
& 
\includegraphics[scale = .3]{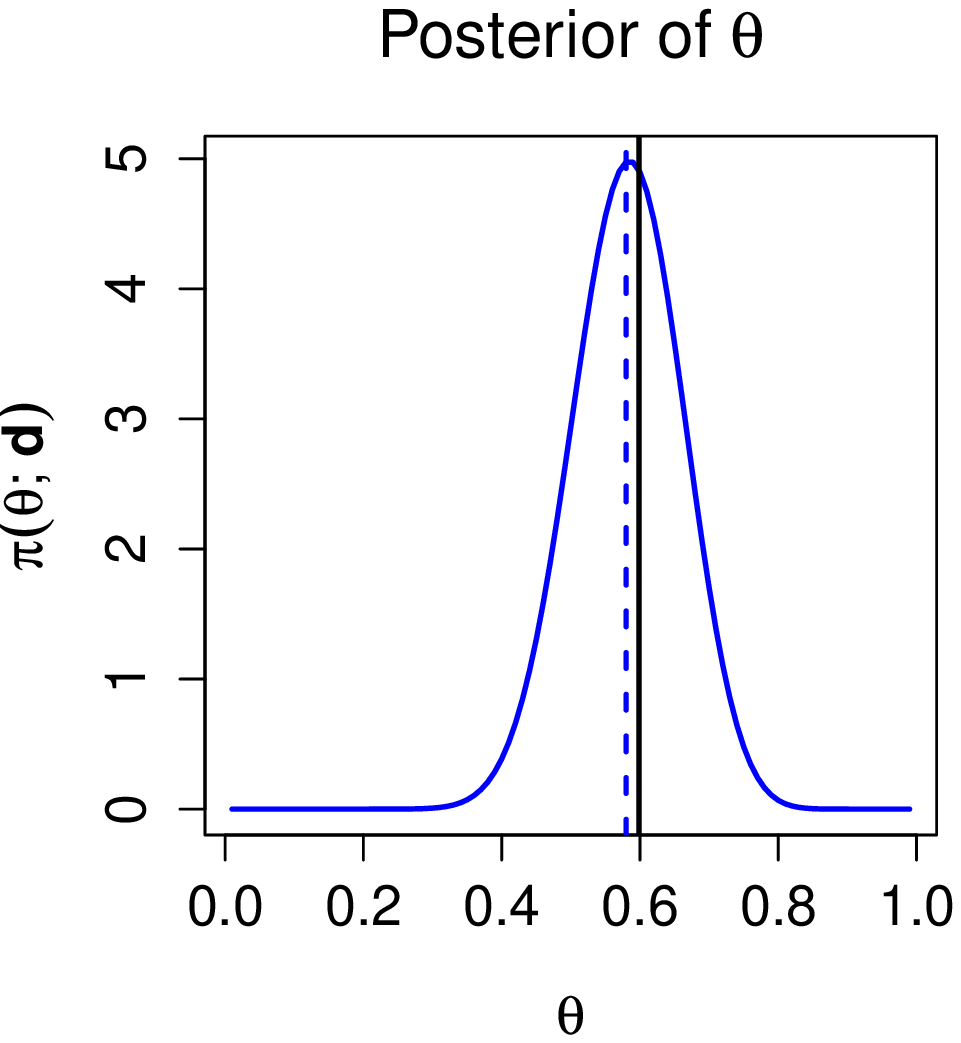} 
&
\includegraphics[scale = .3]{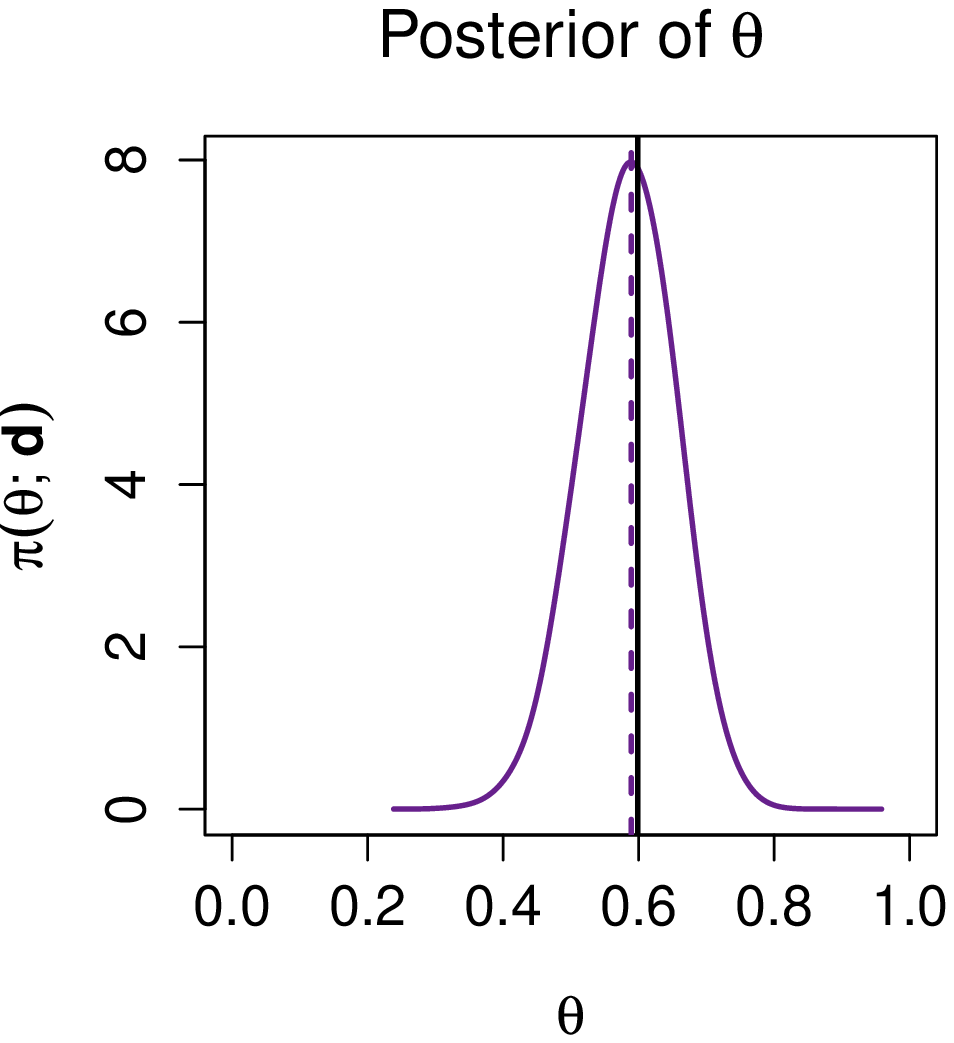}\\ \hline
$n=100$
&
\includegraphics[scale = .3]{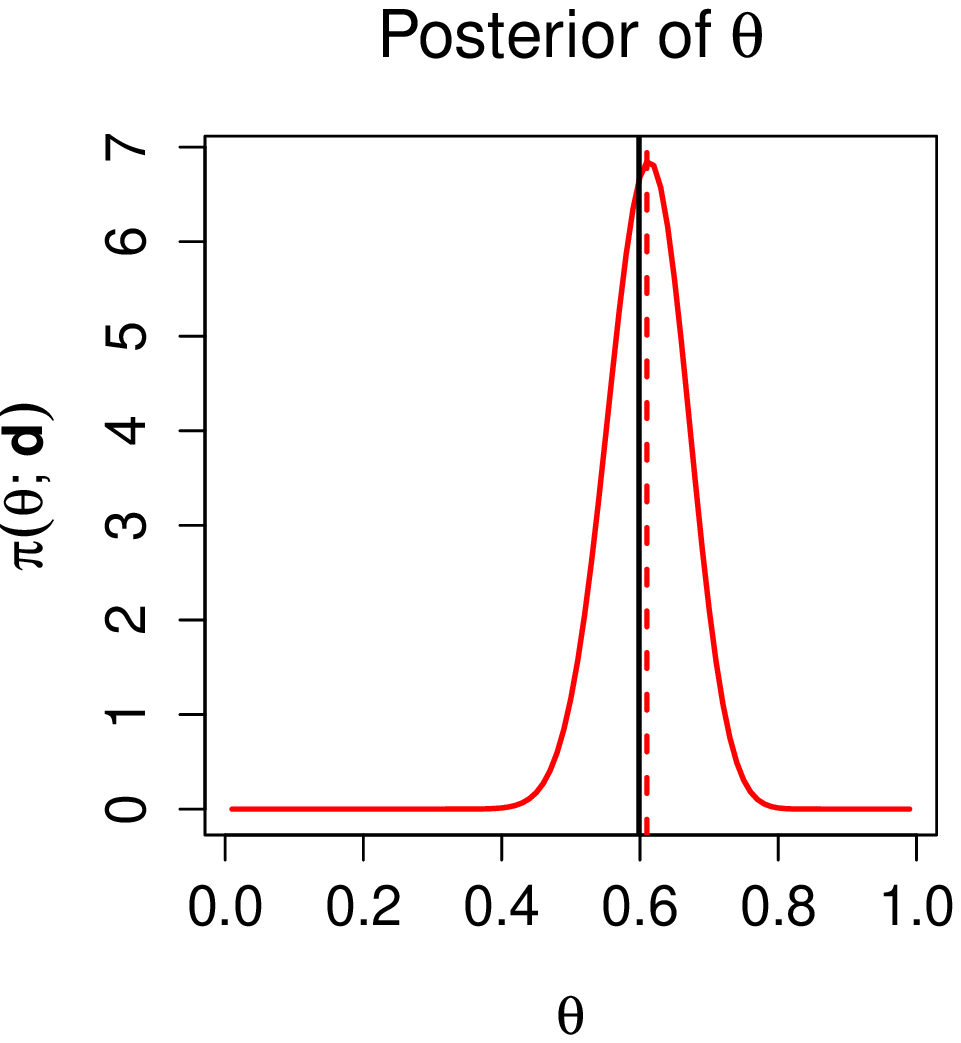}
& 
\includegraphics[scale = .3]{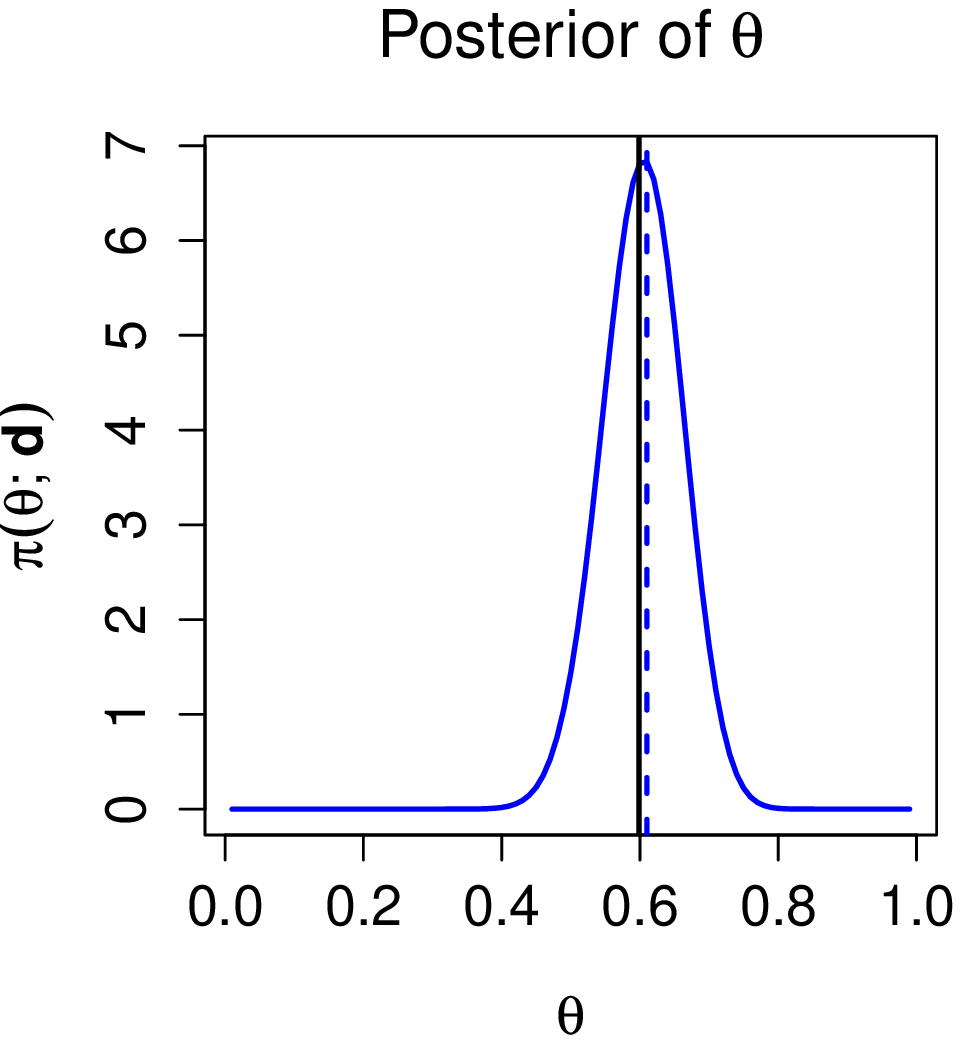} 
&
\includegraphics[scale = .3]{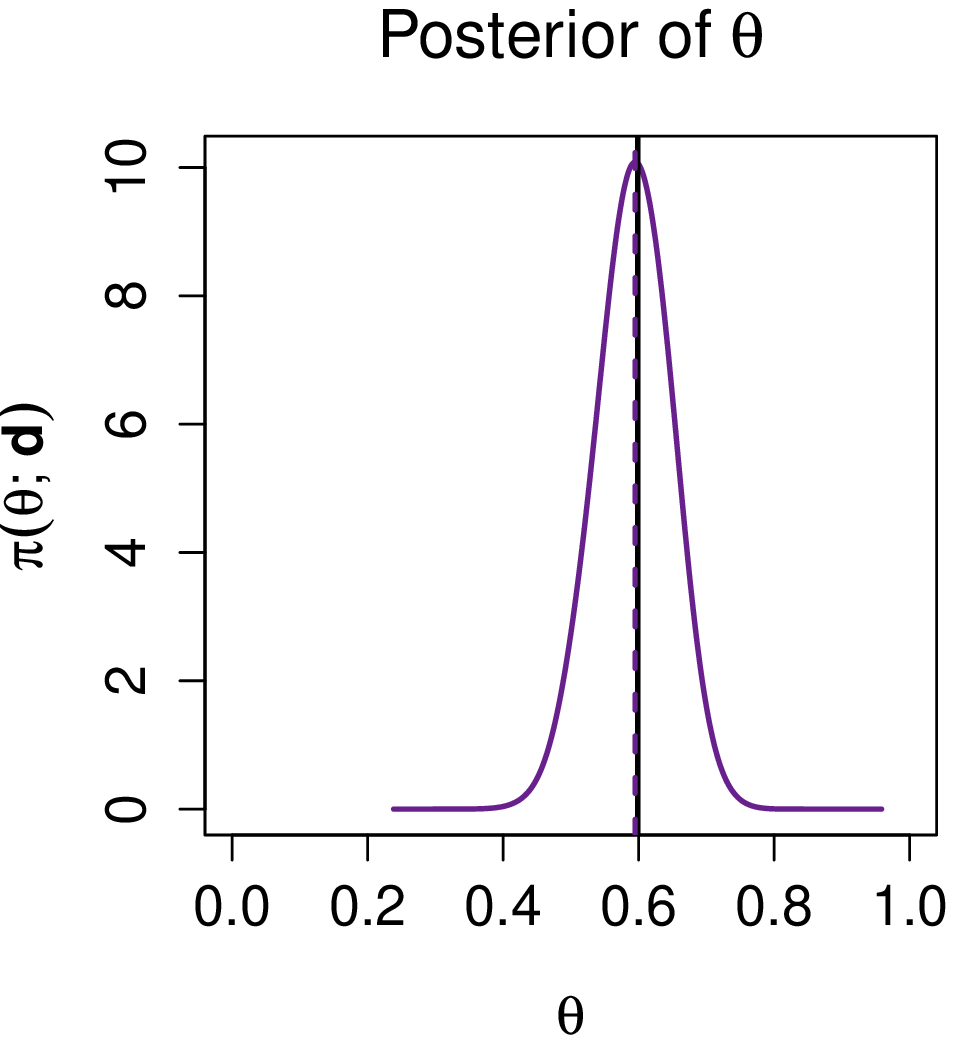}\\ \hline
\end{tabular}
\label{tbl:thetaplots}
\end{table}

\newpage 
{\noindent {\underline {Qualitative Comments on Results}}} 

An examination of the contour plots of Table l shows that the centroids of these contours get closer to their true values as $n$ increases; this is to be expected of any meaningfully construed Bayesian procedure. Furthermore, the contours become more and more concentrated as $n$ increases, with the $AN(5)$ based prior contours showing the greatest movement; this is followed by the $(OL)^{-}$ based contours. Convergence of the posterior centroid to its true value occurs for the $AN(5)$ prior, for $n$ as small as 30. This is followed by the $(OL)^{-}$ based prior. For $n \geq 50$, all three priors appear to give similar results.

The negative correlation of the $AN(5)$ prior is retained by the posterior for all the values of $n$. Surprisingly, the negative correlation of the $(OL)^{-}$ based posterior distribution diminishes as $n$ increases, and almost vanishes when $n = 50$. In general, the centroids of the contour plots generated by the $AN(5)$ prior distribution tend to be closer to the true values of $\eta$ and $\theta$ as compared to the centroids of the contours provided by the other priors.

An examination of the plots of the marginal posterior distributions of $\eta$ given in Table 2 suggests that the modes of the posterior distributions get closer to the true values of $\eta$ as $n$ increases; again, this is to be expected. The dispersion of the posterior distribution of $\eta$ is the smallest for the $AN(5)$ distribution over all values of $n$. Not withstanding the above, it appears that all three prior distributions
provide satisfactory inferences for $\eta$.

An examination of the plots of the marginal posterior distributions of $\theta$ given in Table 3 suggests a behavior that parallels the behavior of the plots in Table 2. 

Based on the above, we are inclined to conclude that for small sample sizes, the $AN(5)$ as a joint prior distribution for $\eta$ and $\theta$ performs better vis-{\`a}-vis closeness of the centroid of the posterior distribution and the modes of the marginal distributions to the true values of $\eta$ and $\theta$, than those under the $(OL)^{-}$ and the {\it independent} priors. Also, the posterior contours are sharper under $AN(5)$
prior distribution than the others. When $n$ is large, say $n = 100$, the choice of the prior does not seem to matter, save for the fact that the $AN(5)$ based prior tends to retain, albeit slightly, the negative dependence between $\eta$ and $\theta$.

\subsection{Assessing Predictive Values of Screening Tests}

With $\eta$ and $\theta$ interpreted as probabilities, and the joint posterior distribution of $\eta$ and $\theta$ together with the posterior of $\pi$ at hand, assessing $\Lambda = \Pr(D|S)$ and $\Psi = \Pr({\bar D}| {\bar S})$ 
can be achieved by numerical methods, using the fact that: 
\begin{eqnarray*}
\Lambda = \frac{\eta \pi}{\eta \pi + (1 - \theta)(1 - \pi)} {\mbox { and }}  \Psi = \frac{\theta (1 - \pi)}{\theta (1- \pi) + (1 - \eta)\pi}. 
\end{eqnarray*} 
However, with $\eta$ and $\theta$ interpreted as propensities, the above relations for $\Lambda$ and $\Psi$ are not meaningful because the probability calculus cannot be invoked on propensities. All the same, if a new item subjected to a screening test reveals an $S$, then interpreting $\eta$ as the likelihood of $D$ with $S$ known, can be used in conjunction with Bayes' Law to obtain the posterior predictive probability of $D$ as 
\begin{eqnarray*}
\Pr(D; S, \bd) & \propto & \sum\limits_{i=1}^{n} \eta_{i} \Pr(\eta_{i}; \bd) \pi^{*}, 
\end{eqnarray*} 
where $\pi^{*} = \Pr(D; \bd)$ is the posterior probability of the disease prevalence.  

Similarly, were the item to reveal ${\bar S}$, then 
\begin{eqnarray*}
\Pr({\bar D}; {\bar S}, \bd) & \propto & \sum\limits_{i=1}^{n} \theta_{i} \Pr(\theta_{i}; \bd) (1 - \pi^{*}). 
\end{eqnarray*} 
The quantities $\Pr(\eta_{i}; \bd)$ and $\Pr(\theta_{i}; \bd)$ are those given in Section 3.3. 

\subsection{Discussion and Summary Remarks on Section 3}

Test sensitivity and test specificity are the two key parameters that determine the quality of diagnostic tests used in a variety of fields. These parameters are adversarial in the sense that an increase in one results in a decrease in the other and vice versa. The aim of this paper was to develop a Bayesian inferential mechanism for assessing the above parameters. To do so one needs as a prior distribution a bivariate distribution on the unit square, where the said distribution should encapsulate a negative dependence. Two such families of bivariate distributions are considered, one due to \cite{arng11} and the other, a new family based on a transformation of one of the variables in a bivariate beta family of distributions developed by \cite{olli03}. Both families of distributions share a common architecture, namely, the ratios of independent gamma distributions.

An analysis of some synthetically generated test diagnosis data is used to illustrate the workings of the approach proposed here. For purposes of comparison, independent priors for sensitivity and specificity are also considered. Whereas the analysis done here does not indicate a dramatically different inferential conclusion between a use of the independent and the negatively dependent priors, it does show that the latter have a edge over the former. More important, since sensitivity and specificity are known to be adversarial, {\it coherence} mandates that the priors encapsulate negative dependence. Of the two negatively dependent bivariate beta distributions considered here, the one by \cite{arng11} gives the sharper results, both with respect to closeness to the true values and narrowest bands of uncertainty.

In principle, all matters pertaining to diagnostics should be addressed from a decision-theoretic perspective. This would call for an assessment of utilities for a correct diagnosis, as well as for false diagnosis. This is a topic in its own right which remains to be addressed. There is a parallel between the work described here and that done by GJR (1991). The main difference lies not merely in the choice of the independent priors used by the above authors versus the negatively dependent priors used by us, but also in experimental design used to collect the data. GJR (1991) use retrospective (field) data and are therefore better grounded in practice than what is discussed here. The starting point for the above authors is the screening test data, whereas for us it is the confirmatory test data. In GJR (1991), confirmatory tests are then given to those who screen positive to see how effective the screening test is. In our case, it is the opposite; screening tests are given to those who are known to be diseased or not diseased. Both approaches suffer from the drawback that the first test given may not result in a diseased (or non-diseased) subject. Which of the two data generation approaches is optimum is another decision-theoretic issue that remains to be addressed. 

Clearly, there is more that needs to be done in the arena of diagnostics than the matter of inference about its adversarial parameters. Paramount among these remaining problems is the choice of the sample size, and the choice of a sampling plan. The proposed work has a broader impact in the sense that it could be of value whenever one needs to estimate competing propensities. 

\section{Amiable Inference in System Performance Assessment}

Consider a two-component series system, with $X^{*}_{i} = 1 (0)$, if the $i$-th component is surviving (failed) at time $t > 0$, $i = 1, 2$. Similarly, let $Y^{*} = 1 (0)$ if the entire system is surviving (failed) at $t$. Let $\tau$ be the now time, normally set equal to zero. Let ${\cal E}$ be the operating environment; we assume that ${\cal E}$ is static and is known to us. We also assume that failed components are not repaired, nor replaced. Let ${\cal H}$ represent all the background information about the components and the system that we have at time $\tau \geq 0$. Thus, ${\cal E}$ and ${\cal H}$ are the assumed known values of the set-up described above. The case of a two component parallel redundant system proceeds along the lines parallel to the ones given below with appropriate modifications. 

We are required, at time $\tau$, to specify our personal probability that at a future time $t > 0$, $Y^{*} = 1$; that is, the survivability of the system under the knowledge of ${\cal E}$ and ${\cal H}$. Formally, we need to access $\Pr^{\tau}(Y^{*} = 1; {\cal E} {\mbox { and }} {\cal H})$. An implicit, but important, point in reliability/survival analysis is that there are {\it two} time indices to keep in mind. One is the ``mission time" $t > 0$, and the other is the time of assessment $\tau \geq 0$. The assessment time is important, because as time marches on, ${\cal H}$ changes, and with an expanding ${\cal H}$, ones personal probability that $Y^{*} = 1$, may change as well. This feature has been carefully articulated by \cite{arja93} in a striking paper on information and reliability; also see \cite{arno84}. 

It is common to suppress $\tau$, ${\cal E}$ and ${\cal H}$, so that the explicit notation $\Pr^{\tau}(Y^{*} = 1; {\cal E} {\mbox { and }} {\cal H})$ is simply written $\Pr(Y^{*} = 1)$. The role of ${\cal E}$ is important, because ${\cal E}$ encapsulates the conditions under which a propensity occurs. With the above in mind, we may invoke the law of total probability by introducing a parameter $\theta \in (0, 1)$, and writing 
\begin{eqnarray}
\nonumber 
\Pr(Y^{*} = 1) & = & \Pr(X^{*}_{1} {\mbox { and }} X_{2}^{*} = 1) \\ 
 & = & \int\limits_{\theta} \Pr(X^{*}_{1} = 1 {\mbox { and }} X_{2}^{*} = 1 | \theta) \Pr(\theta) d\theta,
\label{II1}
\end{eqnarray} 
where $\theta$ is interpreted as the propensity of each component to survive to time $t$, and $\Pr(\theta)$ encapsulates our personal probability of $\theta$. 

The set-up of Equation (\ref{II1}) is meaningful when the two components are judged similar in some sense. For example, the components are randomly picked from a batch of identically manufactured units. This is made explicit by the feature that both $X_{1}^{*}$ and $X_{2}^{*}$ are conditioned on a {\it common} $\theta$. The role of ${\cal E}$ and ${\cal H}$ is to help one specify $\Pr(\theta)$ for which a natural model is a beta distribution with parameter $\alpha$ and $\beta$; that is, $\Pr(\theta)$ is ${\cal B}(\theta; \alpha, \beta)$. 

If the judgment that  $X_{1}^{*}$ and $X_{2}^{*}$ are in some sense similar cannot be justified, then we must introduce two parameters $\theta_{i} \in (0, 1)$, $i = 1, 2$, each interpreted as a propensity of its associated component surviving to $t$. We then invoke the law of total probability to write 
\begin{eqnarray}
\Pr(Y^{*} = 1) & = & \iint \limits_{(\theta_{1}, \theta_{2})} \Pr(X^{*}_{1} = 1 {\mbox { and }} X_{2}^{*} = 1 | \theta_{1}, \theta_{2}) \Pr(\theta_{1}, \theta_{2}) d\theta_{1} d \theta_{2},
\label{II2}
\end{eqnarray} 
where $\Pr(\theta_{1}, \theta_{2})$ encapsulates our joint uncertainty about $\theta_{1}$ and $\theta_{2}$; as before ${\cal E}$ and ${\cal H}$ help to guide its choice. 

The exact specification of $\Pr(\theta_{1}, \theta_{2})$ is facilitated by the materials of Section 2. However, whereas the focus in Section 3 was the adversarial character of $\eta$ and $\theta$, with system survivability, the propensity parameters $\theta_{1}$ and $\theta_{2}$ tends to be amiable. Consequently, our choices for $\Pr(\theta_{1}, \theta_{2})$ is the bivariate beta of \cite{olli03} with positive correlation, namely the $(OL)^{+}$ family, or the $AN(5)$ family of \cite{arng11} with parameters so chosen that the correlation is positive. The question we address is how these choices of distributions impact system survivability? We also need to compare such assessments with those that would be obtained under the assumption of independent propensities, namely, when $\Pr(\theta_{1}, \theta_{2}) = \Pr(\theta_{1}) \Pr(\theta_{2})$, and also when the two components are assumed to be similar; i.e., when $\theta_{1} = \theta_{2} = \theta$, and $\Pr(\theta)$ is ${\cal B}(\theta; \alpha, \beta)$. In making such comparisons, we must ensure fairness via a judicious choice of parameters. 

\subsection{The Case of Exchangeable Lifetimes} 

With reference to Equation (\ref{II1}), suppose that $X_{1}^{*}$ is (conditionally) independent of $X_{2}^{*}$ given $\theta$. Then, 
\begin{eqnarray}
\nonumber
\Pr(X_{1}^{*} = X_{2}^{*} = 1) & = & \int\limits_{\theta} \Pr(X^{*}_{1} = 1 | \theta) \Pr( X_{2}^{*} = 1 | \theta) \Pr(\theta) d\theta \\ 
& = & \int\limits_{\theta} \theta^{2} \Pr(\theta) d\theta = E(\theta^{2}) = \frac{\alpha \beta + \alpha^{2} (\alpha + \beta+ 1)}{(\alpha + \beta)^{2} (\alpha + \beta+ 1)},
\label{II4}
\end{eqnarray} 
when $\Pr(\theta)$ is ${\cal B}(\theta; \alpha, \beta)$. With $\alpha = \beta = 1$, the system's survivability is 1/3, whereas the survivability of each component, is the expected value of $\theta$, $E(\theta; \alpha, \beta) = 1/2$. 
We immediately notice that the answer for the system's survivability is not the conventional 1/4. This is because the components lifetimes $X_{1}^{*}$ and $X_{2}^{*}$ are generated by the same $\theta$, making $X_{1}^{*}$ and $X_{2}^{*}$ {\it exchangeable} in the sense of \cite{defi37}. Under what assumptions will  the conventional answer, namely 1/4, be justified? This matter is articulated next. 

\subsection{The Case of Hierarchically Independent Lifetimes} 

With reference to Equation (\ref{II2}), suppose that given $\theta_{1}$, $X_{1}^{*}$ is independent of both $X_{2}^{*}$ and $\theta_{2}$, and that given $\theta_{2}$, $X_{2}^{*}$ is independent of $\theta_{1}$. Then, we have 
\begin{eqnarray}
\Pr(Y^{*} = 1) & = & \iint \limits_{(\theta_{1}, \theta_{2})} \Pr(X^{*}_{1} = 1 | \theta_{1}) \Pr(X_{2}^{*} = 1 | \theta_{2}) \Pr(\theta_{1}, \theta_{2}) d\theta_{1} d \theta_{2}.
\label{II4}
\end{eqnarray} 

Suppose further that (in the light of ${\cal E}$ and ${\cal H}$), $\theta_{1}$ is independent of $\theta_{2}$. That is, a knowledge of the propensity of one component does not change ones assessment of the propensity of the other component. When such is the case $\Pr(\theta_{1}, \theta_{2}) = \Pr(\theta_{1}) \Pr(\theta_{2})$, and now Equation (\ref{II4}) becomes
\begin{eqnarray*}
\Pr(Y^{*} = 1) & = & \int\limits_{\theta_{1}} \Pr(X^{*}_{1} = 1 | \theta_{1}) P(\theta_{1}) d \theta_{1} \bcdot 
\int\limits_{\theta_{2}} \Pr(X^{*}_{2} = 1 | \theta_{2}) P(\theta_{2})  d \theta_{2}\\ \\
& = & E(\theta_{1}) E(\theta_{2}),
\end{eqnarray*} 
where $E(\theta_{i})$, is the survivability of component $i$, $i = 1, 2$. Now the system survivability is the product of the survivability of each component, and if $\theta_{i} \sim {\cal B}(\theta_{i}; \alpha_{i}, \beta_{i})$, $i = 1, 2$, then 
\begin{eqnarray}
\Pr(Y^{*} = 1) & = & \left(\frac{\alpha_{1}}{\alpha_{1} + \beta_{1}} \right) \left(\frac{\alpha_{2}}{\alpha_{2} + \beta_{2}} \right). 
\label{II5}
\end{eqnarray} 
When $\alpha_{1} = \alpha_{2} = \alpha$, and $\beta_{1} = \beta_{2} = \beta$, the above expression becomes $[\alpha/(\alpha + \beta)]^{2}$, which for $\alpha = \beta = 1$ yields 1/4, the conventional answer. 

Thus the conventional answer can only be justified under a {\it hierarchy} of independence assumptions. First, one assumes the independence of $\theta_{1}$ and $\theta_{2}$, and then the conditional independence of $X_{1}^{*}$ and $X_{2}^{*}$ given $\theta_{1}$ and $\theta_{2}$. This feature of hierarchical independence needs to be highlighted, not only in the reliability and the survival analysis literatures, but also in machine learning and classification. The latter go under the label of ``{\it na{\"i}ve Bayes}", or more colorfully ``{\it idiot's Bayes}" \citep[c.f.][]{hayu01}. In all cases, it is implicitly assumed that it is only the lifetimes $X_{i}$, $i = 1, 2$, that are (conditionally) independent. 

Recall that assessments of system survivability under the assumptions of exchangeability and of hierarchical independence yield $[\alpha/(\alpha + \beta)]^{2} + \alpha \beta/[(\alpha + \beta)^{2}(\alpha + \beta + 1)]$ and $[\alpha/(\alpha + \beta)]^{2}$, respectively, as answers. This means that for series system, hierarchical independence results in a conservative assessment of system survivability. The difference is the term $\alpha \beta/[(\alpha + \beta)^{2}(\alpha + \beta + 1)]$, and this also happens to be the variance of $\theta$, when $\theta \sim {\cal B}(\theta; \alpha, \beta)$. An appreciation of the cause of this difference can be had by looking at the hierarchical construction of the two models for survivability. This will be done later in Section 4.4, once we look at the case of system survivability under interdependent propensities, the scenario which has motivated the materials of Section 2 of this paper. 

\subsection{System Survivability Under Interdependent Propensities} 

Suppose that the assumptions of conditional independence of the lifetimes leading to Equation (\ref{II4}) continue to hold, but the propensities $\theta_{1}$ and $\theta_{2}$ cannot be judged independent. The dependence between $\theta_{1}$ and $\theta_{2}$ is captured in the choice of $\Pr(\theta_{1}, \theta_{2})$ vis-{\`a}-vis either the $(OL)^{+}$ model or the $AN(5)$ model. Thus, our expression for system reliability can be written as: 
\begin{eqnarray*}
\Pr(Y^{*} = 1) & = & \iint \limits_{(\theta_{1}, \theta_{2})} \theta_{1} \theta_{2} \Pr(\theta_{1}, \theta_{2}) d\theta_{1} d \theta_{2}\\ \\
& = & E(\theta_{1} \theta_{2}). 
\end{eqnarray*} 
To evaluate $E(\theta_{1} \theta_{2})$, we  use the relationship 
\begin{eqnarray}
E(\theta_{1} \theta_{2}) = \rho(\theta_{1}, \theta_{2}) \sqrt{ V(\theta_{1}) V(\theta_{2}) } + E(\theta_{1}) E(\theta_{2}),   
\label{II6}
\end{eqnarray}
where $\rho(\theta_{1}, \theta_{2})$ is the correlation between $\theta_{1}$ and $\theta_{2}$, and $V(\theta_{i})$ is the variance of $\theta_{i}$, $i = 1, 2$. Values of  $\rho(\theta_{1}, \theta_{2})$ for selected values of the parameters that go to construct the $(OL)^{+}$ and $AN(5)$ models have been tabulated by \cite{olli03} and by \cite{arng11}, respectively. 

Under the above set-up, the lifetimes $X_{1}^{*}$ and $X_{2}^{*}$ are {\it unconditionally} dependent. This is because $\theta_{1}$ and $\theta_{2}$ which spawns them are dependent. Intuitively, the strength of dependence between  $X_{1}^{*}$ and $X_{2}^{*}$ is a function of the strength of dependence between $\theta_{1}$ and $\theta_{2}$. The dependence is the strongest when a common $\theta$ generates  $X_{1}^{*}$ and $X_{2}^{*}$, because now $\theta_{1} = \theta_{2} = \theta$; thus the exchangeable case is a special case of interdependent propensities. The dependence vanishes when $\theta_{1}$ is independent of $\theta_{2}$. Otherwise, the strength of dependence is intermediate between those two extreme cases. An appreciation of the nature of dependence between $X_{1}^{*}$ and $X_{2}^{*}$ is by examining the illustrations below which show the hierarchical construction of the three scenarios for system survivability discussed above. 

\subsection{Conceptualization of Hierarchically Constructed Lifetimes}

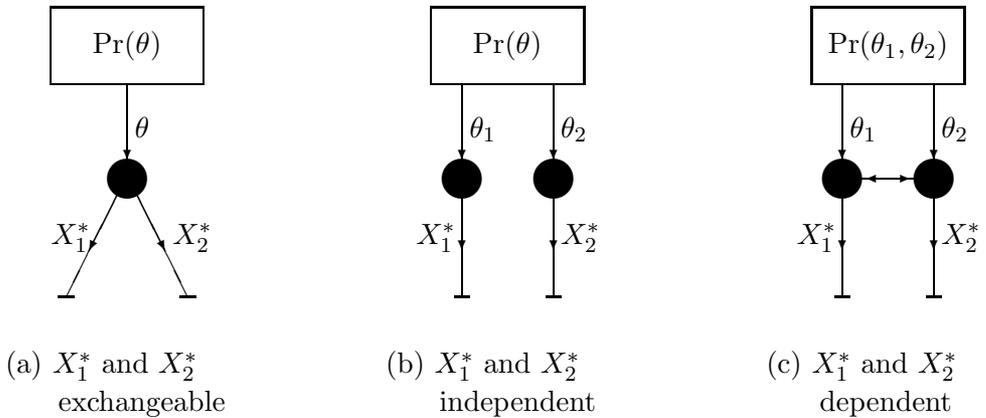
\begin{figure}[h]
\begin{center}
\setlength{\unitlength}{1mm}
\begin{picture}(150,70)(0, 0)
\put(20,50){\framebox(20,10){$\Pr(\theta)$}}
\put(30,50){\vector(0,-1){10}}
\put(31,43){$\theta$}
\put(30,37.5){\circle*{5.5}}
\put(30,38){\line(1,-2){8}}
\put(30,38){\vector(1,-2){5}}
\put(30,38){\line(-1,-2){8}}
\put(30,38){\vector(-1,-2){5}}
\put(20,29){$X_{1}^{*}$}
\put(36,29){$X_{2}^{*}$}
\put(21,22){\line(1,0){2}}
\put(37,22){\line(1,0){2}}
\put(14,12){(a) $X_{1}^{*}$ and $X_{2}^{*}$}
\put(21,7){exchangeable}
\put(70,50){\framebox(20,10){$\Pr(\theta)$}}
\put(74,50){\vector(0,-1){10}}
\put(86,50){\vector(0,-1){10}}
\put(75,43){$\theta_{1}$}
\put(87,43){$\theta_{2}$}
\put(74,37.5){\circle*{5.5}}
\put(86,37.5){\circle*{5.5}}
\put(74,38){\line(0,-1){16}}
\put(74,38){\vector(0,-1){10}}
\put(86,38){\line(0,-1){16}}
\put(86,38){\vector(0,-1){10}}
\put(68,29){$X_{1}^{*}$}
\put(87,29){$X_{2}^{*}$}
\put(73,22){\line(1,0){2}}
\put(85,22){\line(1,0){2}}
\put(64,12){(b) $X_{1}^{*}$ and $X_{2}^{*}$}
\put(71,7){independent}
\put(120,50){\framebox(20,10){$\Pr(\theta_{1}, \theta_{2})$}}
\put(124,50){\vector(0,-1){10}}
\put(136,50){\vector(0,-1){10}}
\put(125,43){$\theta_{1}$}
\put(137,43){$\theta_{2}$}
\put(124,37.5){\vector(1,0){9}}
\put(136,37.5){\vector(-1,0){9}}
\put(124,37.5){\circle*{5.5}}
\put(136,37.5){\circle*{5.5}}
\put(124,38){\line(0,-1){16}}
\put(124,38){\vector(0,-1){10}}
\put(136,38){\line(0,-1){16}}
\put(136,38){\vector(0,-1){10}}
\put(118,29){$X_{1}^{*}$}
\put(137,29){$X_{2}^{*}$}
\put(123,22){\line(1,0){2}}
\put(135,22){\line(1,0){2}}
\put(114,12){(c) $X_{1}^{*}$ and $X_{2}^{*}$}
\put(121,7){dependent}
\end{picture}
\end{center}
\caption{Hierarchical Lifetime Construction.}
\label{fig2} 
\end{figure}

Figure \ref{fig2} provides a visualization of constructing the lifetime of each component under the three scenarios of exchangeability, independence and a dependence that is intermediate to that produced under exchangeability and independence. They describe the different mechanisms via which the two lifetimes are generated. 

In Figure \ref{fig2}(a), one generates a single $\theta$ from the distribution $\Pr(\theta)$, and then using this $\theta$ as a parameter, one generates two independent Bernoulli variable $X_{1}^{*}$ and $X_{2}^{*}$. Because $X_{1}^{*}$ and $X_{2}^{*}$ are based on a common seed, namely $\theta$, they are positively dependent; indeed exchangeable. Conditioned on $\theta$, they are independent. In Figure \ref{fig2}(b), one starts with the same $\Pr(\theta)$ as in Figure \ref{fig2}(a), and generates two independent parameters $\theta_{1}$ and $\theta_{2}$. Next, using $\theta_{1}$ as a parameter, one generates a Bernoulli variable $X_{1}^{*}$, and then using $\theta_{2}$ as a parameter, one generates independently (of $X_{1}^{*}$), another Bernoulli variable $X_{2}^{*}$. Here $X_{1}^{*}$ and $X_{2}^{*}$ are unconditionally independent. In Figure \ref{fig2}(c), one starts with a non-trivial joint distribution of $\theta_{1}$ and $\theta_{2}$, namely $\Pr(\theta_{1}, \theta_{2})$, and then generates the parameter vector $(\theta_{1}, \theta_{2})$. The two-sided arrow linking $\theta_{1}$ and $\theta_{2}$ reveals the fact that $\theta_{1}$ and $\theta_{2}$ are dependent. Next, using $\theta_{1}$ as a parameter, one generates a Bernoulli variable $X_{1}^{*}$, and then using $\theta_{2}$ as the parameter, one generates independently (of $X_{1}^{*}$) a second Bernoulli variable $X_{2}^{*}$. Here, $X_{1}^{*}$ and $X_{2}^{*}$ are unconditionally dependent because $\theta_{1}$ and $\theta_{2}$ are linked. Contrast this with the architecture of Figure \ref{fig2}(b) wherein there is no link between $\theta_{1}$ and $\theta_{2}$. In Figure \ref{fig2}(a), the link between $\theta_{1}$ and $\theta_{2}$ is so strong, that indeed $\theta_{1}$ and $\theta_{2}$ collapse to a single $\theta$. 

In all three constructions described above, $X_{1}^{*}$ and $X_{2}^{*}$ are, given their respective parameters, independent. An interesting generalization of the above constructive scheme would be to generate conditionally dependent Bernoulli vector ($X_{1}^{*}$ and $X_{2}^{*}$), and investigate the survivability of the ensuing series system. 

\subsection{Comparison of Survivability of Series Systems}

In this section, we explore the consequences of assuming the exchangeable, the $(OL)^{+}$, and the $AN(5)$ versions of the interdependent propensities on series system survivability. Our aim is to see if it is possible to draw some general conclusions based on an examination of the empirical results. 

We start with the exchangeable case with $\theta \sim {\cal B}(\alpha, \beta)$, and different choices of $\alpha$ and $\beta$. Recall that here $E(\theta) = \alpha/(\alpha + \beta)$, $V(\theta) = \alpha \beta/[(\alpha + \beta)^{2}(\alpha + \beta + 1)]$, and $\Pr(Y^{*} = 1)$ -- system survivability -- is $[\alpha \beta + \alpha^{2}(\alpha + \beta + 1)]/[(\alpha + \beta)^{2}(\alpha + \beta + 1)]$. Table 4 summarizes the results.

\begin{table} 
\caption{The Case of Exchangeable Lifetimes with ${\cal B}(\theta; \alpha, \beta)$.}  
\begin{tabular}{c c c c}  \hline 
Values of             & $E(\theta)$: Component     & $V(\theta)$: Variance  &  Series System \\ 
$(\alpha, \beta)$     & Survivability              & of $\theta$            &  Survivability \\  \hline 
(1, 1)    & 0.50 & 0.080 & 0.333 \\ 
(3, 1)    & 0.75 & 0.037 & 0.600 \\ 
(10.1, 1) & 0.90 & 0.006 & 0.835 \\ 
(5.5, 6)  & 0.90 & 0.013 & 0.836 \\ 
(3, 0.3)  & 0.90 & 0.020 & 0.845 \\ 
(1, 0.1)  & 0.90 & 0.040 & 0.866 \\ \hline         
\end{tabular} 
\end{table} 

In Table 5, we give results when the interdependence between $\theta_{1}$ and $\theta_{2}$ is described by the $(OL)^{+}$ distribution with the parameters $\alpha_{1}$, $\alpha_{2}$ and $\alpha_{3}$ so chosen that there is a match between the marginal distributions $\theta_{i}$, $i = 1, 2$, and the distribution of $\theta$ in Table 4. 

\begin{table} 
\caption{The Case of $(\theta_{1}, \theta_{2})$ Having the $(OL)^{+}$ Distribution.} 
\begin{tabular}{c c c c c}  \hline 
Distribution of             & $E(\theta_{i})$: Survivability  & $V(\theta)$: Variance  &  $\rho(\theta_{1}, \theta_{2})$:                    &  Series System \\
$\theta_{i}$, $i = 1, 2$    & of component $i$                & of $\theta_{i}$        &  Correlation   &  Survivability \\  \hline 
${\cal B}(1, 1)$    & 0.50 & 0.080 & 0.478 & 0.290 \\ 
${\cal B}(3, 1)$    & 0.75 & 0.037 & 0.861 & 0.595 \\ 
${\cal B}(3, 0.3)$  & 0.90 & 0.020 & 0.859 & 0.845 \\ 
${\cal B}(1, 0.1)$  & 0.90 & 0.040 & 0.681 & 0.855 \\ \hline         
\end{tabular} 
\end{table} 

An examination of the entries in the last four rows of Table 4 suggests that the survivability of the series system increases with the variance of $\theta$. The same is also true when we look at the last two rows of Table 5, and this is despite the fact that the entry for $\rho(\theta_{1}, \theta_{2})$ in row three is larger than the corresponding entry in row four. 

Recall that the survivability of a two-component series system when $\theta_{i} \sim {\cal B}(1, 1)$, $i = 1, 2$, and $\theta_{1}$ and $\theta_{2}$ independent, is 0.25 -- see Equation (\ref{II5}). Comparing this 0.25 with entries in the last column of Tables 4 and 5 suggests that all else being equal, series system survivability increases with the strength of dependence between $\theta_{1}$ and $\theta_{2}$. 

Moving on to the scenario wherein the interdependence between $\theta_{1}$ and $\theta_{2}$ is described by an $AN(5)$ model, suppose that $\alpha_{1} = \alpha_{2} = \alpha_{5} = 10$, and $\alpha_{3} = \alpha_{4} = 0.1$. Then, $\theta_{1} \sim {\cal B}(10.1, 10.1)$ and $\theta_{2} \sim {\cal B}(10.1, 10.1)$, so that $E(\theta_{i}) = 0.5$, $V(\theta_{i}) = 0.012$, $i = 1, 2$, and from Table 1 of \cite{arng11}, $\rho(\theta_{1}, \theta_{2}) = 0.484$. These entities yield 0.255 as the system survivability. The closest match to the above choices of $E(\theta_{i})$, $i = 1, 2$, and $\rho(\theta_{1}, \theta_{2})$, is the first row of Table 5, for which the value of the system survivability is 0.29. This result could prompt the claim that an $AN(5)$ model for interdependence may yield conservative assessments of system survivability as compared to a suitably matched $(OL)^{+}$ model. But any such claim should be tempered by the fact that in the $(OL)^{+}$  case, $V(\theta_{i})$, $i=1, 2$, is 0.08, whereas its value in a corresponding $AN(5)$ case is 0.012, and our previous conjecture that the greater the marginal variance, the larger is the survivability. 

To affirm if the claim of conservative survivability assessment is generally true in all cases entailing the $AN(5)$ distribution, we consider two other combinations of values of the $\alpha_{i}$'s, $i = 1,\ldots, 5$. Specifically, setting $\alpha_{1} = \alpha_{2} = 10$, $\alpha_{3} = \alpha_{4} = 0.1$, and $\alpha_{5} = 1$, yields $E(\theta_{1}) = E(\theta_{2}) = 0.9$, $V(\theta_{1}) = V(\theta_{2}) = 0.007$, and $\rho(\theta_{1}, \theta_{2}) = 0.755$; this is a close match to row three of Table 5. By contrast, setting $\alpha_{1} = 5$, $\alpha_{2} = 10$, $\alpha_{3} = \alpha_{4} = 0.1$ and  $\alpha_{5} = 0.5$, provides a close match to row four of Table 5. The only caveat here is that the marginal distributions of $\theta_{1}$ and $\theta_{2}$ are different; see rows three and four of Table 6, which summarizes the consequences of these choices. A match for row two of Table 5 has not been considered. 

\begin{table} 
\caption{The Case of $(\theta_{1}, \theta_{2})$ Having the $AN(5)$ Distribution.} 
\begin{tabular}{c c c c c}  \hline 
Distribution of             & $E(\theta_{i})$: Survivability  & $V(\theta)$: Variance  &  $\rho(\theta_{1}, \theta_{2})$:                    &  Series System \\
$\theta_{i}$, $i = 1, 2$  & of component $i$                & of $\theta_{i}$        &  Correlation   &  Survivability \\  \hline 
${\cal B}(10.1, 10.1)$    & 0.50 & 0.012 & 0.484 & 0.255 \\ 
${\cal B}(10.1, 1.1)$     & 0.90 & 0.007 & 0.755 & 0.815 \\ 
${\cal B}(5.1, 0.6)$      & 0.89 & 0.014 & 0.675 & 0.840 \\ 
${\cal B}(10.1, 0.6)$     & 0.94 & 0.004 & 0.675 & 0.840 \\ \hline         
\end{tabular} 
\end{table} 

\subsection{Discussion and Summary Remarks on Section 4} 

Comparing the system survivability entry of row two of Table 6 with that of row four of Table 5, and row three (four) of Table 6 with that of row four of Table 5, reaffirms the claim, that all things being equal, the choice of an $AN(5)$ model for the interdependence between the propensities $\theta_{1}$ and $\theta_{2}$ results in conservative assessments of system survivability as compared to the $(OL)^{+}$ model. But this model could be a feature of the fact that the $AN(5)$ distribution tends to be more concentrated than the $(OL)^{+}$ distribution. From the viewpoint of safety and risk, conservative assessments of survivability may be more desirable than those that tend to be on the optimistic side. Thus, the choice of what joint distribution to use, the  $(OL)^{+}$ or the $AN(5)$ depends, in the case of a series system, to ones disposition to the risk of system failure. The $AN(5)$ bivariate beta distribution being more concentrated than a corresponding matched $(OL)^{+}$ distribution tends to yield more conservative assessment of series system reliability. 

The essential import of the material of Section 4 is two-fold. The first is approach for incorporating positive dependence for the survivability assessment of a (two-component) series system using bivariate beta distributions on a unit square. The extension to parallel systems is relatively straightforward. Generalization to multi-component systems and networks is a challenge that remains to be addressed. For this, an essential first step would be to develop families of joint distributions (with interdependence) on a unit cube. Alternatively, one could modularize a system in terms of independent pairs, each pair being a series (parallel) system of interdependent lifetimes. 

The second import of this section is a closer look into the architecture of dependent and independent lifetimes; see Figure \ref{fig2}. The key point here is that any assumption of independent lifetimes has to be justified, at a minimum, via a two-stage argument. This latter point may not have been made transparent in the past.

\section*{Acknowledgements} 

Professor Vijay Nair, the Editor, hit the nail on the head by advising us how to re-organize and streamline the paper; deeply appreciated. The insightful comments of Referee III helped improve  the technical content of the paper. Referee II, though unsupportive of this work, ended up making valuable comments which enabled added verbiage to improve clarity. The positive comments of Referee I who elegantly summarized the gist of the paper and suggested a sharper focus are deeply appreciated. Overall, the review process was both a pleasurable and productive experience. 

Nozer D. Singupurwalla's work was supported by The US Army Research Grant W911NF-09-1-0039, The US National Science Foundation Grant DMS-09-15156, The City University of Hong Kong [Project Numbers 9380068 and 9042083 (Hong Kong GRF)]. 
Hon Keung Tony Ng's work was supported by  a grant from the Simons Foundation (\#280601 to Tony Ng).


\section*{Appendix: Generalized Bivariate Beta: Closure Under Complementation}

For $n = 8$, \cite{arng11} introduced the $AN(8)$ distribution discussed below. However, its closure properties and their relevance for inference in diagnostics were not explored. This section pertains to a discussion of these matters, and in a sense is a contribution to the applied probability aspect of this paper. 

We start our discussion about closure by noting that since $X = U_{1}/(U_{1} + U_{3}) \sim {\cal B}(\alpha_{1}, \alpha_{3})$, $1 - X = U_{3}/(U_{1} + U_{3})$ will also be ${\cal B}(\alpha_{3}, \alpha_{1})$; thus the univariate beta is closed under complementation. However, the univariate beta distribution of the first kind is not closed under reciprocation, because $1/X = 1 + U_{3}/U_{1}$, and $U_{3}/U_{1}$ has what is known as an {\it inverted beta distribution} (or a {\it beta distribution of the second kind}) with parameter $(\alpha_{1}, \alpha_{3})$, hence-forth, denote ${\cal B}^{2}(\alpha_{1}, \alpha_{3})$. In general, the beta distribution of the second kind with parameters $\alpha$ and $\beta$ is the distribution of the ratio of two independent gamma distributed random variables with shape parameters $\alpha$ and $\beta$, respectively, and a common scale parameter. Contrast the  beta distribution of the second kind  with the beta distribution of the first kind which is the ratio of two dependent gamma distributed random variables. The probability density at $x$ of a  beta distribution of the second kind  with parameters $\alpha$ and $\beta$ is 
\begin{eqnarray*} 
f(x; \alpha, \beta) = \frac{\Gamma(\alpha + \beta)}{\Gamma(\alpha) \Gamma(\beta)} x^{\alpha - 1} (1 + x)^{-(\alpha + \beta)}, ~x \geq 0; 
\end{eqnarray*} 
here $\Gamma(\alpha) = \int_{0}^{\infty} s^{\alpha-1} e^{-s} ds$, is the gamma function. Note that since $X = U_{1}/(U_{1} + U_{3}) \sim {\cal B}(\alpha_{1}, \alpha_{3})$, $X/(1 - X) = U_{1}/U_{3} \sim {\cal B}^{2}(\alpha_{1}, \alpha_{3})$, and that $(1 - X)/X = U_{3}/U_{1} \sim {\cal B}^{2}(\alpha_{3}, \alpha_{1})$. Thus, the beta distribution of the second kind is {\it closed} under reciprocation. The above observations enable us to argue that the $AN(8)$ distribution is closed under complementation. But before doing so, it is helpful to give a summarized perspective on the closure properties of the multivariate beta distribution. 

Since the marginal distributions of the $(OL)^{+}$ and the $(OL)^{-}$ family of bivariate beta distributions are a univariate beta, we say that the $(OL)^{+}$ and the $(OL)^{-}$ family of distributions are {\it closed} under {\it marginal complementation}. The same is also true of the joint distribution of $(1-X, 1-Y)$, the bivariate complementation of $(OL)$, henceforth denoted $(OL)^{*}$. 
By complementation of $X$ we mean a consideration of $(1 - X)$. 
A comparison of the joint probability density functions of the $(OL)^{+}$ and the $(OL)^{-}$   distributions shows that they are different. Similarly, with a comparison of the joint probability density functions of $(OL)^{+}$ and $(OL)^{-}$, and $(OL)^{+}$ and $(OL)^{*}$. Such comparisons lead to the claim that the $(OL)^{+}$, $(OL)^{-}$, and the $(OL)^{*}$ are {\it not closed} under {\it coordinate-wise and joint complementation}. Since the $AN(5)$ family of distributions encompasses the $(OL)^{+}$ family, we claim that like the $(OL)^{+}$ family, the $AN(5)$ is  {\it not closed} under {\it coordinate-wise} and {\it joint complementation}.

For $n = 8$, the $AN(8)$ construction begins with random variables $V$ and $W$, where 
\begin{eqnarray*}
V = \frac{U_{1} + U_{5} + U_{7}}{U_{3} + U_{6} + U_{8}}~~{\mbox {and }} W = \frac{U_{2} + U_{5} + U_{8}}{U_{4} + U_{6} + U_{7}}. 
\end{eqnarray*}
Since $V$ and $W$ are ratios of independent gamma distributed random variables, their distributions are beta distributions of the second kind, but since $V$ and $W$ share common elements, their joint distribution is a bivariate beta distribution of the second kind, denoted ${\cal B}^{2}(2)[\alpha_{1}, \ldots, \alpha_{8}]$ or simply ${\cal B}^{2}(2)[\balpha]$. It is easy to see that this bivariate beta distribution of the second kind is {\it closed} under {\it coordinate-wise} and {\it joint complementation}. That is, 
\begin{eqnarray*}
(V, W) \sim {\cal B}^{2}(2)[\balpha] \Longrightarrow (V, W^{-1}) \sim  (V^{-1}, W) \sim  (V^{-1}, W^{-1}) \sim {\cal B}^{2}(2)[\balpha]. 
\end{eqnarray*}
If we now let 
\begin{eqnarray*}
X_{2} & = & \frac{V}{1+V} = \frac{U_{1} + U_{5} + U_{7}}{U_{1} + U_{3} + U_{5} + U_{6} + U_{7} + U_{8}},  \\
{\mbox {and }} 
Y_{2} & = & \frac{W}{1+W} = \frac{U_{2} + U_{5} + U_{8}}{U_{2} + U_{4} + U_{5} + U_{6} + U_{7} + U_{8}},  
\end{eqnarray*}
then the ${\cal B}^{2}(2)[\balpha]$ distribution of $(V, W)$ implies that the joint distribution of $(X_{2}, Y_{2})$ is a bivariate beta, namely, ${\cal B}(2)[\balpha]$. This $AN(8)$ distribution is indeed the generalized bivariate beta $(GBB)$ which titles this section. 

It is easy to see that the marginal distributions of $X_{2}$ and $Y_{2}$ are beta distributions. The $(GBB)$ distribution includes the $(OL)^{+}$, the $(OL)^{-}$, and the $(OL)^{*}$ as special cases, each encompassing the non-negative, the non-positive, and the non-negative dependencies, respectively. This is because the $(OL)^{+}$ family is obtained by setting $\alpha_{3} = \alpha_{4} = \alpha_{5} = \alpha_{7} = \alpha_{8} = 0$, the $(OL)^{-}$ family by setting $\alpha_{2} = \alpha_{3} = \alpha_{5} = \alpha_{6} = \alpha_{7} = 0$, or by setting $\alpha_{1} = \alpha_{4} = \alpha_{5} = \alpha_{6} = \alpha_{8} = 0$, and the $(OL)^{*}$ family by setting $\alpha_{1} = \alpha_{2} = \alpha_{6} = \alpha_{7} = \alpha_{8} = 0$. 

To see why the $(GBB)$ family is {\it closed} under {\it coordinate-wise complementation}, let $Y_{2}^{*} = (1 - Y_{2})$, and note that 
\begin{eqnarray*}
(X_{2}, Y_{2}^{*}) & = & \left(\frac{V}{1 + V}, 1 - \frac{W}{1 + W} \right)\\ 
& = &  \left(\frac{V}{1 + V}, \frac{1}{1 + W} \right) = \left(\frac{V}{1 + V}, \frac{W^{-1}}{1 + W^{-1}} \right), 
\end{eqnarray*}
and this has a bivariate beta distribution because, as noted before, $(V, W^{-1}) \sim {\cal B}^{2}(2)[\balpha]$, and the result that if $Z = X/(1 + X) \sim {\cal B}^{2}(\alpha, \beta)$, then $X = Z/(1 + Z) \sim {\cal B}(\alpha, \beta)$. Using an analogous argument considering $X_{2}^{*} = 1 - X_{2}$, we can see that $(X_{2}^{*}, Y_{2}^{*})$ is also a bivariate beta, and so we may claim that the $AN(8)$ distribution is {\it closed} under both {\it coordinate-wise} and {\it joint complementation}. 

We, therefore, have at hand an eight-parameter family of bivariate beta distributions, abbreviated $(GBB)$, that is closed under complementation and which encompasses both positive and negative dependence (i.e., takes correlations in the full range $[-1, 1]$). It includes the $(OL)^{+}$, the $(OL)^{-}$ and the $(OL)^{*}$  distributions as special cases, and serves as a family of prior distributions for adversarial and amiable inference in a Bayesian setting. Its constructive definition based on ratios of independent gamma variates facilitates ease of simulation. 
Although the $AN(8)$ model encompasses the $AN(5)$ model and the other bivariate beta models, it may be criticized on grounds of parsimony.

Despite the fact that the $AN(8)$ family of distributions has not been used by us in the context of inference in diagnostics, the inclusion of this Appendix, which can be seen as a stand alone piece, is for methodological completeness.

\end{document}